\documentclass[12pt]{article}
\usepackage[pdftex]{graphicx}
\usepackage{amsmath}

\usepackage{amsmath}
\usepackage[svgnames,hyperref]{xcolor}
\usepackage{empheq}
\usepackage[svgnames,hyperref]{xcolor}
\usepackage{bm}
\usepackage{caption}
\usepackage[font=footnotesize,labelfont=bf]{caption}
\usepackage{algorithm}
\usepackage[noend]{algpseudocode}
\usepackage{calrsfs}
\usepackage[svgnames,hyperref]{xcolor}
\usepackage[svgnames,hyperref]{xcolor}
\usepackage{hyperref}
\usepackage{caption}
\usepackage{algorithm}
\usepackage[noend]{algpseudocode}
\usepackage{calrsfs}
\usepackage{amssymb}
\usepackage{color,wrapfig}
\usepackage{soul}
\usepackage{dirtytalk}

\usepackage[super,compress,comma]{natbib}

\setcitestyle{authoryear,open={(},close={)},sort} 

\usepackage{appendix}
\usepackage{titlesec}

\usepackage{moresize}
\usepackage{soul}
\usepackage{color}

\usepackage{subcaption}
\newcommand{\figuremacro}[5]{
    \begin{figure}[#1]
        \centering
        \includegraphics[width=#5\columnwidth]{#2}
        \caption[#3]{\textbf{#3}#4}
        \label{fig:#2}
    \end{figure}
}
\newcommand{\tablemacro}[5]{
    \begin{table}[#1]
        \centering
        \caption[#3]{\textbf{#3}#4}
        \includegraphics[width=#5\columnwidth]{#2}
        \label{tab:#2}
    \end{table}
}

\begin{document}

\pagenumbering{arabic}

\begin{center}
{\large \bf{A foundational framework for the mesoscale modeling of dynamic elastomers and gels}}
\end{center}

\renewcommand*{\thefootnote}{\fnsymbol{footnote}}

\begin{center}
 {\small \textbf{Robert J. Wagner$^1$\footnote{Correspondence to: Robert.J.Wagner@Binghamton.edu} \& Meredith N. Silberstein$^2$} \\ 
 \ssmall $^1$Department of Mechanical Engineering, State University of New York at Binghamton, Binghamton, NY, USA  \\
 \ssmall $^2$Sibley School of Mechanical \& Aerosspace Engineering, Cornell University, Ithaca, NY, USA
}
\end{center}
\renewcommand*{\thefootnote}{\arabic{footnote}}
\setcounter{footnote}{0}

\textit{Discrete mesoscale network models, in which explicitly modeled polymer chains are replaced by implicit pairwise potentials, are capable of predicting the macroscale mechanical response of polymeric materials such as elastomers and gels, while offering greater insight into microstructural phenomena than constitutive theory or macroscale experiments alone. However, whether such mesoscale models accurately represent the molecular structures of polymer networks requires investigation during their development, particularly in the case of dynamic polymers that restructure in time. We here introduce and compare the topological and mechanical predictions of an idealized, reduced-order mesoscale approach in which only tethered dynamic bonding sites and crosslinks in a polymer's backbone are explicitly modeled, to those of molecular theory and a Kremer-Grest, coarse-grained molecular dynamics approach. We find that for short chain networks ($\sim$12 Kuhn lengths per chain segment) at intermediate polymer packing fractions, undergoing relatively slow loading rates (compared to the monomer diffusion rate), the mesoscale approach reasonably reproduces the chain conformations, bond kinetic rates, and ensemble stress responses predicted by molecular theory and the bead-spring model. Further, it does so with a 90\% reduction in computational cost. These savings grant the mesoscale model access to larger spatiotemporal domains than conventional molecular dynamics, enabling simulation of large deformations as well as durations approaching experimental timescales (e.g., those utilized in dynamic mechanical analysis). While the model investigated is for monodisperse polymer networks in theta-solvent, without entanglement, charge interactions, long-range dynamic bond interactions, or other confounding physical effects, this work highlights the utility of these models and lays a foundational groundwork for the incorporation of such phenomena moving forward.}

\section{Introduction}

Synthetic polymers are one of the most diversely applied classes of materials available to engineers. Due to their extensive sets of design parameters (e.g.,  molecular weight; polydispersity; composition; crosslink concentration/type; targeted incorporation of supramolecular interactions; swelling with solvent; etc.), polymers exhibit a broad set of emergent mechanical structures and properties appropriate for diverse applications including in structural composites \citep{chung_review_2019}, renewable energy systems \citep{rydz_chapter_2021}, packaging \citep{vallejos_classical_2022,ibrahim_need_2022}, adhesives \citep{heinzmann_supramolecular_2016}, and biomaterials \citep{chen_biomedical_2022}. A particular trait of interest in polymeric design is the inclusion of dynamic bonds. Polymers with dynamics bonds include covalently adaptable polymers \citep{kloxin_covalent_2010} such as vitrimers \citep{yue_vitrimerization_2020,shen_nonsteady_2021,hubbard_vitrimer_2021,hubbard_creep_2022}, and supramolecular polymers \citep{brunsveld_supramolecular_2001,yount_strong_2005} such as physically bonded elastomers and gels \citep{vidavsky_tuning_2020,mordvinkin_rheology_2021,xu_thermosensitive_2022}. The dissociation of such bonds from highly stressed states, followed by their re-attachment into lower energy configurations is a dissipative mechanism that can grant dynamic polymers exceptional toughness \citep{haque_lamellar_2011,gong_high-strength_2016,bai_fatigue_2018,li_role_2022}, extensibility \citep{tuncaboylu_tough_2011,jeon_extremely_2016,zhang_superstretchable_2019,cai_highly_2022}, self-healing capabilities \citep{kersey_hybrid_2006,brochu_self-healing_2011,li_highly_2016,li_shape-memory_2021}, and even mechanosensitive response \citep{wojtecki_using_2011,eom_mechano-responsive_2021,doolan_next-generation_2023}. Furthermore, the additional design parameters introduced by the inclusion of dynamic bonds (e.g., tether length for telechelic bonds \citep{ge_strong_2015,mordvinkin_rheology_2021}, coordination number of charged species in metallopolymers \citep{yount_small-molecule_2005,zhang_bridging_2020,vidavsky_tuning_2020}, etc.) provide polymers with enhanced tunability. For instance, simply changing the molar concentration or type of dynamically bonding species in a polymer may yield elastic moduli spanning multiple orders of magnitude \citep{xu_thermosensitive_2022,huang_structural_2022}. Therefore, predictive models that can map the highly varied emergent mechanical properties of such polymers from their \textit{ab initio} compositions (prior to synthesis and experimental testing) may greatly benefit researchers developing new materials requiring certain traits. However, the same rich sets of constituents and compositional design choices responsible for dynamic polymers' highly varied properties also imbue them with hierarchical length scales and relaxation timescales spanning from $10^{-10}$ to $10^{-3}$ m and  $10^{-15}$ to $10^{6}$ s, respectively, making this mapping process challenging. 

At the macroscale -- well above the thermodynamic limit and length scales of heterogeneities -- continuum approaches \citep{james_theory_1943,flory_molecular_1985,tanaka_viscoelastic_1992,arruda_three-dimensional_1993,wu_improved_1993,miehe_micro-macro_2004,vernerey_statistically-based_2017} are suitable for mechanical property predictions, provided proper implicit representation of the pertinent first order physics such as single-chain mechanics \citep{james_theory_1943,miehe_micro-macro_2004,saleh_perspective_2015}, swelling \citep{flory_thermodynamics_1942,hong_inhomogeneous_2009,bouklas_swelling_2012}, bond dynamics \citep{leibler_dynamics_1991,stukalin_self-healing_2013}, macroscopic damage \citep{bouklas_effect_2015,shen_rate-dependent_2020,lee_finite_2023}, etc. 
Although such approaches have been used extensively to successfully predict polymeric properties and indirectly deduce microstructural origins of observed traits with a wide breadth of physical effects \citep{xu_thermosensitive_2022,bosnjak_modeling_2022}, they generally rely on some combination of homogenization assumptions, affine deformations, and mean-field approximations \citep{cpicu_mechanics_2011}. As a result, they are limited in their ability to directly map composition to microstructure, and then microstructure to global mechanical properties, as is needed for the predictive design of newly developed materials. 

With the advent and increasing prevalence of powerful computational resources, as well as the broadening accessibility of open-source software, researchers have been able to circumvent the need for extensive homogenization by using high-fidelity, discrete modeling approaches such as molecular dynamics (MD) \citep{bergstrom_deformation_2001,somasi_brownian_2002,doyle_brownian_2005} in frameworks such as LAMMPS \citep{thompson_lammps_2022}. Such methods can easily accommodate dynamic bond kinetics via either deterministic activation of bonds due to heuristic rules \citep{goodrich_enhanced_2018} or, more commonly, incorporation of Monte Carlo methods that randomly sample the formation or dissociation of dynamic bonds based on transition state theory \citep{evans_dynamic_1997,hoy_thermoreversible_2009,stukalin_self-healing_2013,amin_dynamics_2016,perego_volumetric_2020,zhao_molecular_2022,wagner_catch_2024-1}. MD simulations have proven valuable for linking microstructure to bulk properties of dynamic polymers, and have recently been used to study features such as enhanced transport \citep{goodrich_enhanced_2018,huang_dynamic_2023,taylor_smoother_2024}, microstructural relaxation \citep{yang_molecular_2015,amin_nonlinear_2020}, self-healing \citep{zheng_molecular_2021}, and polymer reprocess-ability \citep{zhao_molecular_2022}. However, the number of discrete particles that must be modeled to capture representative volume elements (RVE) of dynamic polymers using MD, combined with the small vibrational timescales of said particles (on the order of picoseconds) restricts such models to nanometer and nanosecond domains \citep{agrawal_prediction_2016,liu_accurate_2020,zhang_understanding_2023}. These spatiotemporal scales regularly reach $10^2$ nanometers and $10^2$ nanoseconds with the use well-established coarse-graining methods such as bead-spring representation of Kuhn segments in a polymer chain \citep{kremer_dynamics_1990,somasi_brownian_2002} or Brownian dynamics \citep{doyle_brownian_2005} to implicitly model solvent. Yet, the computational cost of such simulations remains high and there still exists several spatiotemporal orders of magnitude between the molecular scales of chemical structure (i.e., {\AA}ngstroms to nanometers and femtoseconds to nanoseconds), and the macroscale at which experiments are mostly conducted and continuum models are commonly applied (e.g., upwards of millimeters and milliseconds). 

To address this spatiotemporal gap, many researchers have begun investigating polymeric materials at $10^2$ to $10^3$ nanometer scales using of a class of ``discrete network models" (DNMs). In these models, crosslinks within polymer networks are explicitly represented as ``nodes", while the constitutive chains that link them together are captured implicitly using statistically derived force-extension relations \citep{hernandez_cifre_brownian_2003,sugimura_mechanical_2013,kothari_mechanical_2018,wagner_network_2021,wyse_jackson_structural_2022}. Since DNMs avoid representing every atom or lumped Kuhn segment in a polymer chain, they reduce the number of explicitly tracked particles and therefore computational cost of simulated polymer networks. Vernerey and coworkers have recently developed one such set of DNMs for dynamic polymers that reproduce the viscoelastic mechanical predictions of Transient Network Theory (TNT) \citep{vernerey_statistically-based_2017,wagner_network_2021}, as well as the experimentally measured mechanical stresses and microstructural traits of both stable \citep{wagner_mesoscale_2022} and dynamically crosslinked gels \citep{wagner_coupled_2023}. However, these DNMs neglect the non-equilibrium effects of frictional drag due to inter-chain and solvent-chain interactions on the basis of quasi-static loading conditions. Additionally, these DNMs implicitly model dynamic bonding sites, often referred to as ``stickers" \citep{leibler_dynamics_1991,mordvinkin_rheology_2021}, so that they cannot investigate the phenomenon of bond lifetime renormalization put forth by \cite{stukalin_self-healing_2013} whereby stickers may break and reform bonds with the same partner multiple times before undergoing partner exchange. Without capturing these transient effects, it remains unclear to what extent the predictions of prior dynamic polymer DNMs agree with those of nanoscale MD and molecular theory, as well as over what timescales these DNMs are effectively applicable. 

To address these shortcomings and explore the applicable conditions for dynamic polymer DNMs, we here introduce an MD-consistent DNM (\textbf{Fig. \ref{fig: Branched Dynamic Network Graph}A}) greatly expanding on that of \cite{wagner_network_2021}. This DNM, hereafter referred to as simply the ``mesoscale model", explicitly tracks only two types of ``nodes" to represent polymers (\textbf{Fig. \ref{fig: Branched Dynamic Network Graph}B}). These are (i) the anchoring crosslinks (representing the locations at which chains are grafted to the polymer backbones of the network), and (ii) the distal stickers that reversibly associate with neighboring stickers. Stickers may represent either reversible covalently bonding \citep{kloxin_covalent_2010,richardson_hydrazone_2019,yue_vitrimerization_2020,wagner_coupled_2023} or physically interacting sites \citep{vidavsky_tuning_2020,zhang_bridging_2020,xu_thermosensitive_2022,cai_highly_2022}. As in prior DNMs, node-to-node interactions are captured via idealized implicit pairwise bond potentials derived from statistical mechanics. We use this model to investigate how input parameters such as polymer chain lengths, dynamic bond activation energies, polymer concentrations, and externally applied loading rates mediate distal sticker exploration and binding kinetics, which in turn govern network-scale topologies and mechanical stress responses. We examine these mappings as predicted by not only the mesoscale model, but also a conventional bead-spring approach in which every Kuhn segment comprising a polymer chain is explicitly modeled as a node attached to adjacent segments via a finitely extensible, nonlinear elastic potential \citep{kremer_dynamics_1990,somasi_brownian_2002,cruz_review_2012,sliozberg_computational_2013} (\textbf{Fig. \ref{fig: Branched Dynamic Network Graph}C}). In doing so, we identify the applicable regimes and limitations of the mesoscale model, quantify its computational cost savings over conventional coarse-grained MD, and explore its ability to model the mechanics of larger material domains.

    \figuremacro{H}{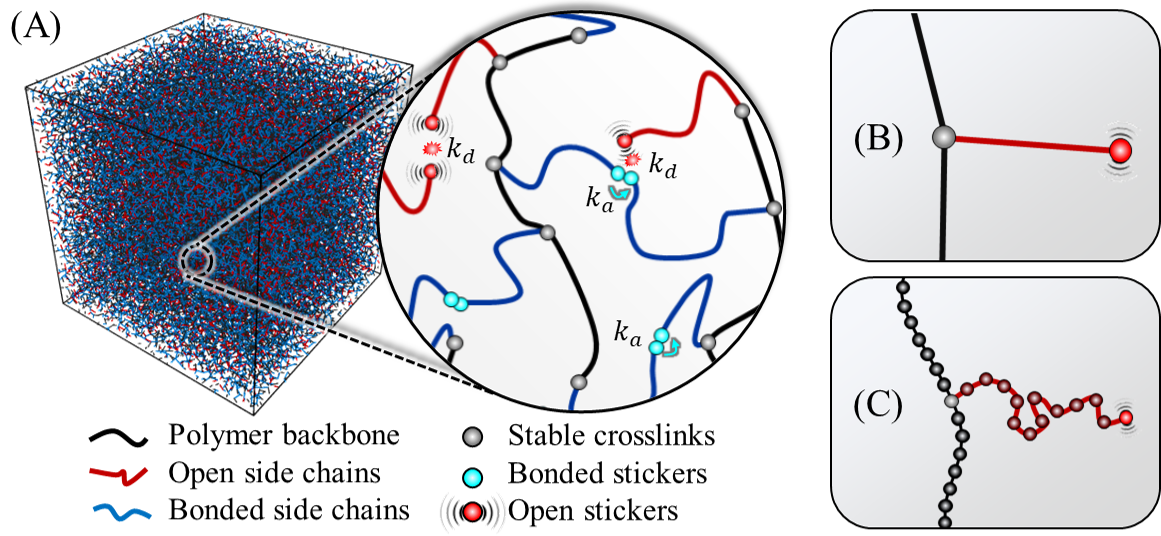}{}{\textbf{Discrete network modeling of dynamic polymers. (A)} 3D RVE and close-up 2D schematic of a branched polymeric network in which backbone chains are depicted as black, while open and bonded branching side chains are red and blue, respectively. Attachment and detachment events of stickers are characterized by their respective rates, $k_a$ and $k_d$. \textbf{(B)} Schematic of the mesoscale modeling approach. Only the crosslinks that tether side groups to the network (grey) and stickers (red) are explicitly modeled as particles. \textbf{(C)} Schematic of a conventional bead-spring model. Every Kuhn segment is modeled as a discrete particle attached to neighboring segments by nonlinear springs. Backbone Kuhn segments are black, branched side-chain Kuhn segments are dark red, stable crosslinks tethering the branched chains to the network are grey, and open stickers are red. \label{fig: Branched Dynamic Network Graph}}{0.9}
    
In the remainder of this work, we introduce the idealized bead-spring and mesoscale modeling methods (\textbf{Section \ref{sec: Discrete modeling approaches}}); examine the single-chain exploration and bond dynamics predictions, which shed light on statistical pairwise bond association theory (\textbf{Section \ref{sec: Results of single-chain}}); compare and contrast the network-scale mechanical predictions and computational costs of the bead-spring and mesoscale models (\textbf{Section \ref{sec: Results of network-scale}}); and then demonstrate how the mesoscale approach may be employed to probe larger spatiotemporal scales than are easily accessible using conventional MD (\textbf{Section \ref{sec: Spatiotemporal Extrapolation}}).

\section{Discrete modeling methods}
\label{sec: Discrete modeling approaches}

Here we detail the methods of both the newly introduced mesoscale model and the analogous bead-spring model used for validation. Both models are implemented using the MD framework LAMMPS  \citep{thompson_lammps_2022} built with the ``Transient Network Theory" package recently introduced by \cite{wagner_network_2021,wagner_catch_2024-1}. Network initiation and data post-processing are conducted using custom scripts within MATLAB 2022a. Below, we first introduce the equation of motion used to update both models' particles' positions in time as a function of the forces exerted on them (\textbf{Section \ref{sec: Equation of Motion}}). We then detail the pairwise bond association/dissociation rules used to update both models' attachment/detachment kinetics (\textbf{Section \ref{sec: Bond Kinetics}}). For detailed numerical initiation procedures of both single-chain and network-scale studies, as well as lists of pertinent unit conversions and parameters, see \textbf{Appendices \ref{Appendix - initiation procedures}-\ref{sec: Parameters and Unit Normlization}}. 

\subsection{Equation of motion}
\label{sec: Equation of Motion}

In both the bead-spring and mesoscale models, we update the position of each explicitly modeled particle, $\alpha\in [1,\mathcal{N}]$, in time, $t$, using conventional Brownian dynamics. The effects of solvent are implicitly captured by stochastic forces (that represent thermal fluctuations due to solvent interactions), and solvent-induced drag forces (proportionate to the particle's velocity). Based on relatively low particle masses and high viscosities of the surrounding mediums (i.e., solvent or adjacent polymers), the position, $\bm x^\alpha$, of each particle $\alpha$ is updated according to an overdamped Langevin equation of motion: 
\begin{equation}
    \gamma^\alpha \frac{d \bm x^\alpha}{dt} = \sum_\beta \bm f^{\alpha \beta} + \sqrt{2\gamma^\alpha k_b T} \bm \eta^\alpha.
    \label{eq: Brownian Equation of Motion}
\end{equation}
The term on the left-hand side of Eq. (\ref{fig: Branched Dynamic Network Graph}) constitutes drag force where $\gamma^\alpha$ is a friction coefficient related to node $\alpha$'s untethered diffusion coefficient, $D^\alpha$, through the Einstein relation (i.e, $D^\alpha = k_b T/\gamma^\alpha$) \citep{einstein_uber_1905,rubinstein_polymer_2003}. The first term on the right-hand side of Eq. (\ref{eq: Brownian Equation of Motion}) is the unbalanced force due to all pairwise polymer interactions between node $\alpha$ and its interaction neighbors, $\beta$. The final term on the right-hand side of Eq. (\ref{eq: Brownian Equation of Motion}) captures thermal fluctuations due to solvent interactions \citep{rubinstein_polymer_2003}, where $k_b T$ is the thermal energy, $k_b=1.38 \times 10^{-23}$ J/K is the Boltzmann constant, $T$ is absolute temperature, and $\bm \eta$ stochastically prescribes Gaussian-distributed thermal noise with a variance $dt^{-1}$. This noise, $\bm \eta^\alpha$, satisfies the conditions:
\begin{equation}
    \langle \bm \eta^\alpha \rangle_\mathcal{N} = 0,
    \label{eq: Gaussian Symmetry}
\end{equation}
\begin{equation}
    \langle \bm \eta^\alpha (t_1) \cdot \bm \eta^\alpha (t_2) \rangle_\mathcal{N} = \delta(t'-t).
    \label{eq: Markovian}
\end{equation}
where the operator $\langle \Box \rangle_\mathcal{N}$ denotes ensemble averaging over all $\mathcal{N}$ nodes. Eqs. (\ref{eq: Gaussian Symmetry}) and (\ref{eq: Markovian}) respectively convey that the mean of $\bm \eta$ is zero (i.e., no net thermal force) and that thermal fluctuation is a delta-correlated stationary process in time (i.e., there is no temporal correlation between the thermal fluctuation applied on a given node at any arbitrary time $t$ and subsequent time $t'$). Eq. (\ref{eq: Brownian Equation of Motion}) may be used to explicitly estimate each particle’s discrete displacement, $\delta \bm x$, over a discrete time step, $\delta t$. To achieve numerical stability, $\delta t$ was set to $4 \times 10^{-4} \tau_0$ for the bead-spring model, and $4 \times 10^{-3}\tau_0$ for the mesoscale model, where $\tau_0 = b^2/D_0$ is the prescribed monomer diffusion timescale \citep{einstein_uber_1905} and $b$ is the Kuhn length characterizing the size of a mer. Note that no equation of motion is provided for rotational degrees of freedom, as all particles are treated as point masses, consistent with ideal chain assumptions of no rotational penalty between bonded segments \citep{rubinstein_polymer_2003,doi_soft_2013}.

While both models update their nodes' positions through Eq. \eqref{eq: Brownian Equation of Motion}, the pairwise forces their nodes experience ($\bm f^{\alpha \beta}$) due to both bonded and non-bonded polymer interactions must be captured distinctly. For the mesoscale model, bonded interactions consist primarily of the entropic tensile forces of implicit chains, resulting from their reduced conformational degrees of freedom as they elongate. Entropic tension of the chain connecting node $\alpha$ to its $\beta^{th}$ neighbor is computed as $-\partial \psi_c/\bm r^{\alpha \beta}$, where $\bm r^{\alpha \beta}$ is said chain's end-to-end vector, and $\psi_c$, is its Helmholtz free energy. Free energy, $\psi_c$, is prescribed via the Pad\'e approximation of ideal Langevin chains as:
\begin{equation}
    \psi_c = k_b T \left\{ \frac{(\bm \lambda_c^{\alpha \beta})^2}{2} -N \log \left[ N- (\bm \lambda_c^{\alpha \beta})^2 \right] \right\},
    \label{eq: Langevin force}
\end{equation}
where $N$ is the number of Kuhn segments in the chain and $\bm \lambda_c^{\alpha \beta}=\bm r^{\alpha \beta}/(\sqrt N b)$ is the chain's stretch \citep{cohen_pade_1991}. To simplify the mesoscale approach, we invoke the ideal chain assumption such that all non-bonded polymer interactions (e.g., excluded volume repulsion, depletion forces, and long-range interactions) may be neglected.  

For the bead-spring model, the primary bonded interactions are the forces cohering the covalent bonds comprising Kuhn segments. These are computed as $-\partial \psi_b/\bm r^{\alpha \beta}$ where the free energy of each Kuhn segment, $\psi_b$, is captured using the ``nonlinear" bond style of LAMMPS given by: 
%
%
\begin{equation}
    \psi_b = \frac{E (\bm r^{\alpha \beta}-b)^2}{L^2-(\bm r^{\alpha \beta}-b)^2}.    
    \label{eq: FENE}
\end{equation}
Here $E$ is an energy scale that modulates bond stiffness, $L$ is the finite length by which the bond may be stretched or compressed from equilibrium, $b$ (the Kuhn length) is the finite rest length of the bond, and $\bm r^{\alpha \beta}$ remains the bond's end-to-end vector \citep{rector_simulation_1994}. We found that setting $E=800 k_b T$, and $L=b$ provides ample numerical stability while accurately reproducing the theoretically predicted end-to-end distributions of entropic chains (\textbf{Section \ref{sec: Single Chain End-to-end Results}}), Kuhn segment length distributions within $\pm 10\%$ of the prescribed rest length ($b$), and force-extension relations in agreement with $\bm f(\bm r)=-\partial \psi_c/\partial \bm r$ (\textbf{Appendix \ref{Appendix - bead-spring potential}}). Dynamic bonds, which form at the length scale $b$, were also modeled using Eq. \eqref{eq: FENE} for both modeling approaches, but with $E$ lowered to $100 k_b T$ to bolster numerical stability. Careful consideration should be given when prescribing $L$ and $E$ for material-specific dynamic bond types in future applications of this method. To represent non-bonded, excluded volume interactions in the bead-spring model, we included a Lennard-Jones potential of the form:
\begin{equation}
    \psi_{LJ}=4 k_b T \left[\left( \frac{\sigma_0}{d^{\alpha \beta}} \right)^{12} - \left( \frac{\sigma_0}{d^{\alpha \beta}} \right)^6 \right],\; (d^{\alpha \beta} \leq d_c)
    \label{eq: Lennard Jones}
\end{equation}
where $\sigma_0=2^{-1/6}b$ is the equilibrium distance between two interacting nodes, and $d^{\alpha \beta}$ is the distance between neighboring nodes $\alpha$ and $\beta$ within cutoff distance $d_c = 2\sigma_0$ of each other. 


In addition to their differences between pairwise polymer interactions, the bead-spring and mesoscale approaches must also have different damping coefficients, $\gamma^\alpha$, prescribed to each node. The damping coefficient prescribed to a particle in the bead-spring model must account for only one Kuhn segment and may therefore be approximated as $\gamma_0 = k_b T/D_0$, where $D_0$ is the diffusion coefficient of an untethered monomer.\footnote{Here, $D_0$ is taken as a sweeping parameter using reasonable values based on experimental literature \citep{shimada_precise_2005,kravanja_diffusion_2018,shi_molecular_2021} (see \textbf{Appendix \ref{sec: Parameters and Unit Normlization}}). Through the prescription of $D_0$ and $b$, the characteristic time, $\tau_0 = b^2/D_0$, it takes a monomer to diffuse distance $b$, is established. This, in turn, dictates the timescale of the model.} In contrast, the coefficient of a node in the mesoscale model must also account for the friction of adjacent Kuhn segments in its attached chain(s). Rouse scaling theory predicts that the frictional coefficient of a tethered chain with $N$ Kuhn segments scales as $\gamma^\alpha \approx N \gamma_0$, \citep{rouse_theory_1953,rubinstein_polymer_2003}. We find that in networks of freely diffusing chains, this relationship holds (\textbf{Sections \ref{sec: Results of network-scale}} and \textbf{\ref{sec: Spatiotemporal Extrapolation}}). However, in studies in which one end of a tethered chain is fixed, we instead find that the mesoscale model most closely approximates the Rouse sub-diffusion of the bead-spring model when $\gamma^\alpha = N^{2/3}\gamma_0$ for the mesoscale nodes (\textbf{Section \ref{sec: Results of single-chain}}). 

\subsection{Bond kinetics}
\label{sec: Bond Kinetics}

Dissociation between attached stickers is prescribed according to Eyring's theory \citep{eyring_activated_1935} at a rate of:
\begin{equation}
    k_d = \tau_0^{-1} \exp \left( - \frac{\varepsilon_d}{k_bT} \right),    
    \label{eq: Detachment rate}
\end{equation}
\noindent where $\tau_0 = b^2/D_0$ is the characteristic time it takes a monomer (i.e., detached sticker) to diffuse its Kuhn length, $b$, and $\varepsilon_d$ is the activation energy of bond dissociation \citep{eyring_activated_1935}. Analogously to dissociation, bond attachment is prescribed with an intrinsic rate:
\begin{equation}
    k_a = \tau_0^{-1} \exp \left( - \frac{\varepsilon_a}{k_b T} \right),
    \label{eq: Attachment rate}
\end{equation}
where $\varepsilon_a$ is the activation energy of association. However, unlike dissociation, bond association is intrinsically predicated on the ability of detached stickers to encounter one another through their stochastic diffusion, and so will inherently be influenced by traits such as the open sticker concentration \citep{stukalin_self-healing_2013} and tethered chain length as discussed in \textbf{Section \ref{sec: Single Chain End-to-end Results}}. Nonetheless, assuming no long-range interactions \citep{rubinstein_polymer_2003}, encounters are defined as occurring when two stickers diffuse within one Kuhn length, $b$, of each other \citep{stukalin_self-healing_2013}. To determine if an attached set of stickers dissociates, or an ``encountering" pair of open stickers associates, a memoryless Poisson process is assumed so that the probability of reaction, $P$, evolves in time according to:
\begin{equation}
    dP_i = 1-\exp \left(-k_i t \right) dt
    \label{eq: Poisson process}
\end{equation}
where the index ``$i$" denotes either attachment, ``$a$", or detachment, ``$d$" \citep{wagner_network_2021,wagner_coupled_2023}. Integrating Eq. (\ref{eq: Poisson process}) over the discrete time interval $[t,t+\delta t]$ and checking the resulting probability against a random number in the uniformly distributed range 0 to 1, permits the capture of stochastic reactions in both models. 

Average effective attachment and detachment rates over the total simulated time (from $t=0$ to $t=t_f$) are computed as:
\begin{equation}
    \bar k_a = t_f^{-1} k_s \int_0^{t_f} N_a \left[ (c-c_a) V\right]^{-1} dt,
    \label{eq: Measured Attachment Rate}
\end{equation}
and:
\begin{equation}
    \bar k_d = t_f^{-1} k_s \int_0^{t_f} N_d (c_a V)^{-1} dt,
    \label{eq: Measured Detachment Rate}
\end{equation}
respectively. Here, $k_s$ is the sampling frequency; $N_a$ and $N_d$ are the discrete numbers of attachment and detachment events at time $t$, respectively; $c$, $c_a$, and $c-c_a$ are the total, attached, and detached chain concentrations at time $t$, respectively; and $V$ is the domain volume. To ensure adequate temporal resolution, $k_s$ was set to twenty times that of the prescribed monomer oscillation frequency (i.e., $k_s = 20 \tau_0^{-1}$). We found that increasing $k_s$ from $15\tau_0^{-1}$ to $20\tau_0^{-1}$ influenced neither the diffusive behavior of tethered chains nor the binding kinetics between them for either discrete modeling approach (see \textbf{Appendix \ref{Appendix - dt convergence}}). 

\section{Single-chain and bond kinetics validation studies}
\label{sec: Results of single-chain}

Before utilizing the mesoscale model for network-scale mechanical predictions, we must first verify that the single-chain statistics of the mesoscale model agree with both the predictions of the bead-spring model, as well as prevailing statistical theory for ideal chains \citep{rubinstein_polymer_2003,doi_soft_2013}. Additionally, we must ensure that the bond association, dissociation, and partner exchange kinetics predicted by the mesoscale model reasonably agree with those predicted by the bead-spring model. Model validation results for single chain exploration, pairwise bond kinetics, and partner exchange kinetics in ensembles of chains are presented in \textbf{Sections \ref{sec: Single Chain End-to-end Results}-\ref{sec: Bond Exchange Results}} below.

\subsection{Tethered chains follow Gaussian statistics and approximate Rouse sub-diffusion}
\label{sec: Single Chain End-to-end Results}
 
We conducted tethered single-chain studies to corroborate that individual chains' end-to-end conformations and tethered diffusion characteristics agree between models. To explore the effects of chain length, the number of Kuhn segments was swept over $N = \{12, 18, 24, 30, 36 \}$. The lower limit of $N=12$ was set adequately high to still observe Gaussian statistics. Meanwhile, the upper limit of $N=36$ was set to three times the lower limit to ensure that a comparably high chain length was explored without modeling chains significantly longer than the entanglement length ($N=35$) predicted by \cite{kremer_dynamics_1990}. To also explore the effects of modulating the monomer diffusivity (which governs the timescale of the model per the relation $\tau_0=b^2/D_0$), the diffusion coefficient was swept over $D_0 = \{0.125, 0.25, 0.5, 1, 2, 4, 8\} \times 10^{-10}$ m$^2$ s$^{-1}$. The central value of $D_0 = 10^{-10}$ m$^2$ s$^{-1}$ was selected based on a realistic value for diffusion of poly(ethylene glycol) oligomers with low degrees of polymerization (on the order of two mers, which corresponds to one Kuhn segment) at ambient temperatures in good solvent \citep{shimada_precise_2005}. 

To achieve adequate statistical sampling, ensembles of $n_p = 1331$ and $n_p = 125$ non-interacting polymer chains were generated for the mesoscale and bead-spring models, respectively. The significantly larger ensemble of chains for the mesoscale model was set arbitrarily high and was easily enabled by the model's reduced computational cost, while the sample size for the bead-spring model was set adequately high to observe convergence in predicted results. Chains were generated by first initializing their fixed tethering nodes in a 3D grid at the position set $\{ \bm x^0 \}$.\footnote{For single-chain, tethered diffusion studies, no inter-chain bond kinetics between the chains' free ends were included so that the initial spacing between the positions $\{ \bm x_0 \}$ is arbitrary. However, to leverage the spatial parallelization of LAMMPS for higher throughput sampling, the tethering sites were separated on their grid by a distance of $6 Nb$.} Simulated polymer chains were then randomly generated from each tethering site using random-walk generation procedures according to Eqs. (\ref{eq: Random Walk - bead-spring}) and (\ref{eq: Initiating Mesoscale Chain}) for the bead-spring and mesoscale iterations of the model, respectively. Once all chains were initiated, the positions of all non-fixed nodes were updated according to Eq. (\ref{eq: Brownian Equation of Motion}), and then the end-to-end vectors of all $m\in \left[1,n \right]$ chains were measured over time. End-to-end vectors are defined as:
\begin{equation}
    \{ \bm r_m \} = \sum_\alpha^\Lambda \{ \bm x_m^{\alpha+1} - \bm x_m^\alpha \},    
\end{equation}
where the maximum node number, $\Lambda$, for each molecule is the number of Kuhn segments ($\Lambda = N$) for the bead-spring model and two ($\Lambda = 2$) for the mesoscale model (see \textbf{Fig. \ref{fig: Single-chain analysis}A}). 

\textbf{Fig. \ref{fig: Single-chain analysis}B-C} depict the probability distribution functions (PDFs) of finding a chain at a given end-to-end stretch, $\lambda_c=r/(\sqrt N b)$, as predicted by the mesoscale model (grey histogram), the bead-spring model (black diamonds), and the analytically derived joint PDF of Eq. (\ref{eq: Gaussian Joint PDF}) (black curve) for Gaussian chains with $N=12$ (\textbf{Fig. \ref{fig: Single-chain analysis}B}) and $N=36$ (\textbf{Fig. \ref{fig: Single-chain analysis}C}) Kuhn segments. Recall that stretch is defined as the end-to-end length, $r$, of a chain normalized by its mean expected length, $\sqrt N b$, (as predicted by Gaussian, random-walk statistics) so that the functional form of the joint PDF of Eq. (\ref{eq: Gaussian Joint PDF}) with respect to $\lambda_c$ is:
\begin{equation}
    P(\lambda_c) = 4\pi \left( \frac{3}{2\pi N b^2} \right)^{\frac{3}{2}} \lambda_c^2 \exp \left(-\frac{\lambda_c^2}{2} \right),
    \label{eq: Gaussian Distribution-stretch}
\end{equation}
whose variance is unity and is thus independent of $N$ or $b$ \citep{doi_soft_2013}. Therefore, providing end-to-end distributions in terms of stretch induces collapse of the PDFs with respect to the two chain lengths for a normalized comparison. Note that probabilities of \textbf{Fig. \ref{fig: Single-chain analysis}B-C} are also re-normalized as $p =  P(\lambda_c)/\int P(\lambda_c) d\lambda_c$ so that the PDF integrates to unity over the range $\lambda_c \in [0,\infty)$ and thus aligns vertically with the discrete distributions. Results indicate that the mesoscale model's predicted end-to-end distributions are in excellent agreement ($R^2 \geq 0.99$) with the analytically derived Gaussian PDF. Additionally, the bead-spring model agrees with both the mesoscale and analytical models when the characteristic energy scale from Eq. \eqref{eq: FENE} is set to $E=800 k_b T$. 

    \figuremacro{H}{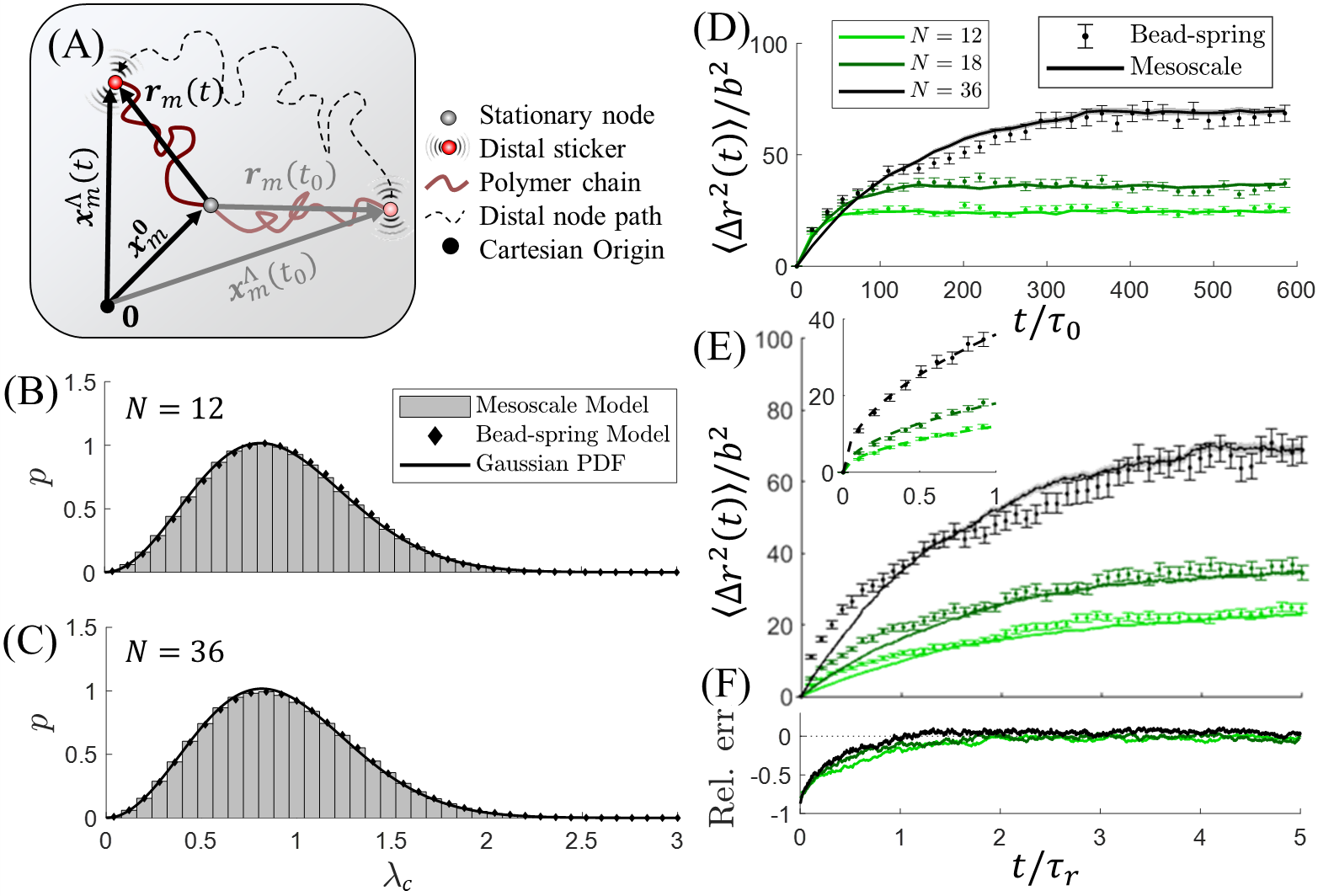}{Tethered diffusion characteristics. }{\textbf{(A)} Illustration of a tethered chain, anchored to a fixed node (grey). The sticker (red) diffuses over the time interval $[t_0,t]$, from positions $\bm x^\Lambda_m(t_0)$ to $\bm x^\Lambda_m(t)$. \textbf{(B,C)} PDFs of chain end-to-end stretch, $\lambda_c$, predicted by the mesoscale model (grey histogram, $n_p = 1331$ chains), bead-spring model (discrete diamonds, $n_p = 125$ chains), and analytical Gaussian distribution of Eq. (\ref{eq: Gaussian Distribution-stretch}) (black curve), when $N=12$ and $N=36$. \textbf{(D)} Normalized MSD, $\langle \Delta r^2\rangle /b^2$, versus time for $N \in \{ 12,18,36 \}$ as predicted by the bead-spring (discrete data) and mesoscale (solid curves) models. Total time, $t \approx 600 \tau_0$, is above the Rouse time, $\tau_r$ so that steady state is reached for all $N$. \textbf{(E)} MSD with respect to time at the order of the Rouse timescale, $0< t < 5\tau_r$, for $N \in \{ 12,18,36 \}$, predicted by the bead-spring (discrete data) and mesoscale (solid curves) models. Inset depicts bead-spring data (discrete data) for $0<t<\tau_r$ along with fits from Eq. (\ref{eq: Rouse Model}) (dashed curves). \textbf{(F)} Relative error between the bead-spring and mesoscale models (bead-spring model as reference) with respect to time. Results of \textbf{(B-F)} are for the median value of $D_0 = 10^{-10}$ m$^2$ s$^{-1}$ only, as $D_0$ has no meaningful effect on end-to-end conformations or MSD with respect to normalized time (\textbf{Fig. \ref{fig: Extended MSD results - effect of damper}}). All error bars represent standard error of the mean (SE). \label{fig: Single-chain analysis}}{0.9}
    
While the PDFs of \textbf{Fig. \ref{fig: Single-chain analysis}B-C} indicate excellent agreement between the instantaneous ensemble of chain conformations from both discrete models and statistical mechanics, they reveal nothing of the exploratory behavior of individual chains as they diffuse through space. To characterize exploration in time, we also compute the mean-square displacement (MSD) of the chains' distal ends according to:
\begin{equation}
    \langle \Delta r^2 (t) \rangle = \langle (\bm x_m^\Lambda (t) - \bm x_m^\Lambda (t_0))^2 \rangle  
    \label{eq: MSD}
\end{equation}
where $\Lambda$ denotes the freely diffusing node at the end of the chain, $t_0$ is the reference time from which MSD is measured, and the operator $\langle \square \rangle$ denotes ensemble averaging over all $m \in [1,n_p]$ chains. \textbf{Fig. \ref{fig: Single-chain analysis}D} compares the MSD measured for stickers in the mesoscale model (continuous curves) to those of the bead-spring model (discrete data) over a duration of $t \approx 600 \tau_0$. Over this relatively longer timescale, the MSDs plateau at values that agree between models, regardless of diffusion coefficient (\textbf{Appendix \ref{Appendix - extended single-chain diffusion}}, \textbf{Fig. \ref{fig: Extended MSD results - effect of damper}}). However, for the eventual purposes of investigating bond kinetics that are mediated by the frequency at which two binding sites enter within short distances (i.e., $b=\sqrt{D_0 \tau_0}$) of each other, it is paramount to observe exploration behavior over shorter timescales and smaller spatial areas.
        
\textbf{Fig. \ref{fig: Single-chain analysis}E} displays the same MSD data from \textbf{Fig. \ref{fig: Single-chain analysis}D}, but over the much shorter timescale of $0<t<5\tau_r$ where $\tau_r$ is the Rouse time, or the time it takes a polymer chain to diffuse its own characteristic area, $Nb^2$. It is well established that in the regime $\tau_0 < t < \tau_r$, tethered chains explore 3D space according to:
\begin{equation}
    \langle \Delta r^2 (t) \rangle \approx b^2 \left( \frac{t}{\tau_s} \right)^{1/2}
    \label{eq: Rouse Model}
\end{equation}
where $\tau_s=\tau_r N^{-2}$ is the time it takes a distal sticker at the end of a chain of $N$ Kuhn segments to diffuse its own characteristic size, $b^2$ \citep{hult_advances_1999,rubinstein_polymer_2003,stukalin_self-healing_2013}. Observing \textbf{Fig. \ref{fig: Single-chain analysis}E}, we see that the mesoscale and bead-spring models are in relatively good agreement at timescales on the order of $1<t< 5\tau_r$ (with less than 10$\%$ relative error per \textbf{Fig. \ref{fig: Single-chain analysis}F}). From the inset of \textbf{Fig. \ref{fig: Single-chain analysis}E}, we also see that the bead-spring model is in excellent agreement with the Rouse model ($R^2>0.95$) over the time range $0<t<\tau_r$ when the Rouse time, $\tau_r$, is taken as the time at which measured MSD first exceeds $Nb^2$ and the sticker diffusion timescale, $\tau_s$, is treated as a fitting parameter (\textbf{Fig. \ref{fig: Extended MSD results - Rouse model}}).\footnote{We find $\tau_s/\tau_0 \approx 0.07$, regardless of $N$ or $D_0$ (\textbf{Fig. \ref{fig: Extended MSD results - effect of damper}.H}).} However, below the Rouse timescale the mesoscale model under-predicts the MSD predicted by the bead-spring and Rouse models, with a relative error upwards of 50$\%$ (\textbf{Fig. \ref{fig: Single-chain analysis}F}). This high magnitude of relative error is partially due to the fact that MSD approaches zero when $t \rightarrow 0$. Yet, that the relative error is always negative, indicates that the mesoscale model consistently under-predicts distal sticker diffusion over short timescales.

Underprediction of the MSD by the mesoscale model, as compared to the bead-spring model, could be due to a disagreement in radial MSD (i.e., MSD due to displacement along the end-to-end direction of the chain), tangential MSD (i.e., MSD due to displacements normal to the end-to-end direction of the chain), or some combination of both. The radial and tangential components of MSD are respectively defined as mean-square change in end-to-end length:
\begin{equation}
    \langle \Delta r_r^2(t) \rangle = \langle \left[ r_m(t) - r_m (t_0) \right]^2 \rangle,
    \label{eq: Radial MSD}
\end{equation}
and mean-square circumferential distance swept by the distal end of the chain:
\begin{equation}
    \langle \Delta r_t^2(t) \rangle = \langle \left[ r_m (t_0) \theta_m \right]^2 \rangle,
    \label{eq: Tangential MSD}
\end{equation}
over the time interval $t\in [t_0,t]$. Here, $\bm r_m(t) = \bm x^\Lambda_m(t) - \bm x^0_m(t)$ and $\bm r_m(t_0) = \bm x^\Lambda_m(t_0) - \bm x^0_m(t_0)$ are the end-to-end vectors of chain $m$ at times $t$ and $t_0$, respectively; and $\theta_m = \cos^{-1} \left[ \frac{\bm r_m(t_0) \cdot \bm r_m(t)}{r_m(t_0) r_m(t)}\right]$ is the angle between the end-to-end vectors at time $t_0$ and $t$ (see \textbf{Fig. \ref{fig: Radial and tangential  MSD}A}). Examining \textbf{Fig. \ref{fig: Radial and tangential  MSD}B-C}, we see that the cumulative radial and tangential MSDs of the mesoscale model closely mirror those of the bead-spring model, rarely deviating by more than 10$\%$. However, over the time period at which dynamic bonding is sampled in future studies ($\tau_0/20$), the mesoscale stickers explore approximately $0.95\pm 0.10$ $b^2$ and $2.17 \pm 0.06$ $b^2$ less than their bead-spring counterparts in the radial and tangential directions, respectively (\textbf{Fig. \ref{fig: Radial and tangential  MSD}D-E}). This is true regardless of chain length. The agreement of long-term ($t \geq \tau_r$) exploratory behavior between models supports that the mean paths of the stickers in both models traverse similar characteristic trajectories. However, the discrepancy in MSD below the Rouse time ($t<\tau_r$) indicates that the stickers of the bead-spring model do so with a higher vibrational amplitude on the order of $b^2$, and this is true of both their tangential and end-to-end vibration modes. 

Notably, the error is more pronounced for the tangential diffusion mode, meaning that there are greater deviations between the models' vibrational sticker amplitudes in directions normal to chain end-to-end vectors than in-line with them. This is likely because radial movement is constrained so that magnitudes of radial MSD are smaller than those of tangential MSD (\textbf{Fig. \ref{fig: Radial and tangential  MSD}B-C}). However, it may also arise from the fact that tangential diffusion is mediated entirely by drag and Brownian forces on the stickers. Both of these depend on the damping coefficient, which must be set higher for the mesoscale model to achieve similar MSDs past $\tau_r$. In contrast, radial diffusion is also checked by the entropic chain forces, which are independent of the damping coefficient and in good agreement between models (\textbf{Fig. \ref{fig: Appendix_bond_potential}.B}), thus perhaps reducing error. In any case, to ensure that discrepancies in diffusive behavior do not affect bond kinetics, or -- by extension -- topological network reconfiguration and network mechanics, we next investigate the emergent bond attachment and detachment rates of these two models as not only a function of their intrinsically assigned kinetic rates through Eqs. \eqref{eq: Detachment rate} and \eqref{eq: Attachment rate}, but also the exploratory behavior of their chains.

    \figuremacro{H}{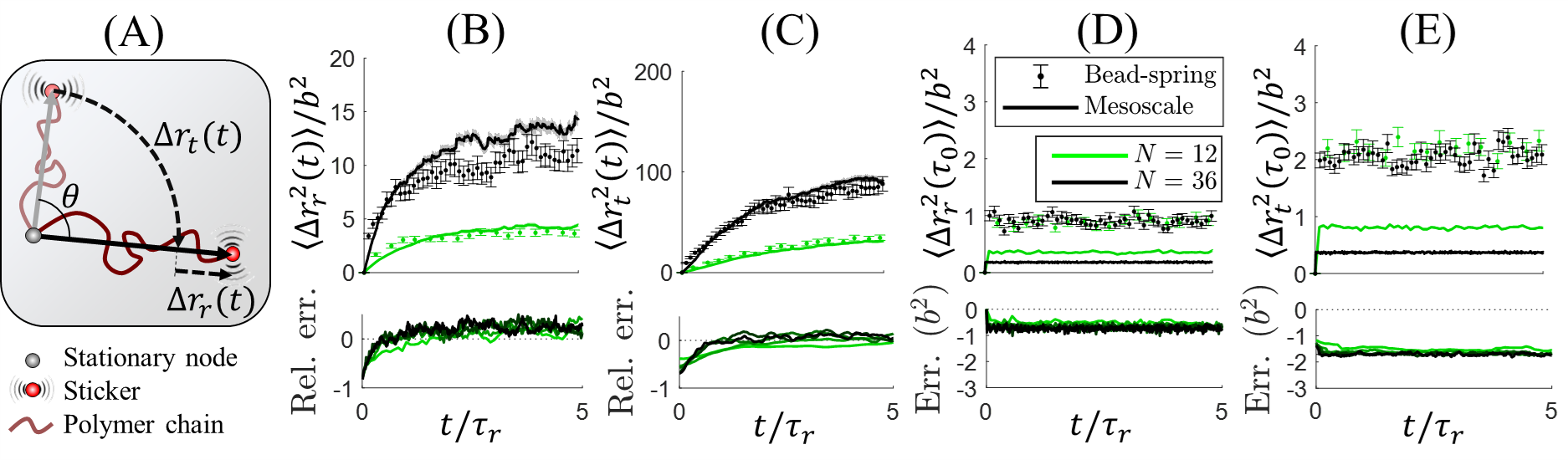}{Radial and tangential tethered diffusion. }{\textbf{(A)} Illustration of a tethered chain anchored to a fixed node (grey) whose sticker (red) diffuses from end-to-end vector $\bm r_m(t_0)$ (grey arrow) to $\bm r_m (t)$ (black arrow). Geometric definitions of $\Delta r_t(t)$ (dashed arc) and $\Delta r_r(t)$ (dashed arrow) are illustrated. \textbf{(B)} Radial and \textbf{(C)} tangential components of MSD (\textbf{top}) with respect to time, $t/\tau_r$, for the bead-spring model (discrete data) and mesoscale model (solid curves). Relative error (\textbf{bottom}) between models (bead-spring model serves as reference). \textbf{(D)} Radial and \textbf{(E)} tangential components of average square displacement (\textbf{top}) over the duration $[t,t+\tau_0/20]$, with respect to normalized time for the bead-spring model (discrete data) and mesoscale model (solid curves). Absolute error (\textbf{bottom}) between models (bead-spring model serves as reference). \textbf{(B-E)} Error bars represent SE. \label{fig: Radial and tangential  MSD}}{1}
      
\subsection{Chain attachment mirrors Bell's model for detachment}
\label{sec: Pairwise attachment}

While the kinetic association rate, $k_a^{ap}$, prescribed \textit{a priori} between two stickers within distance $b$ of each other is set according to \textbf{Section \ref{sec: Bond Kinetics}} (see \textbf{Appendix \ref{Appendix - validating bond kinetics}} and \textbf{Fig. \ref{fig: Kinetic Rates Validations}} for validation), the actual rate of attachment must also depend on the probability, $P_e$, with which said stickers encounter one another such that the emergent attachment rate is $k_a \propto k_a^{ap} P_e$. However, an analytical form of this probability is not immediately available from the literature. Therefore, to investigate how this probability evolves in each of the two discrete models while also interrogating agreement between their associative kinetics, we conducted studies in which $n_p=216$ tethered chains with stickers at their distal ends were positioned near fixed stickers. The distal stickers and fixed stickers were then allowed to bond/unbond with each other in mutually exclusive pairs (\textbf{Fig. \ref{fig: Single chain bond kinetics}A-C}). To probe the effects of distance, the separation length, $d_{ts}$, between each chain's tethering site and its paired fixed sticker was swept over the range $d_{ts} \in [0.0125,0.5]Nb$. Neighboring pairs were separated from each other by a distance greater than $2Nb$, so that no two chains' stickers could come within bonding distance, $b$, of each other. Instead, each tethered chain was only within reach of the fixed sticker belonging to its pair. 

Once all fixed nodes were positioned, the chains were initiated as described in \textbf{Section \ref{sec: Single Chain Initiation}} through Eqs. (\ref{eq: Random Walk - bead-spring}) to (\ref{eq: Initiating Mesoscale Chain}). After initiation, all non-fixed nodes' positions were updated according to Eq. (\ref{eq: Brownian Equation of Motion}). Stickers that came within distances less than $b$ of each other were checked for attachment according to Eqs. (\ref{eq: Attachment rate}-\ref{eq: Poisson process}) and the methods of \textbf{Section \ref{sec: Bond Kinetics}}. The time-averaged rates of attachment and detachment were then computed according to Eqs. (\ref{eq: Measured Attachment Rate}) and (\ref{eq: Measured Detachment Rate}), respectively. Besides sweeping the separation distance, $d_{ts}$, the number of Kuhn segments, $N$, and associative activation energies, $\varepsilon_a$, were also swept over the ranges $N=\{12, 18, 36 \}$ and $\varepsilon_a = \{0.01, 0.1, 1 \} k_b T$ to elucidate the effects of chain lengths and intrinsically prescribed binding rates, respectively. The upper limit of $\varepsilon_a = k_b T$ was selected because over computationally viable time domains (on the order of $10^{3} \tau_0$ to $10^{5} \tau_0$ for the bead-spring model with longer chains) we found that the number of discrete attachment events no longer provided adequate statistical sampling sizes when $\varepsilon_a \sim 10 k_b T$. Meanwhile, the lower limit of $\varepsilon_a=0.01 k_b T$ was selected because, at this activation energy scale, the intrinsic attachment rate approaches the monomer diffusion frequency ($k_d^{ap} \rightarrow \tau_0^{-1}$) so that sampling lower activation energies ($\varepsilon_a<0.01 k_b T$) becomes redundant. The bond kinetics sampling rate, $k_s$ and data output frequency were both set to $20\tau_0^{-1}$.  

\textbf{Fig. \ref{fig: Single chain bond kinetics}D-E} indicates that the ensemble-averaged attachment rates, $\bar k_a$, measured from both the bead-spring and mesoscale models are in reasonable agreement with one another across separation distances (here characterized by the chain stretch, $\lambda_c = d_{ts}/(\sqrt N b)$, required for attachment), chain lengths (through $N$)\footnote{Results from three unevenly spaced values of $N$ are presented based on an observed nonlinear relation between $k_a$ and $N$, which reveals that as $N$ increases the sensitivity of $k_a$ to $N$ decreases.}, and for two disparate activation energies ($\varepsilon_a = \{ 0.01,1 \} k_b T$). As expected, increasing the bond activation energy reduces the emergent attachment rate as indicated by the lower values of $\bar k_a$ from \textbf{Fig. \ref{fig: Single chain bond kinetics}E} when compared to those of \textbf{Fig. \ref{fig: Single chain bond kinetics}D}. Intuitively, increasing chain length (by increasing $N$) diminishes the attachment rate. This is attributed to the fact that longer chains have larger available exploration volumes and therefore are statistically less likely to encounter neighboring sticker sites at any given moment. Note that the models' predicted values of $\bar k_d$ are consistently in excellent agreement with the value set \textit{a priori} through Eq. (\ref{eq: Detachment rate}), which remains constant with respect to $\lambda_c$ since $k_d$ is intentionally made independent of chain stretch in Eq. (\ref{eq: Detachment rate}). 

    \figuremacro{H}{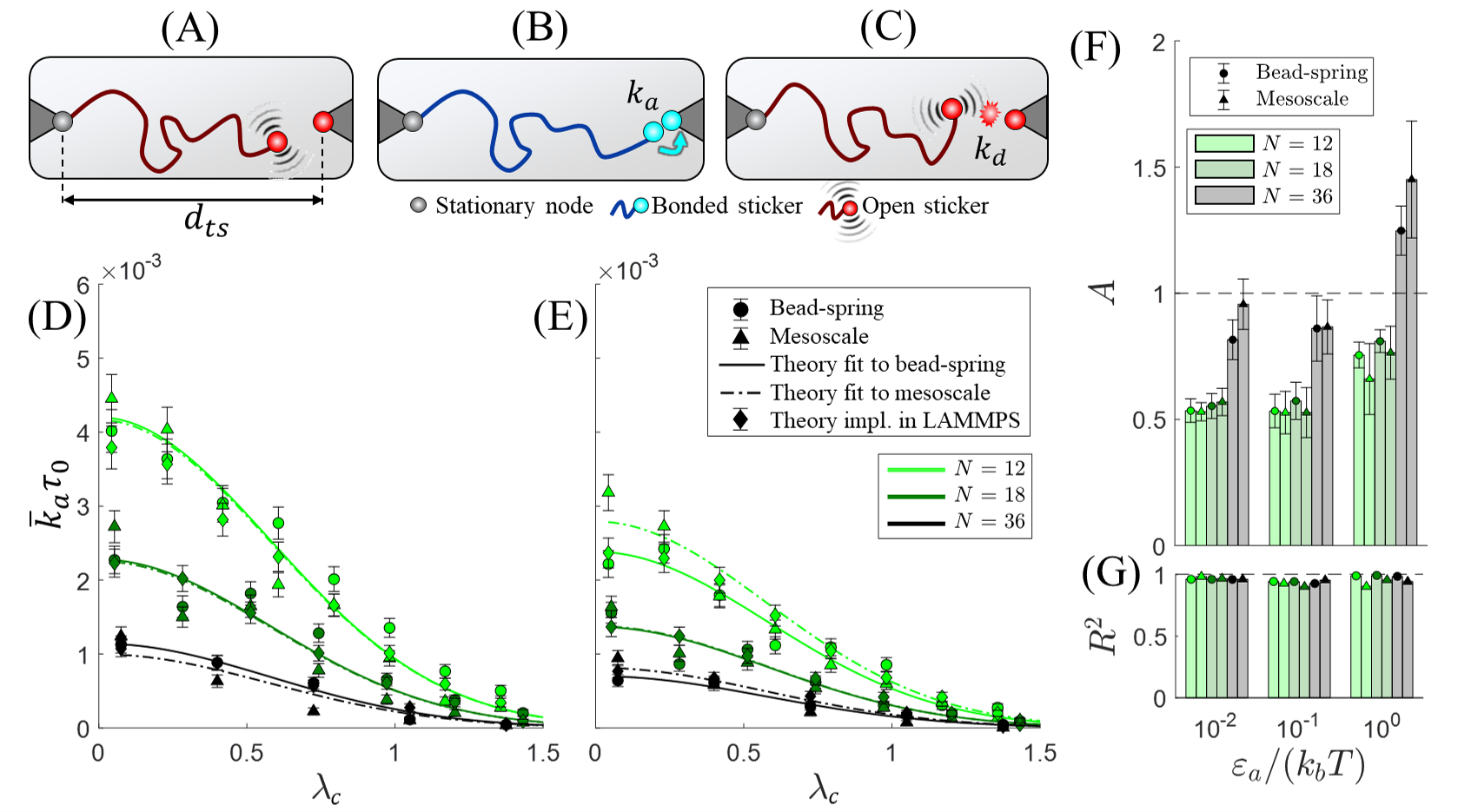}{Bond kinetics of a single tethered chain. }{\textbf{(A-C)} Illustration of a tethered chain whose free end is a sticker (red) that may bind or unbind with a fixed sticker (fixed red node to right). Distance between the fixed tethering node (grey) and fixed sticker (red) is denoted $d_{ts}$. This sticker pair is shown \textbf{(A)} in an arbitrary detached initial state, \textbf{(B)} immediately after an attachment event, and \textbf{(C)} after the subsequent detachment event.  \textbf{(D-E)} Average normalized attachment rates, $\bar k_a \tau_0$, with respect to normalized separation distance, $\lambda_c = d_{ts}/(\sqrt N b)$  when the associative bond energy is \textbf{(D)} $\varepsilon_a = 0.01 k_b T$ and \textbf{(E)} $\varepsilon_a = k_b T$. Data is provided for both the bead-spring (circles) and mesoscale (triangles) models. Error bars represent SE. Best fits of Eq. \eqref{eq: Attachment rate scaling theory} for the bead-spring (solid curves) and mesoscale (dashed curves) data are also displayed, treating prefactor, $A$, as a fitting parameter. Data is also shown for the LAMMPS implementation of Eq. \eqref{eq: Attachment rate scaling theory} when $A$ is taken as the average value from the fits to the bead-spring and mesoscale models. \textbf{(F)} Prefactor, $A$, for all activation energies and chain lengths. Error bars represent the 95$\%$ confidence interval. \textbf{(F)} Goodness of fit between the discrete models and scaling theory, characterized by $R^2$. \label{fig: Single chain bond kinetics}}{1}
    
Interestingly, the average attachment rates of \textbf{Fig. \ref{fig: Single chain bond kinetics}D-E} follow Gaussian relations with respect to separation distance, $d_{ts}$. Based on this observation, we postulate that the encounter probability, $P_e$, between the two stickers is proportionate to the probability, $P(d_{ts})$, of finding a Gaussian chain at end-to-end length $d_{ts}$ through Eq. (\ref{eq: Gaussian Joint PDF}). We also logically postulate that $P_e$ scales directly with the characteristic volume, $b^3$, within which a sticker ``encounters" and may therefore bind to a neighbor. Thus, the encounter rate scales as:
\begin{equation}
    P_e \propto b^3 \left( \frac{2}{3} \pi N b^2 \right)^{-\frac{3}{2}} \exp \left( -\frac{3}{2} \frac{d_{ts}^2}{Nb^2} \right).
    \label{eq: Encounter rate scaling}
\end{equation}
Substituting Eq. \eqref{eq: Encounter rate scaling}, along with the definition of $k_a^{ap}$ through Eq. \eqref{eq: Attachment rate}, into the postulated relation $k_a \propto k_a^{ap} P_e$, writing the relation in terms of chain stretch, $\lambda_c = d_{ts}/(\sqrt N b)$, and then simplifying predicts that the emergent attachment rate scales as:
\begin{equation}
    k_a \propto \tau_0^{-1} \left( \frac{2}{3} \pi N\right)^{-\frac{3}{2}} \exp \left[ - \frac{\varepsilon_a + \psi_a(\lambda_c)}{k_b T} \right],
    \label{eq: Attachment rate scaling theory}
\end{equation}
where $\psi_a(\lambda_c) \approx \frac{3}{2} {k_b T} \lambda_c^2$ is the Helmholtz free energy of an ideal chain at relatively low end-to-end stretches, $\lambda_c \sim 1$ (i.e., within the Gaussian regime). 

Significantly, Eq. \eqref{eq: Attachment rate scaling theory} predicts that the attachment rate scales with the inverse exponent of not only the intrinsic bond activation energy, $\varepsilon_a$, but also the Helmholtz free energy, $\psi_a$, of the entropic chain that would exist if attachment occurred. While we here prescribe force-independent bond dissociation for simplicity, Eq. \eqref{eq: Attachment rate scaling theory} also mirrors Bell's model for force-dependent slip-bond dissociation \citep{bell_models_1978} which states that the rate of bond detachment is given by: 
\begin{equation}
    k_d^{Bell} = K \exp \left[ -\frac{\varepsilon_d - \psi_a(\lambda_c)}{k_b T} \right]
    \label{eq: Bell's detachment rate}
\end{equation}
where $\psi_a$ is the Helmholtz free energy of the already attached chain and the prefactor $K$ is an attempt frequency generally taken as $\tau_0^{-1}$ \citep{evans_dynamic_1997,song_force-dependent_2021,wagner_coupled_2023}. In essence, Eq. \eqref{eq: Attachment rate scaling theory} predicts that the stretch-dependent Helmholtz free energy of a polymer chain additively increases the effective energy barrier of attachment for a binding site located at its distal end. Meanwhile, Eq. \eqref{eq: Bell's detachment rate} states that the same stretch-dependent Helmholtz free energy subtractively decreases the effective energy barrier for bond dissociation. These relations between kinetic rates and single-chain Helmholtz free energies emerge from statistical mechanics when forward and reverse reaction rates are functionally derived using the end-to-end state and transition state partition functions of polymer chains \citep{buche_chain_2021}. Furthermore, they are consistent with the experimental and theoretical findings of investigators such as \cite{guo_association_2009} or \cite{bell_kinetics_2017} who studied the association of polymer-tethered ligands to fixed receptor sites. 

The theoretical scaling predictions from Eq. \eqref{eq: Attachment rate scaling theory} are also in good agreement with both models investigated here if we multiply $k_a$ from Eq. \eqref{eq: Attachment rate scaling theory} by a dimensionless prefactor, $A$, which is treated as a fitting parameter (see \textbf{Fig. \ref{fig: Single chain bond kinetics}F}). Values of $A$ fitted to the mesoscale model are in agreement with those fitted to the bead-spring model for every combination of activation energy and chain length investigated (\textbf{Fig. \ref{fig: Single chain bond kinetics}F}) further illustrating agreement between the two discrete approaches. While values of $A$ range from $0.53\pm0.08$ to $1.45\pm 0.23$, they are generally less than unity suggesting that Eq. \eqref{eq: Attachment rate scaling theory} tends to over-predict measured attachment rates from the discrete approaches. Furthermore, there are consistent trends whereby the largest chain length and the highest activation energy correspond to greater values of $A$. The former trend may indicate that the scaling proportionality $k_a \propto N^{-3/2}$, which states that $k_a$ is proportionate to the cubic inverse of the maximum allowable stretch, $\lambda_c^{max}=N^{1/2}$, is incomplete in Eq. \eqref{eq: Attachment rate scaling theory}. Alternatively, it may suggest that there exists some unknown source of error (common to both discrete models), which is more pronounced for shorter chain lengths due to their higher sensitivity of $k_a$ to $N$. Meanwhile, the latter trend (that $A$ increases to above unity as the $\varepsilon_a$ increases) indicates that for systems with higher activation energy, the scaling theory trends towards under-predicting the attachment rate. We hypothesized that this trend results from bond-history dependent sticker exploration. Namely, that bonds that recently detached are statistically more likely to undergo repeat attachment with the same neighbors over shorter timescales, a notion put forth by \cite{stukalin_self-healing_2013} and investigated as it pertains to partner exchange rates in \textbf{Section \ref{sec: Bond Exchange Results}}. However, independently plotting the rate of first-time attachment rates, $\bar k_{a,1}$, versus overall attachment rates, $\bar k_a$, (including repeat events) reveals little change in the observed kinetics for the simulation times here (\textbf{Fig. \ref{fig: ka_vs_ka1}}).

While the exact physics necessitating the inclusion of prefactor, $A$, remain unclear, no fitting is required to match the variance of Eq. \eqref{eq: Attachment rate scaling theory} (see also Eq. \eqref{eq: Gaussian Joint PDF}). Indeed, the high $R^2$ values of \textbf{Fig. \ref{fig: Single chain bond kinetics}G} (all $>0.9$) confirm that the scaling theory can be fit appropriately across all activation energies and chain lengths simply by modulating $A$. This is true irrespective of chain length or activation energy (\textbf{Fig. {\ref{fig: Single chain bond kinetics}D-E}}). This also remains true when a set of two tethered chains capable of binding at their distal ends are modeled within various tether-to-tether distances, $d_{tt}$, of one another (\textbf{Fig. \ref{fig: Pairwise kinetics analysis}}) and allowed to bond/unbond. In this case, the number of Kuhn segments in the would-be chain after attachment is simply $2N$ so that the Helmholtz free energy in terms of end-to-end (i.e., tether-to-tether) distance becomes $\psi_a = \frac{3}{4} \frac{k_b T}{Nb^2} d_{tt}^2$, but remains $\psi_a = \frac{3}{2} k_b T \lambda_c^2$ in terms of chain stretch per Eq. \eqref{eq: Attachment rate scaling theory}. Substituting this free energy into Eq. \eqref{eq: Attachment rate scaling theory} and then fitting only $A$ nicely reproduces the average two-chain, pairwise attachment rates measured from both discrete models (\textbf{Fig. \ref{fig: Pairwise kinetics analysis}}). Importantly, this lack of needed adjustment to the variance confirms the key interpretation of Eq. \eqref{eq: Attachment rate scaling theory} that the rate of attachment is penalized by the Helmholtz free energy that a chain must attain in order to stretch sufficiently for bond attachment. 

Evidently, Eq. \eqref{eq: Attachment rate scaling theory} provides additional means for coarse-graining whereby only the backbone sites (at which side chains are grafted into a polymer network) are explicitly modeled. Then, the bonds formed between tether sites by the telechelic association of distal stickers can be implicitly captured as energetic potentials via Eq. \eqref{eq: Langevin force} that act with probabilities set through Eqs. \eqref{eq: Poisson process} and \eqref{eq: Attachment rate scaling theory}. Furthermore, the maximum bond interaction length may be set to the maximum stretch, $\lambda_c\approx 1.5$, at which chains observably associate (observe \textbf{Fig. \ref{fig: Single chain bond kinetics}D-E} and \textbf{Fig. \ref{fig: Pairwise kinetics analysis}.D} for one- and two-chain systems, respectively). We gauged this prospective coarse-graining method by implementing Eq. \eqref{eq: Attachment rate scaling theory} into LAMMPS and conducting an analogous study in which only the fixed nodes are modeled and the attachment between them is checked through Eq. \eqref{eq: Attachment rate scaling theory}. This study reproduces the Gaussian relation between $k_a$ and $\lambda_c$ observed for the bead-spring and mesoscale model predictions (\textbf{Fig. \ref{fig: Single chain bond kinetics}D-E}, discrete diamonds), with considerable reduction in computational cost (Table \ref{Table 3}). Since it circumvents the need to track stickers' oscillations within and outside of distance $b$ from one another, this implicit approach allows for dynamic bonding check frequencies, $k_s$ on the order of chain oscillation rates (i.e., $k_s\sim \tau_r^{-1} \sim N^{-2} \tau_0^{-1} $) instead of $k_s\sim 20\tau_0^{-1}$. In this study, $k_s$ is arbitrary since the tethers are fixed, but here we set $k_s=\tau_0^{-1}$ yielding two to three orders of magnitude reduction in computational run time as compared to the bead-spring model, and one to two orders of magnitude reduction as compared to the mesoscale model (Table \ref{Table 3}), without observed deviation in emergent kinetic rates on a pairwise basis (\textbf{Fig. \ref{fig: Single chain bond kinetics}D-E}). 

Despite the significant cost savings of this scaling theory-based approach, it is limited in a number of ways. First, it requires calibration of $A$ using either the bead-spring or mesoscale model. Second, $k_s$ is still limited by the Rouse diffusion rate, $\tau_r^{-1}$, it takes for chains to migrate within or outside of $\sim 1.5 \sqrt N b$ of each other. Finally, this implicit method loses information about the exact position of open stickers. Therefore, it is fundamentally unable to reproduce predictions from the bead-spring and mesoscale models about partner exchange or the phenomena of renormalized bond lifetime as discussed in \textbf{Section \ref{sec: Bond Exchange Results}}. Implementation of Eq. \eqref{eq: Attachment rate scaling theory} should therefore be relegated to modeling systems with dilute dynamic bond concentrations for which partner exchange is rare; stickers should otherwise be explicitly modeled.

\subsection{Bond dynamics and partner exchange kinetics agree between discrete models}
\label{sec: Bond Exchange Results}

Having confirmed good agreement between the bead-spring and mesoscale models' predictions of pairwise bond kinetics for single and two-chain systems, we next investigate whether the complex bond kinetics emerging within ensembles of chains agree between models. Once again, we investigate the rates of bond attachment and detachment. However, attachment events for an ensemble of chains (e.g., a network) at equilibrium can be further sub-divided into two types. The first type is repeat attachment whereby two stickers dissociate and then reattach (without finding new partners during the interim) after some time $\tau_{rpt}$ (\textbf{Fig. \ref{fig: Bond lifetime schematic}}, from $t_1$ to $t_2$). The second type is partner exchange whereby stickers bond to new partners after detaching from old ones after some comparatively longer detached lifetimes $\bar \tau_{exc}$ (\textbf{Fig. \ref{fig: Bond lifetime schematic}}, from $t_3$ to $t_4$). Based on this distinction, \cite{stukalin_self-healing_2013} introduced the renormalized bond lifetime, $\bar \tau_{rnm}$, or the time from when a pair of stickers newly bonds (\textbf{Fig. \ref{fig: Bond lifetime schematic}}, $t_0$) to when one or both of the stickers in the partnership attaches to a new neighbor (\textbf{Fig. \ref{fig: Bond lifetime schematic}}, $t_4$). Evidence suggests that it is the renormalized bond lifetime or inverse exchange rate (not just the detachment rate) that dictates the reconfigurational stress relaxation time in dynamic polymers \citep{stukalin_self-healing_2013}.

    \figuremacro{H}{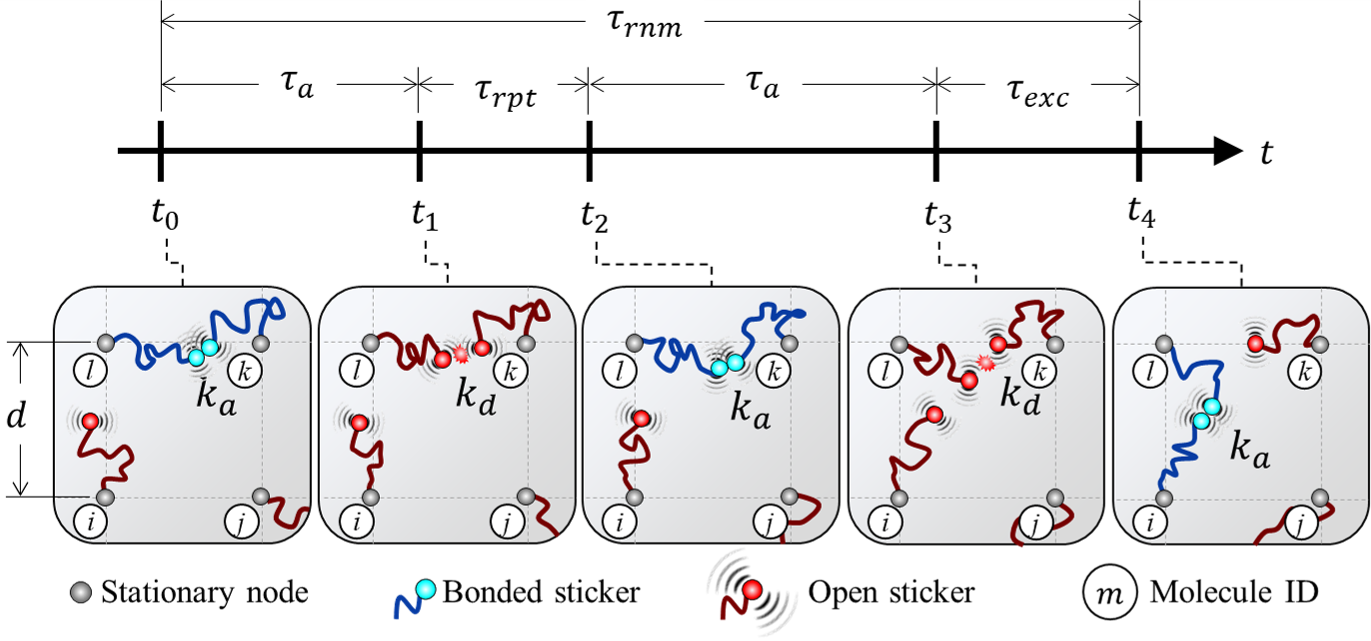}{Illustration of bond lifetimes. }{A schematic of four tethered chains (labeled ``\textit{i}"-``\textit{l}"), whose anchoring nodes (grey) are configured in a fixed grid (with nearest neighbor separation distance, $d$), displays detachment, repeat attachment, and partner exchange events. The definitions of attached bond lifetime, $\tau_a$, detached bond lifetime prior to repeat attachment, $\tau_{rpt}$, detached bond lifetime prior to partner exchange, $\tau_{exc}$, and renormalized bond lifetime, $\tau_{rnm}$ are also illustrated. \label{fig: Bond lifetime schematic}}{0.9}
        
To probe partner exchange kinetics and investigate agreement between the models, we simulated $n_p=343$ tethered polymer chains with stickers at their free ends. Their fixed ends were positioned in a cubic lattice within a periodic RVE. Prior work has demonstrated that sticker concentrations, $c$, significantly influences exchange kinetics \citep{stukalin_self-healing_2013}. To investigate this effect, the lattice constant (i.e., separation distance, $d=c^{-1/3}$, between tether nodes) was swept, which is equivalent to the average separation distance between neighboring stickers, $\bar d$, in an ensemble of isotropically oriented chains. To ensure realistic values of $d$ for a given chain length, $d$ was set based on the prescribed chain length, via $N$, and polymer packing fraction, $\phi$, according to $d = b [\pi N/(6 \phi)]^{1/3}$, assuming that each Kuhn segment occupies an approximate volume of $\pi b^3/6$. To evaluate whether chain length influences exchange kinetics, the number of Kuhn segments per chain was swept over $N=\{12,18,36 \}$ for consistency with \textbf{Section \ref{sec: Results of single-chain}}.  Meanwhile, $\phi$ was swept over the range $\phi \in [0.01, 0.52]$ to capture a breadth of values likely encompassing those of real gels and elastomers \citep{,rubinstein_polymer_2003,marzocca_physical_2013,wagner_mesoscale_2022}. The upper limit of $\phi$ was set based on empirical estimates of polymer packing in systems at ambient conditions (see \textbf{Appendix \ref{Appendix - polymer packing}} for details), as well as our present models' limiting ideal chain assumption, which invokes that chains do not interact via volume exclusion or cohesive forces (e.g. Van der Waals). In reality, as polymer free volume is decreased, a higher resultant concentration of inter-chain interactions will constrain the conformations available to each chain, likely decreasing their initial attachment and partner exchange rates due to greater subdiffusion, while increasing their effective stiffness due to reduction in entropy. Therefore, we impose that $\phi \leq 0.52$ based on the understanding that the validity of the ideal chain assumption is improved at lower packing fractions with fewer inter-chain interactions.

With $d$ computed for each combination of $N$ and $\phi$, the tether nodes were positioned and then chain initiation was conducted per the methods of \textbf{Section \ref{sec: Single Chain Initiation}}. Once initiated, the nodes comprising the tethered chains were stepped through time according to Eq. \eqref{eq: Brownian Equation of Motion} and their distal ends were allowed to bond/unbond according to Eqs. (\ref{eq: Detachment rate}-\ref{eq: Poisson process}). Stickers were only allowed attach to one neighbor at a time, thus enforcing the simple conditions of mutually exclusive pairwise binding. Attachment activation energy was set to $\varepsilon_a = 0.01 k_b T$ (a low value), to induce rapid associative kinetics (i.e., that $k_a^{ap}\approx \tau_0^{-1}$) and therefore interrogate the agreement between models with high temporal resolution demands through $k_s$. The activation energy for detachment was set to $\varepsilon_d = 7 k_b T$ (a relatively high value so $k_d \approx 10^{-3}\tau_0^{-1}$) to ensure that bond lifetimes were comparable to the detached lifetimes, but not so long that detachment (and by extension repeat attachment/partner exchange) was rarely observed over the simulated duration of $8 \times 10^3 \tau_0$. Data for this study was output with a frequency of $10\tau_0^{-1}$, which we found ensures adequate detection of repeat attachment events. Adequate sampling of average kinetic rates ($n\geq 100$) was achieved under these conditions. Reasonable sampling of bound and unbound sticker lifetimes was achieved at most chain concentrations. However, as few as $n<10$ fully attached bond lifetimes were observed for very low chain concentrations. This is because attachment/detachment kinetic rate measurements require observation of only one state transition, whereas full bond lifetime measurements require the observation of two (i.e., start and end state transitions). Additionally, bond lifetimes are power law distributed so that mean lifetimes are sensitive to outliers. For these reasons we emphasize interpretation of rates in our discussion below, reserving mean bond lifetimes for qualitative comparison across event types.   

To characterize bond kinetics, average attachment rates, $\bar k_a$, were computed according to Eq. \eqref{eq: Measured Attachment Rate} (\textbf{Fig. \ref{fig: Associative kinetics}A}).\footnote{ First-time attachment events are excluded from the pool of attachment events from which $\bar k_a$ was computed, as their inclusion here significantly alters the relationship between $\bar k_a$ and $d/(Nb)$. In contrast their exclusion led to consistent $\bar k_a$ versus $d/(Nb)$ relations, regardless of whether sampling was conducted over the entire simulation duration, or only the simulation after approximately steady state fractions of attached/detached chains were reached.} The attachment rates were further partitioned into partner exchange rates, $\bar k_{exc}$ (\textbf{Fig. \ref{fig: Associative kinetics}B}), and repeat attachment rates, $\bar k_{rpt}$ (\textbf{Fig. \ref{fig: Associative kinetics}C}). These were also computed using Eq. \eqref{eq: Measured Attachment Rate}, except the number of attachment events, $N_a$, was simply replaced by the number of exchange events, $N_{exc}$, or repeat events, $N_{rpt}$, respectively. Values of average detached lifetime, $\bar \tau_d$, average detached lifetime prior to partner exchange, $\bar \tau_{exc}$, and average detached lifetime prior to repeat attachment, $\bar \tau_{rpt}$, were directly measured from both models (insets of \textbf{Fig. \ref{fig: Associative kinetics}}). The average detachment rate, $\bar k_d$ (computed using Eq. \eqref{eq: Measured Detachment Rate}) and average fractions of attached and detached chains at steady state ($f_a$ and $f_d$, respectively) are provided for reference (\textbf{Fig. \ref{fig: Extended bond exchange results 1}}), with steady state defined as occurring once $df_a/dt = -df_d/dt < 10^{-4} \tau_0^{-1}$. Average values of attached bond lifetime, $\bar \tau_a$, and renormalized bond lifetime, $\bar \tau_{rnm}$ are also provided for reference (\textbf{Fig. \ref{fig: Extended bond exchange results 2}}). Importantly, when plotted with respect to chain concentration, $c=d^{-3}$, the curves for all outputs reported in \textbf{Fig. \ref{fig: Associative kinetics}} approximately collapse into single curves irrespective of chain length (\textbf{Fig. \ref{fig: Plot timescales wrt conc}}). This is consistent with the predictions of \cite{stukalin_self-healing_2013} and provides additional evidence that bond kinetics within ensembles of chains are governed by sticker concentration, independent of chain length. Distances, $d$, may be also be interchanged with chain stretches, $\lambda_c = d/(\sqrt{2N}b)$, in \textbf{Fig. \ref{fig: Associative kinetics}} to facilitate a more direct comparison to the results of \textbf{Section \ref{sec: Results of single-chain}}. However, we here visualize results in terms of $d/(Nb)$, which most clearly elucidates reasonable agreement between the bead-spring and mesoscale models across distinct chain lengths.
    
\textbf{Figs. \ref{fig: Associative kinetics}} and \textbf{\ref{fig: Extended bond exchange results 1}}-\textbf{\ref{fig: Extended bond exchange results 2}} demonstrate that the bead-spring and mesoscale models' predictions of kinetic rates, steady state bond fractions, and bond lifetimes are in reasonable agreement across all tether separation distances, $d$, and chain lengths, $N$. As in prior sections, the dissociation rates universally match the value set \textit{a priori} (\textbf{Fig. \ref{fig: Extended bond exchange results 1}.A}). However, for the purposes of this study we focus on the associative kinetics, which divulge additional information about repeat attachment versus partner exchange. The magnitudes of the associative kinetic rates are generally on the order of $10^{-3} \tau_0^{-1}$, while corresponding detached bond lifetimes are on the order of $10\tau_0$ to $10^3 \tau_0$, confirming that adequate bond kinetic sampling was established per \textbf{Appendix \ref{Appendix - dt convergence}}.  As expected, attachment rates decrease with increasing chain separation and chain length. However, attachment rates no longer exhibit Gaussian scaling with respect to separation distance as they did on a pairwise attachment basis through Eq. \eqref{eq: Attachment rate scaling theory} in \textbf{Section \ref{sec: Pairwise attachment}}. Instead, the attachment rates for ensembles of chains evolve as:
\begin{equation}
    \bar k_i = k_i^{semi} \left(\frac{d_{max}}{d} - 1 \right),
    \label{eq: empirical ensemble attachment}
\end{equation}
where $d_{max}$ is the maximum empirically predicted tether-to-tether separation distance at which attachment occurs, $k^{semi}_i$ is the attachment rate when $d=d_{max}/2$, and $k_i$ denotes either overall attachment rate, $k_a$, partner exchange rate, $k_{exc}$, or repeat attachment rate, $k_{rpt}$. Fit curves are displayed in \textbf{Fig. \ref{fig: Associative kinetics}}, and empirical values of maximum chain stretch, $\lambda_c^{max}=d_{max}/( \sqrt{2N}b)$, minimum chain concentration, $c_{min} b^3=(b/d_{max})^{3}$, and minimum packing fraction, $\phi_{min}=\pi N b^3/(6 d_{max}^3)$, at which associative kinetics occur, are provided in \textbf{Fig. \ref{fig: emperical kinetics fits}.A-C}. \textbf{Fig. \ref{fig: emperical kinetics fits}.D-F} provides the characteristic attachment rates $k_a^{semi}$, $k_{exc}^{semi}$, and $k_{rpt}^{semi}$, for each event type. 
   
    \figuremacro{H}{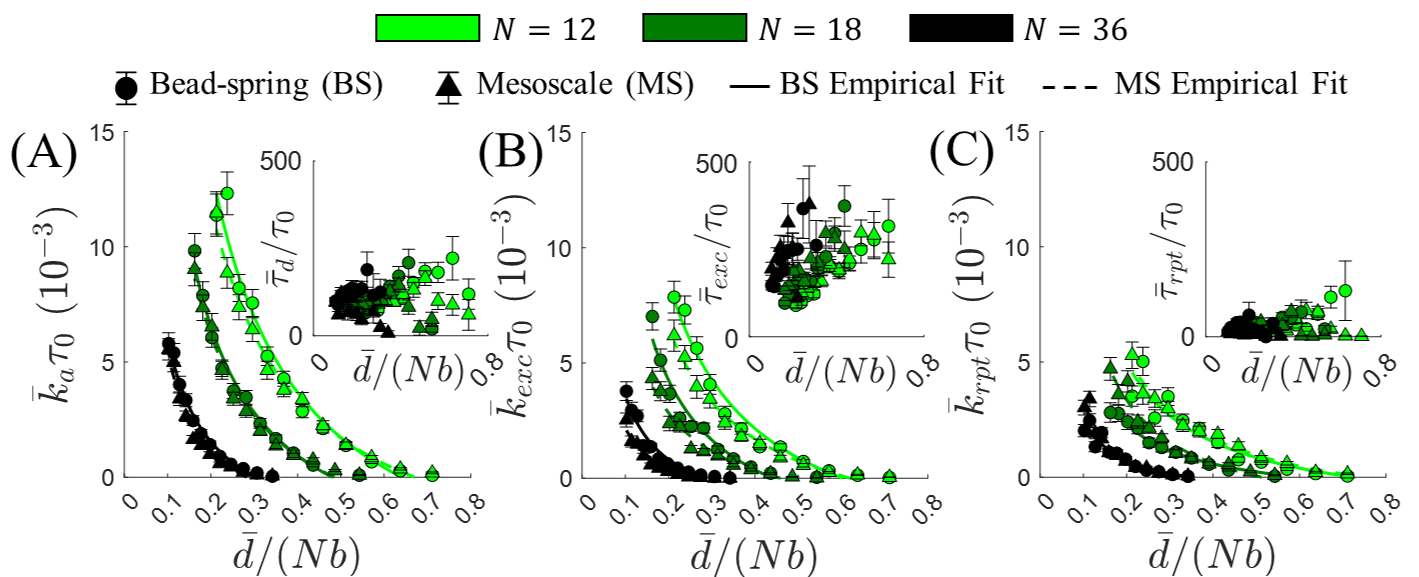}{Associative ensemble bond kinetics. }{\textbf{(A)} overall bond attachment, $\bar k_a$, \textbf{(B)} exchange attachment, $\bar k_{exc}$, and \textbf{(C)} repeat attachment, $\bar k_{rpt}$, rates with respect to normalized chain separation, $\bar d/(Nb)$. Insets in \textbf{(A-C)} display average detached lifetimes, $\bar \tau_d$, detached lifetimes prior to partner exchange, $\bar \tau_{exc}$, and detached lifetimes prior to repeat attachment, $\bar \tau_{rpt}$, respectively. Error bars represent standard error of the mean. The empirical model of Eq. \eqref{eq: empirical ensemble attachment} is fit to all sets of bead-spring and mesoscale data (solid and dashed curves, respectively).  \label{fig: Associative kinetics}}{0.9}
    
Maximum chain stretch at which attachment occurs is generally on the order of $\lambda_{max} \approx 1.5$, consistent with the findings of \textbf{Sections \ref{sec: Pairwise attachment}}. Neither minimum chain concentration nor minimum packing fraction at which attachment kinetics occur appear to vary significantly with chain length (within the 95\% confidence interval), nor do the empirically fit values of $k_a^{semi}$, $k_{exc}^{semi}$, and $k_{rpt}^{semi}$ (\textbf{Fig. \ref{fig: emperical kinetics fits}}). Again, this supports the notion that sticker concentration dictates bond kinetics \citep{stukalin_self-healing_2013}. While any effects of chain length on these empirical parameters are muted, agreement between models appears strongest for the shortest chain length, which is visually expressed in the attached and detached chain fractions (\textbf{Fig. \ref{fig: Extended bond exchange results 1}.B-C}). Although reasonable agreement between models is achieved, the steady state fraction of attached chains is consistently higher for the bead-spring model than mesoscale model at lower chain concentrations (i.e., higher values of $d$). This reveals a disparity in the attachment rates at low chain concentrations that is not otherwise obvious (due to their low magnitudes in \textbf{Fig. \ref{fig: Associative kinetics}A}). Specifically, it suggests that the attachment rates of the bead-spring model outstrip those of the mesoscale model for long chains at low concentrations, which has meaningful network-scale consequences discussed further in \textbf{Section \ref{sec: Results of network-scale}}.  

Aside from facilitating comparison of the two models, the results of \textbf{Fig. \ref{fig: Associative kinetics}} also highlight how {\AA}ngstrom and nanometer binding length scales -- if not calibrated on a bond-specific basis -- could drive inaccuracies that propagate upwards in length and time for either of these approaches. For instance, the exchange rates at lower separation distances (or higher concentrations) in \textbf{Fig. \ref{fig: Associative kinetics}B} are greater than the corresponding repeat attachment rates in \textbf{Fig. \ref{fig: Associative kinetics}C}. This result is unexpected \citep{stukalin_self-healing_2013}, but has been carefully confirmed and is an artifact of our models' assumptions. It arises from the fact that the maximum sticker-to-sticker attachment distance is set equal to the equilibrium length of dynamic bonds (both are $b$). Dynamic bond lengths are normally distributed according to \textbf{Fig. \ref{fig: Appendix_bond_potential}.A}. Their length is nominally $b$ plus or minus the standard deviation of this distribution, so that their dissociation does not necessarily leave their stickers within immediate reattachment range ($\leq b$) of each other. Since the number of free stickers available for partner exchange is typically greater than the number of stickers available for repeat attachment (one), exchange kinetics occur more frequently and this result is more pronounced at higher chain concentrations. Thus, when applying these models to evaluate long-term network structures or stress responses, an initial, high-resolution modeling approach (e.g., density functional theory) or appropriate leveraging of prior work is strongly recommended to justify the prescription of associative length scales. Prescription of a variable, short-ranged binding probability based on the competition between binding potential and kinetic energy may also be considered. 

\section{Network-scale Mechanical Response}
\label{sec: Results of network-scale}

To evaluate whether the bead-spring and mesoscale models return reasonable predictions of network-scale mechanical stress response, we simulated polymer networks undergoing uniaxial extension and stress relaxation while monitoring the stress response. All simulations described in this section were run on the the 80-core Ice Lake (ICX) compute nodes of Stampede3, using the built-in MPI capability of LAMMPS. First, we generated networks of $n_p=60$ branched polymeric chains inside of periodic RVEs. Each polymer has five branched side groups of length $Nb$ tethered to its backbone at even intervals of $Nb$ (\textbf{Fig. \ref{fig: Network model summary}A}). Branches also terminate the chains as illustrated in \textbf{Fig. \ref{fig: Network model summary}A}. For detailed network initiation procedures see \textbf{Appendix \ref{sec: Network Initiation}}. Once initiated, each networks' nodes were stepped through time according to \textbf{Section \ref{sec: Equation of Motion}} and the distal stickers of their side chains were allowed to dynamically attached to and detach from one another according to \textbf{Section \ref{sec: Bond Kinetics}}. In order to reasonably reproduce the MSDs, bond kinetics, and initial network topologies of the bead-spring model, we find that the damping coefficient must be set to the expected value of $\gamma = Nz \gamma_0/2$, based on Rouse theory \citep{rouse_theory_1953,rubinstein_polymer_2003}. Here, $z$ is the number of chains attached to each crosslink ($z=3$ for tethering crosslinks and $z=1$ for stickers) and the multiple of $1/2$ enforces that each node of the mesoscale model encapsulates half of the Kuhn segments of each chain it is attached to. 

Once the networks were formed and equilibrated for some time, $t_{eq}$, the RVEs were uniaxially extended to a maximum network stretch of $\lambda = 3$, at a constant true strain rate, $\dot \varepsilon$. Stretch was applied by deforming the boundaries of the RVE according to the volume conserving deformation gradient $\bm F(t) = $ diag$(e^{\dot \varepsilon t},e^{-\dot \varepsilon t/2},e^{-\dot \varepsilon t/2})$, where $\dot \varepsilon$ denotes the true strain rate in the direction of stretch. Once full extension was reached, the deformation gradient $\bm F(t) = $ diag$(3,1/\sqrt 3, 1/\sqrt 3)$ was sustained for $t_{rlx}= 220 \tau_0$ so that stress relaxation responses could be evaluated. The loading history in the direction of extension is plotted in \textbf{Fig. \ref{fig: Network model summary}B}. To attain adequate sampling, the results of $n=15$ simulations were ensemble averaged for every parameter combination described below. These samples consisted of five initial network structures equilibrated over three different durations, $t_{eq} = \{220,275,330 \}\tau_0$, for each parameter combination.

    \figuremacro{H}{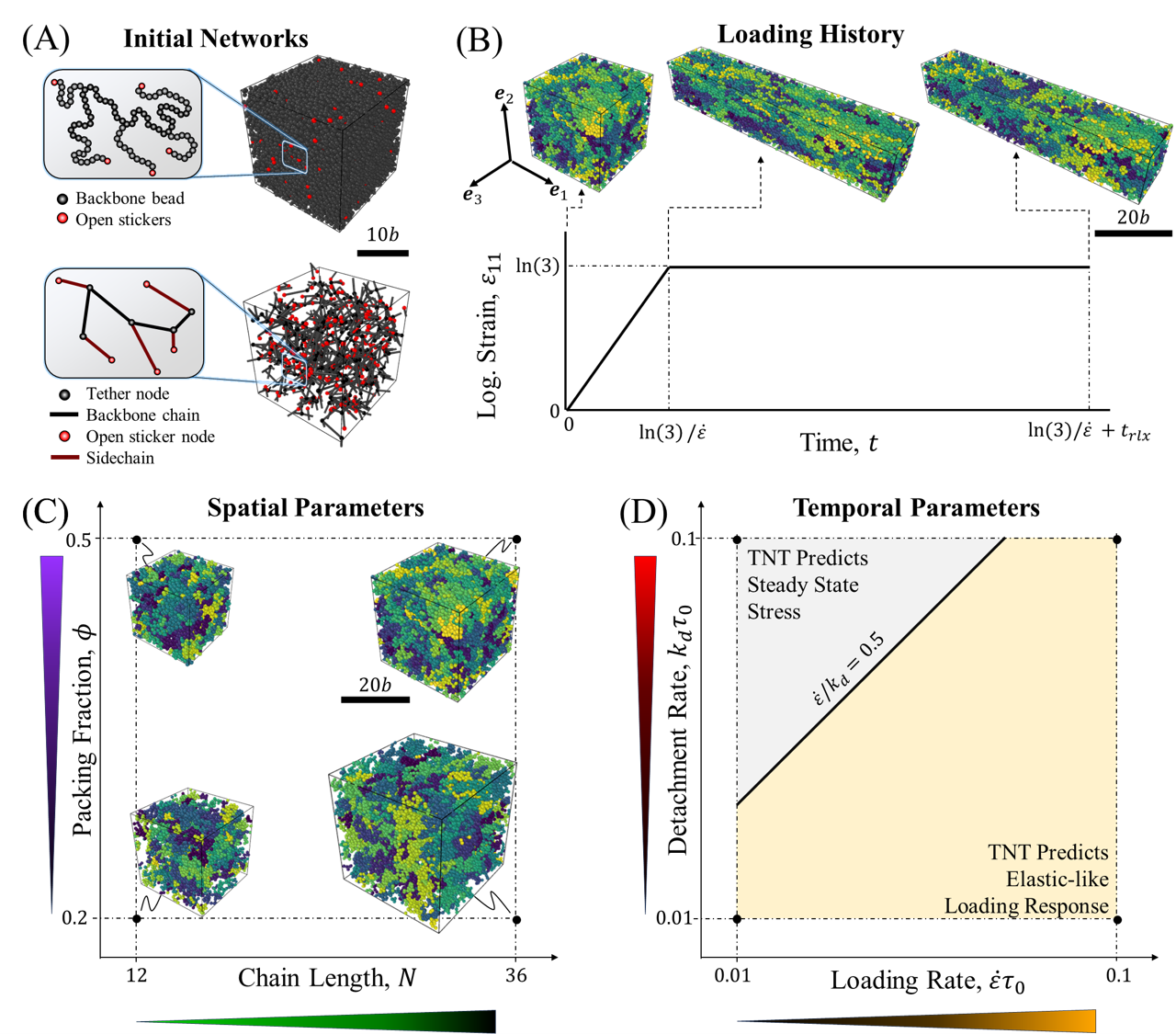}{\textbf{Network-scale parameter space. }}{\textbf{(A)} Initiated bead-spring (top) and mesoscale (bottom) model RVEs comprised of $n_p=60$ polymers where red nodes indicate the positions of stickers. \textbf{(B)} Loading history applied to the RVEs with respect to time. Snapshots of the bead-spring RVE (\textbf{top}) visualize the loading in time. The orthonormal basis $\{\bm e_1, \bm e_2, \bm e_3 \}$ is denoted with the principle direction of stretch as $\bm e_1$. \textbf{(C-D)} Parameter spaces indicating the ranges of investigated \textbf{(C)} spatially and \textbf{(D)} temporally related inputs. Snapshots of the network in \textbf{(C)} illustrate simulated RVEs of the bead-spring model at the extreme chain lengths ($N=\{12,36\}$) and packing fractions ($\phi=\{0.2,0.5\}$). The grey and yellow regions in \textbf{(D)} indicate where linear TNT with constant bond kinetic rates predicts steady state stress onset and elastic-like response during loading, respectively. \label{fig: Network model summary}}{1}
       
To evaluate mechanical response, Cauchy stress was computed throughout simulations as the potential energy-governed component of the virial stress:
\begin{equation}
    \bm \sigma = V^{-1} \sum_\alpha \sum_\beta \bm r^{\alpha \beta} \otimes \bm f^{\alpha \beta} - \pi \bm I.
    \label{eq: virial stress}
\end{equation}
where $V$ is the total RVE volume, $\bm r^{\alpha \beta}$ denotes the end-to-end vector spanning from node $\alpha$ to its attached neighbor $\beta$, $\bm f^{\alpha \beta} = - \partial \psi / \partial \bm r^{\alpha \beta}$ is the force between $\alpha$ and $\beta$, and the isotropic pressure term $\pi \bm I$ represents the volume conserving hydrostatic stress from excluded volume interactions. Since a Lenard-Jones potential is included in the bead-spring model but no such interactions are modeled between the implicit chains of the mesoscale model, $\pi \bm I$ is neither expected nor intended to necessarily match between approaches. Rather, the focus of this study is on each model's prediction of the change in stress due to reorientation, stretch, and reconfiguration of its chains in response to loading. Thus, after equilibration, initial stress is taken as $\bm \sigma = \bm 0$ so that $\pi \bm I$ is equal and opposite to the first term from Eq. \eqref{eq: virial stress}. For both models, $\alpha\beta$ pairs are treated as the polymer chains comprised of $N$ Kuhn segments that chemically attach tether-tether or tether-sticker pairs so that virial stress is computed at the network or ``mesh-scale''. Thus, the free energy of chains is represented by Eq. \eqref{eq: Langevin force}, based on the models' agreement between force extension relations in \textbf{Fig. \ref{fig: Appendix_bond_potential}.B}. It is also possible to directly measure the virial stress in the bead-spring model by carrying out the sum in Eq. \eqref{eq: virial stress} over all Kuhn segments (\textbf{Fig. \ref{fig: Virial Lammps vs mesoscale}}). This approach delivers good agreement with mesh-scale estimates of virial stress for the bead-spring model (except at very high loading rates when $\dot \varepsilon=0.1\tau_0^{-1}$ per \textbf{Fig. \ref{fig: Virial Lammps vs mesoscale}.C}). However, to reduce noise and improve clarity for discussion, we present results within this section using the mesh-scale virial stress.

These network models maintain the assumptions of ideal, monodisperse chains without long-range interactions. For these simplified conditions, we investigate the effects of four key parameters that may influence model agreement. The first two parameters -- related to spatial occupancy of the chains -- are chain length, via $N = \{12,36 \}$, and polymer packing fraction, $\phi = \{0.2,0.5\}$ (\textbf{Fig. \ref{fig: Network model summary}C}). The limits of $N$ and upper limit of $\phi$ were selected for consistency with \textbf{Section \ref{sec: Results of single-chain}}. The lower limit of $\phi=0.2$ was selected because percolated networks were not always observed when $\phi<0.2$. The other two swept parameters -- related to timescales of the simulations -- are the true strain rate, $\dot \varepsilon = \{ 0.01, 0.1 \} \tau_0^{-1}$, and detachment rate, $k_d = \{0,0.1\} \tau_0^{-1}$ (\textbf{Fig. \ref{fig: Network model summary}D}). The range of $\dot \varepsilon$ was selected because it results in microsecond-scale simulations, but still permits ensemble sampling using the bead-spring model at reasonable computational cost. The range of $k_d$ was selected because setting $k_d=0$ begets near-permanent network structures after equilibration (allowing only for attachment of residually dangling stickers), while setting $k_d = 0.1 \tau_0^{-1}$ induces near-complete stress relaxation over relaxation times comparable to the loading times. The intrinsic attachment rate was held constant at a high value, $k_a \approx \tau_0^{-1}$, to ensure that solid-like networks structures were maintained.
    
We began by simulating both models at the extremes of these four parameter ranges. For all cases investigated, the mesoscale model results in at least an 86\% reduction in computational runtime (that scales with the number of polymer chains as $\% \Delta t_{cpu}=1-0.81 n_p^{-0.44}$), and an approximately 90\% reduction in data storage requirements (\textbf{Fig. \ref{fig: Computational cost}}). These savings immediately highlight the utility of the mesoscale model for exploring larger spatiotemporal domains (discussed further in \textbf{Section \ref{sec: Spatiotemporal Extrapolation}}). To compare the models' predictions, \textbf{Figs. \ref{fig: N12}} and \textbf{\ref{fig: N36}} display the normal Cauchy stress, $\sigma_{11}$, in the direction of loading with respect to time for every combination of $\phi$, $k_d$, and $\dot \varepsilon$ when $N=12$ and $N=36$, respectively. Results are presented in SI units of MPa with respect to $\mu$s, demonstrating that reasonable orders of magnitude stress are predicted by these models when compared to rubbery elastomers \citep{qi_durometer_2003,vatankhah-varnosfaderani_mimicking_2017} or small-mesh gels \citep{shibayama_precision_2019}.\footnote{Magnitudes of stress may be renormalized by simply adjusting the Kuhn length per \textbf{Appendix \ref{sec: Parameters and Unit Normlization}}.} Error between models is expressed with respect to time directly beneath each plot as the difference between mesoscale and bead-spring model stress, normalized by the peak bead-spring model stress. This measure (used throughout the remainder of this section) avoids overemphasizing error at small values of stress (e.g., prior to loading or after relaxation), while still allowing for meaningful error comparison between systems with different magnitudes of absolute stress (e.g., networks at higher versus lower packing fractions). Positive error indicates that the mesoscale model predicts higher values of stress than the bead-spring model. For permanent networks loaded at $\dot \varepsilon = 0.01\tau_0^{-1}$ and comprised of chains with $N=12$ Kuhn lengths between crosslinks (\textbf{Fig. \ref{fig: N12}A}), the models are in reasonable agreement at both packing fractions. Peak error of $16\%$ between the models' stress predictions occurs at the end of the loading phase and the models converge to similar values of elastic stress with less than $2\%$ error between them. This suggests similar initial network structures after equilibration.

    
Even in the absence of bond detachment (\textbf{Fig. \ref{fig: N12}A-B}), stress relaxation is still observed in both models due to frictional drag that retards conformational chain relaxation. Due to drag during loading, both models predict elastic-like responses of monotonically increasing stress with no reduction in Young's modulus (\textbf{Fig. \ref{fig: N12}C}), even in the regime in which linear TNT with constant bond kinetics predicts onset of steady state stress (i.e., when the Weissenberg number, $\dot \varepsilon/k_d < 0.5$, \textbf{Fig. \ref{fig: Network model summary}D}). This is attributed to the fact that loading rates are relatively high ($\dot \varepsilon \not \ll \tau_0^{-1}$) and higher loading rates impart greater particle velocities that increase drag effects. Indeed, increasing the loading rate from $\dot \varepsilon=0.01\tau_0^{-1}$ to  $\dot \varepsilon=0.1\tau_0^{-1}$ while maintaining $k_d=0$ and $N=12$ (\textbf{Fig. \ref{fig: N12}B}), imparts higher peak mechanical stresses. Importantly, this increase in loading rate also increases peak error between the approaches from $16\%$ (\textbf{Fig. \ref{fig: N12}A}) to $68\%$ (\textbf{Fig. \ref{fig: N12}B}). This implicates drag forces as a major potential source of error between models and warrants further investigation into the effects of not only loading rate, but also chain length since damping coefficient depends on $N$. 

Observing \textbf{Fig. \ref{fig: N36}}, the mesoscale model under-predicts the stress response of the bead-spring model at all combinations of $\dot \varepsilon$, $k_d$, and $\phi$ when $N=36$ (\textbf{Fig. \ref{fig: N36}}). Notably, very little change in mechanical response is observed as $k_d$ is increased from $0$ to $0.1\tau_0^{-1}$, suggesting that the stress response is not topologically governed, but is rather almost entirely governed by drag for these longer-chained networks (\textbf{Fig. \ref{fig: N36}}). Therefore, this disparity between models is likely attributable to differences in the localization of drag forces within each approach that are exacerbated as $N$ is increased. Since drag forces increase with $N$ in Eq. \eqref{eq: Brownian Equation of Motion} (i.e., $\gamma^\alpha d\bm x^\alpha/dt \propto N$) but entropic tensile forces decrease with $N$ (i.e., $\bm f^{\alpha \beta} \propto N^{-1}$), any differences between models' drag effects may become more pronounced for long-chained networks.
Differences inevitably arise because the bead-spring model distributes drag forces evenly along the lengths of its chains, whereas the mesoscale model concentrates these forces at crosslink sites. To better understand these effects while mapping the regimes in which the mesoscale model is best suited, the remainder of this section provides deeper investigation into the effects of chain length (\textbf{Section \ref{sec: Chain Length Sweep}}), loading rate (\textbf{Section \ref{sec: Loading Rate Sweep}}), and detachment rate (\textbf{Section \ref{sec: Detachment rate sweep}}). The polymer packing fraction is not investigated in further detail since little difference is observed in the error between models when $\phi$ is increased from $0.2$ to $0.5$ (\textbf{Fig. \ref{fig: N12}}). 
  
    \figuremacro{H}{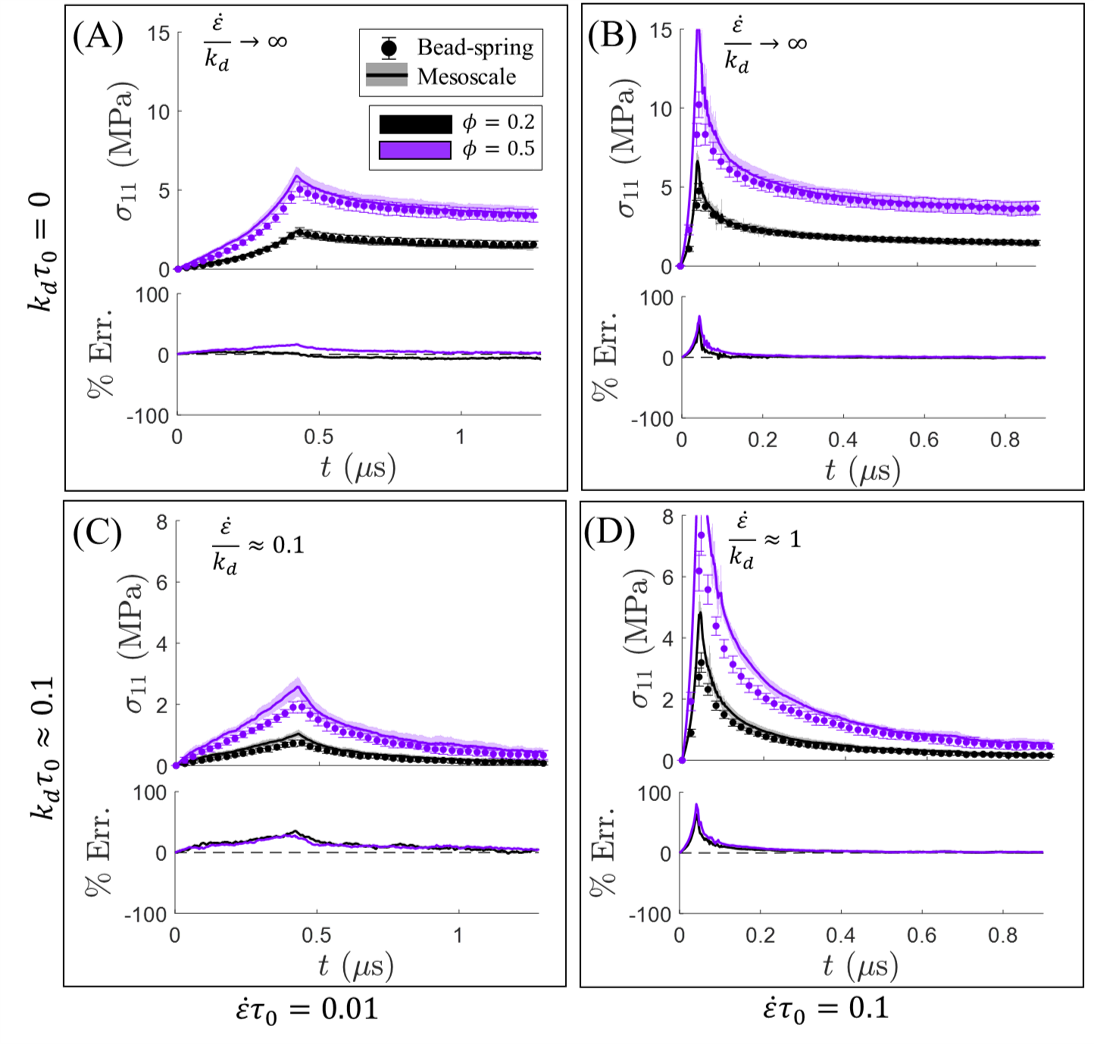}{\textbf{Mechanical response at the parameter extremes when $\bm{N=12}$. }}{Normal Cauchy stress in the direction of loading for the bead-spring and mesoscale models and a modified measure of relative error (mesoscale model stress minus bead-spring model stress as a percentage of peak stress bead-spring model stress), are plotted with respect to time at both polymer packing fractions ($\phi = \{0.2,0.5\}$), loading rates ($\dot \varepsilon = \{0.01,0.1\}\tau_0^{-1}$), and bond detachment rates ($k_d=\{0,0.1\}\tau_0^{-1}$) when $N=12$. Horizontal dashed lines on error plots denote zero error. \label{fig: N12}}{0.95} 
    
    

\subsection{Short chains impart topological and mechanical agreement}
\label{sec: Chain Length Sweep}

To investigate the effects of chain length, we initiated and deformed permanent networks ($k_d=0$) comprised of chains with $N=\{12,18,24,30,36\}$ Kuhn segments according to the procedure and loading described above when $\dot \varepsilon=0.01\tau_0^{-1}$ and $\phi=0.2$. Cauchy stress and relative error are plotted with respect to time in \textbf{Fig. \ref{fig: Chain length sweep}A-B}. \textbf{Fig. \ref{fig: Chain length sweep}B} reveals that error is generally lowest (never exceeding 8\% and dropping to less than 5\% after stress relaxation) when $N=12$, and magnitudes of relative error increase with $N$.

    \figuremacro{H}{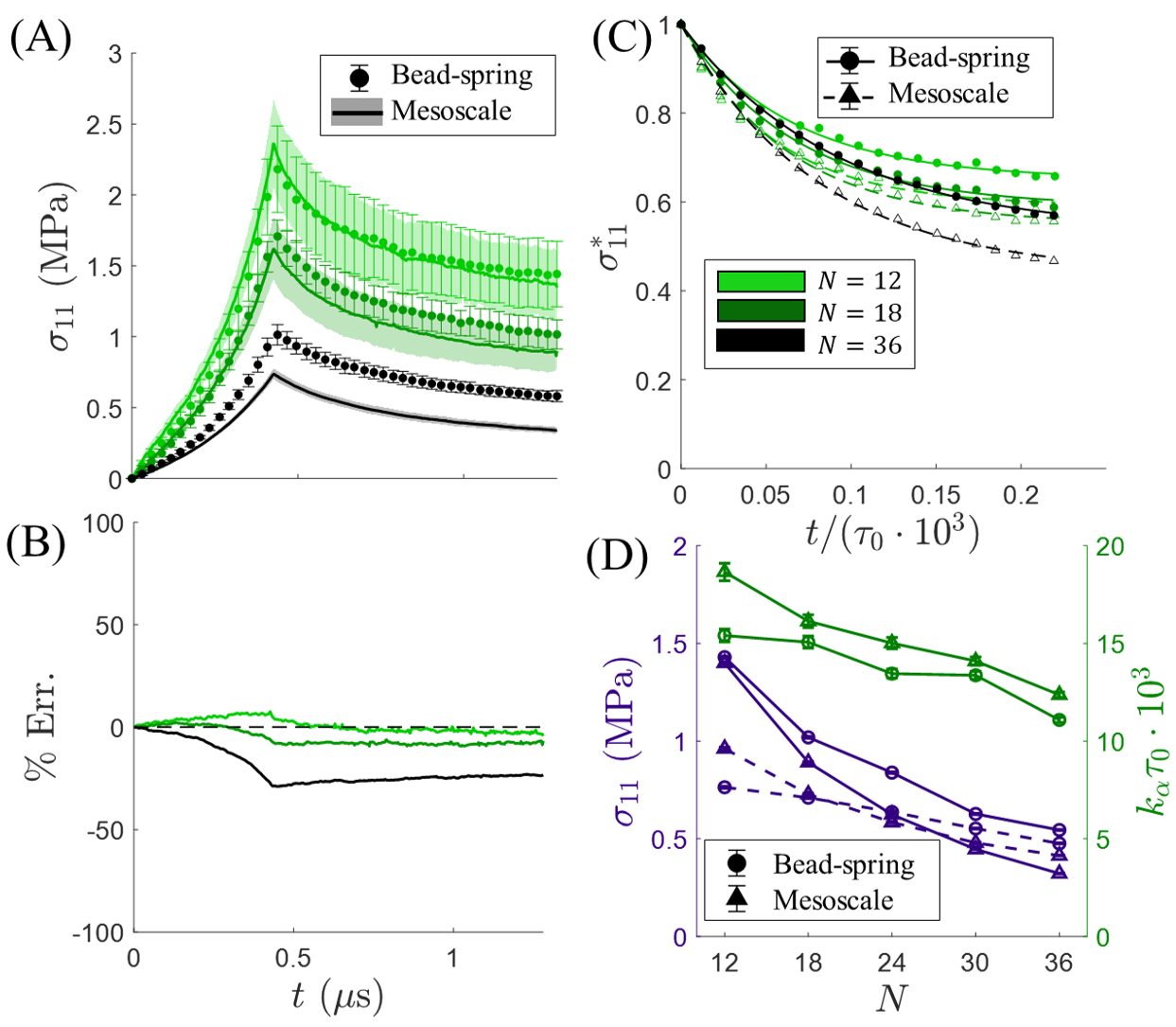}{\textbf{Effects of chain length. }}{\textbf{(A)} Cauchy stress in the direction of loading and \textbf{(B)} error between the mesoscale and bead-spring models (relative to the peak error of the bead-spring model) with respect to time when $\phi=0.2$, $k_d = 0$, $\dot \varepsilon =0.01 \tau_0^{-1}$ and $N=\{12,18,36\}$. \textbf{(C)} Cauchy stress (normalized by peak stress) with respect to time. Solid and dashed curves represent best fits of Eq. \eqref{eq: Maxwell} to the bead-spring and mesoscale data, respectively. \textbf{(D)} Stored stress, $\sigma_e$ (solid purple), and dissipated stress, $\sigma_d$ (dashed purple), along with $\alpha$-relaxation rate, $k_\alpha$, (solid green) for the mesoscale model (triangles) and bead-spring model (circle) with respect to $N$. Error bars represent the 95\% confidence interval from nonlinear least-squares regression analysis.  \label{fig: Chain length sweep}}{0.8}
    
Since no bond dissociation is allowed here ($k_d=0$), error in stress responses between models must originate from differences in initial topology, frictional damping, or both. Observing \textbf{Fig. \ref{fig: Chain length sweep}A}, the mesoscale model under-predicts bead-spring model stress when $N=36$ even near the end of stress relaxation, strongly suggesting differences in network structure. We find that mesoscale networks of long chains ($N=36$) -- although containing the same relative fractions of attached/detached bonds as bead-spring networks (\textbf{Fig. \ref{fig: Chain length attached fracs}A}) -- express far higher average degrees of non-load transmitting bonds between branches of the same molecules (\textbf{Fig. \ref{fig: Chain length attached fracs}.B}). This is attributed to much slower diffusion of the mesoscale (versus bead-spring) nodes during network initiation and equilibration (see \textbf{Fig. \ref{fig: Chain length MSDs}} for MSD data). It appears that the slower diffusion of the stickers and tethers in the mesoscale model predisposes the networks comprised of the highest chain lengths to clustering and intra-molecular bonding, and that this effect is sharply more pronounced when $N=36$ (\textbf{Fig. \ref{fig: Chain length attached fracs}.B}). Fewer percolated load paths in the mesoscale model results in reduced strain energy storage \citep{you_model_2024} and under-predicted mechanical stress as compared to the bead-spring model. This emphasizes the importance of replicating associative bond kinetics between models and affirms network structure as a secondary effect of frictional damping.

To better understand the effects of chain length on frictional damping, we isolate the stress relaxation response. \textbf{Fig. \ref{fig: Chain length sweep}C} displays the stress relaxation response normalized by peak stress, $\sigma_{11}^{max}$. We may fit a set of two parallel Maxwell elements to this response following the form:
\begin{equation}
    \sigma_{11} = \sigma_{11}^{max} \left[ f_\alpha \exp(-k_\alpha t) + (1-f_\alpha) \exp(-k_r t) \right] 
    \label{eq: Maxwell}
\end{equation}
where $f_\alpha$ represents the relative fraction of stress dissipated due to chain relaxation or ``$\alpha$-relaxation'', $k_\alpha$ represents the $\alpha$-relaxation rate mediated by drag forces (not to be confused with attachment rate, $k_a$), and $k_r$ represents the rate of reconfigurational relaxation driven by $k_d$. Since $k_d=0$, $k_r$ must also be zero so that Eq. \eqref{eq: Maxwell} simplifies to a Zener element model where $\sigma_d = \sigma_{11}^{max} f_\alpha$ becomes the dissipated stress, $\sigma_e = \sigma_{11}^{max}(1-f_\alpha)$ becomes the stored or ``equilibrium" stress, and $k_\alpha$ is definitively mediated by drag forces alone. Thus, deviations in $k_\alpha$ between models expose differences in the characteristic chain relaxation rates due to drag. Meanwhile, differences in $\sigma_e$ broadcast differences in the initial, permanent network structures, whereby fewer load carrying, inter-molecular attachments results in smaller values of $\sigma_e$ \citep{you_model_2024}. \textbf{Fig. \ref{fig: Chain length sweep}D} displays fit values of $\sigma_d$, $\sigma_e$, and $k_\alpha$ with respect to $N$. The $\alpha$-relaxation rates, $k_\alpha$, are consistently higher for the mesoscale model than bead-spring model. This faster conformational relaxation indicates that the mesoscale model, which lumps $Nz/2$ Kuhn segments into every node, under-represents frictional resistance of intermediate bead-spring chains, despite its slower diffusive exploration of crosslinks and stickers due to localization of drag forces. However, these effects appear to minimally effect stress response when $N=12$, for which diffusion kinetics (\textbf{Fig. \ref{fig: Chain length MSDs}}) and thus network topologies (see $\sigma_e$ in \textbf{Fig. \ref{fig: Chain length sweep}D}) are in reasonable agreement. Therefore, recommended practice is to limit the mesoscale model's implicit chains to contour lengths around $12b$.

\subsection{Slow loading bolsters mechanical agreement}
\label{sec: Loading Rate Sweep}

To investigate the effects of loading rate, we initiated and deformed networks comprised of chains with $N=12$ and $\phi=0.2$, at loading rates of $\dot \varepsilon = \{ 0.01, 0.02, 0.03, 0.06, 0.1\} \tau_0^{-1}$. To isolate the effects of loading rate on frictional $\alpha$-relaxation we again disable dissociation by setting $k_d=0$. Cauchy stress and relative error between models are plotted with respect to time in \textbf{Fig. \ref{fig: Loading rate sweep_kd0}A-B}. As expected based on prior sections, peak absolute error always occurs at the end of loading and increases monotonically with respect to loading rate. The predictions of both models always converge to values of equilibrated stress within 5\% error of one another under these conditions, further substantiating agreement in initial network topologies between approaches. 

To explore clear loading rate-dependence of the stress relaxation behavior, we isolate and fit the normalized stress relaxation response (\textbf{Fig. \ref{fig: Loading rate sweep_kd0}C-D}) with a stretched Kohlrausch-Williams-Watts exponential decay function \citep{richardson_hydrazone_2019} of the form:
\begin{equation}
    \sigma_{11} = \sigma_{11}^{max} \left\{ f_\alpha \exp \left[ -(k_\alpha t)^\chi)\right] + (1-f_\alpha) \exp(-k_r t) \right\}, 
    \label{eq: KWW}
\end{equation}
where $k_r$ remains zero, $k_\alpha$ is the nominal $\alpha$-relaxation rate, and the parameter $\chi \in [0,1]$ quantifies the degree of variable relaxation rate in time. If $\chi=1$ there is no variability, while if $\chi<1$ the relaxation rate decreases with time. The stretched exponential fit is chosen because networks loaded at faster rates experience greater chain stretches that correspond to disproportionately high entropic tension of their nonlinear chains (\textbf{Fig. \ref{fig: Appendix_bond_potential}.B}). These disproportionately high forces drive the networks' crosslinks into a relaxed state more quickly, resulting in a faster initial relaxation response that slows with time. 

    \figuremacro{H}{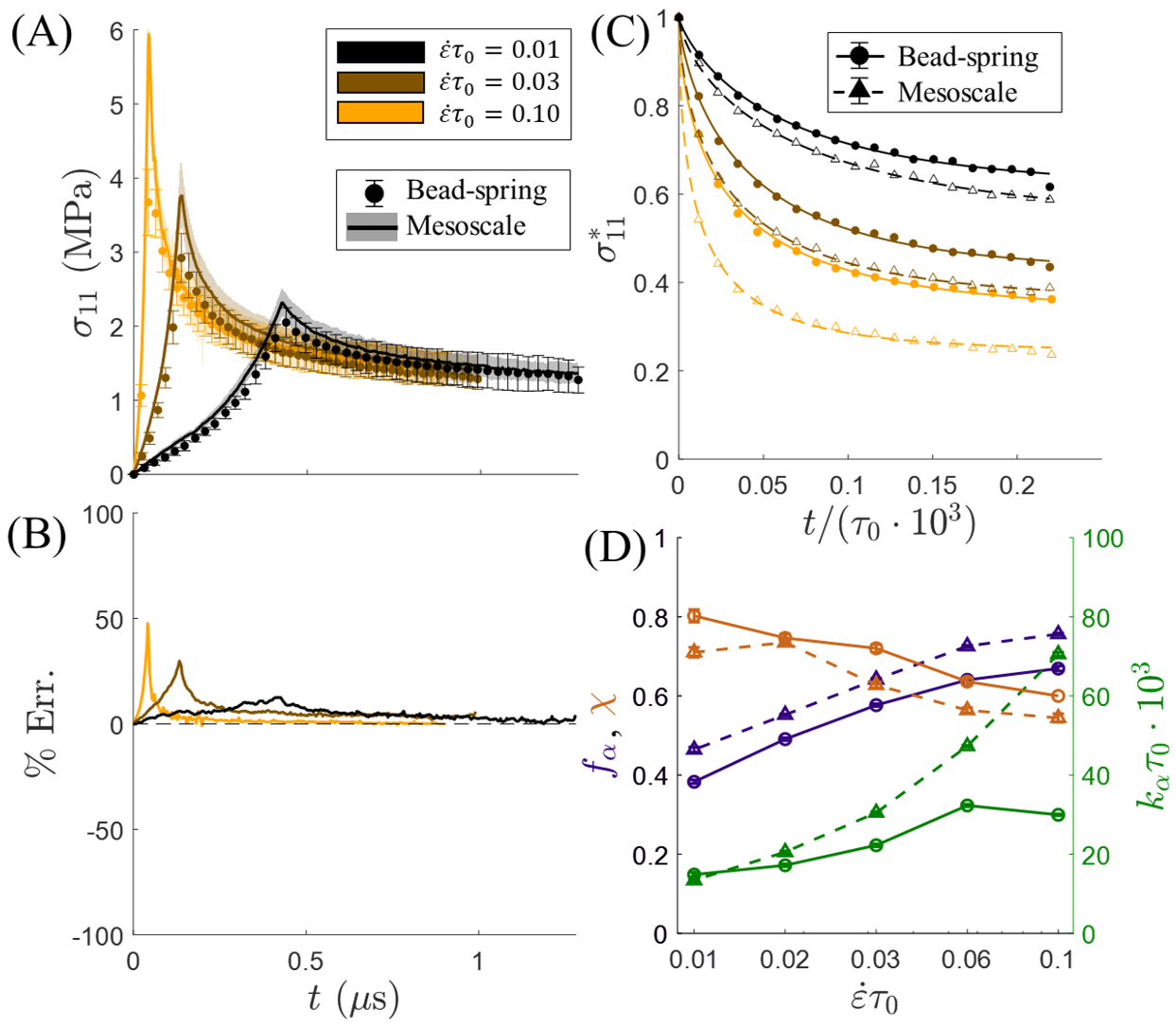}{\textbf{Effects of loading rate. }}{\textbf{(A)} Normal Cauchy stress in the direction of loading and \textbf{(B)} percent relative error between the bead-spring and mesoscale models plotted with respect to time when $\phi=0.2$, $N=12$, $k_d = 0$, and $\dot \varepsilon = \{ 0.01, 0.03, 0.1 \} \tau_0^{-1}$. \textbf{(C)} Cauchy stress, normalized by peak stress, during stress relaxation with respect to time. \textbf{(D)} Fit values of $f_\alpha$ (purple), $\chi$ (orange), and $k_\alpha$ (green) with respect to $\dot \varepsilon$ for the bead-spring (circles) and mesoscale (triangles) models. \label{fig: Loading rate sweep_kd0}}{0.8}

For both models, $\chi$ decreases with increasing loading rates, as expected, indicating more variable $\alpha$-relaxation rates when $\dot \varepsilon$ is higher. The relative fractions of dissipated stress, $f_\alpha$, and nominal relaxation rates, $k_\alpha$, both increase with respect to loading rate, also as expected. Comparing the modeling approaches, we see that the bead-spring model consistently exhibits lower relative fractions of dissipated stress, $f_\alpha$, consistent with the observation that peak stresses of the mesoscale model are greater than those of bead-spring model despite subsequently reaching close equilibrium stresses. While the nominal relaxation rates, $k_\alpha$, of the models are in good agreement at the slowest loading rate, $k_\alpha$ for the mesoscale model increase more dramatically with respect to strain rate. This rate dependence is consistent with the mesoscale model's under-prediction of frictional resistance to conformational change. These results suggests that the mesoscale model should be reserved for loading cases in which $\dot \varepsilon \leq 0.01\tau_0^{-1} $, which is both realistic and necessary for most polymers given that molecular oscillations are typically on the order of $10^{12}$ to $10^{14}$ Hz \citep{herzberg_molecular_1955}. Since slower loading rates require longer loading times, they are also more readily achieved by the mesoscale model than bead-spring approach due to its lower computational cost (see \textbf{Section \ref{sec: Spatiotemporal Extrapolation}}).
        
\subsection{Model agreement is achieved across detachment rates}
\label{sec: Detachment rate sweep}

To investigate the effects of detachment rate, we initiated and deformed networks comprised of chains with $N=12$ and $\phi=0.2$, while sweeping $k_d = \{0, 0.001, 0.003, 0.01, 0.03, 0.1\} \tau_0^{-1}$. The applied loading rate was held at $\dot \varepsilon = 0.01 \tau_0^{-1}$ based on the results of \textbf{Section \ref{sec: Loading Rate Sweep}}. Cauchy stress and relative error between models are plotted with respect to time in \textbf{Fig. \ref{fig: Detachment rate sweep}A-B}. Generally, the models are in reasonable agreement across detachment rates (\textbf{Fig. \ref{fig: Detachment rate sweep}A}), although the frictional effect that causes slight over-prediction of mesoscale model stress remains present. While relative error appears to increase with $k_d$ (because magnitudes of stress also decrease with $k_d$), maximum absolute values of error between models remain around 0.2 MPa regardless of $k_d$ (\textbf{Fig. \ref{fig: Detachment rate sweep}B}). These results suggest that $k_d$ does not drive significant deviation between the models.

To quantify the effect of $k_d$ on relaxation rates, we once more fit Eq. \eqref{eq: KWW} to the normalized stress relaxation response (\textbf{Fig. \ref{fig: Detachment rate sweep}C}). However, $k_r$ is now treated as a fitting parameter that is expected to scale directly with $k_d$. Fit parameters are plotted with respect to $k_d$ in \textbf{Fig. \ref{fig: Detachment rate sweep}D-E}. Examining \textbf{Fig. \ref{fig: Detachment rate sweep}D}, $0.8<\chi\leq 1$ for all $k_d$ of both models, indicating that chain stretch causes little variability in the relaxation rate for the prescribed loading rate\footnote{Since $k_d$ is deliberately made force-independent here, variability in relaxation rate cannot arise from force-dependent detachment and can instead only be attributed to stretch-dependent conformational relaxation.} Additionally, $\chi$ is uncorrelated with $k_d$, consistent with findings from \textbf{Section \ref{sec: Pairwise attachment}} that chains attach to one another within the linear regime of force-stretch (i.e., $\lambda < 1.5$). While the reconfigurational relaxation rate, $k_r$, scales directly with $k_d$ (\textbf{Fig. \ref{fig: Detachment rate sweep}E}, $p<0.01$ for both models), the $\alpha$-relaxation rate, $k_\alpha$, is uncorrelated with $k_d$ ($p>0.1$ for both models) since it is mediated only by the damping coefficient. Nonetheless, one might expect the fraction, $f_\alpha$, of stress dissipated due to conformational relaxation to decrease as $k_r$ increases. However, $f_\alpha$ is also independent of $k_d$ (\textbf{Fig. \ref{fig: Detachment rate sweep}D}, $p>0.2$ for both models). This underscores the coupling between configurational and conformational network relaxation that both models capture, whereby every detachment event results in new conformational degrees of freedom for chain relaxation \citep{stukalin_self-healing_2013,yu_effects_2014,wanasinghe_dynamic_2022,wagner_coupled_2023}. Despite the complex coupling between $k_r$ and $k_\alpha$, \textbf{Fig. \ref{fig: Detachment rate sweep}} broadly suggests that modulating $k_d$ does little to alter agreement in stress response between models. In the following section, we leverage this finding, as well as the reduced computational cost of the mesoscale model (\textbf{Fig. \ref{fig: Computational cost}}), to simulate large spatiotemporal domains and explore the regime in which $k_d \ll \tau_0^{-1}$ as is generally the case in real materials at operational temperatures \citep{richardson_hydrazone_2019,chen_rapid_2019}.

    \figuremacro{H}{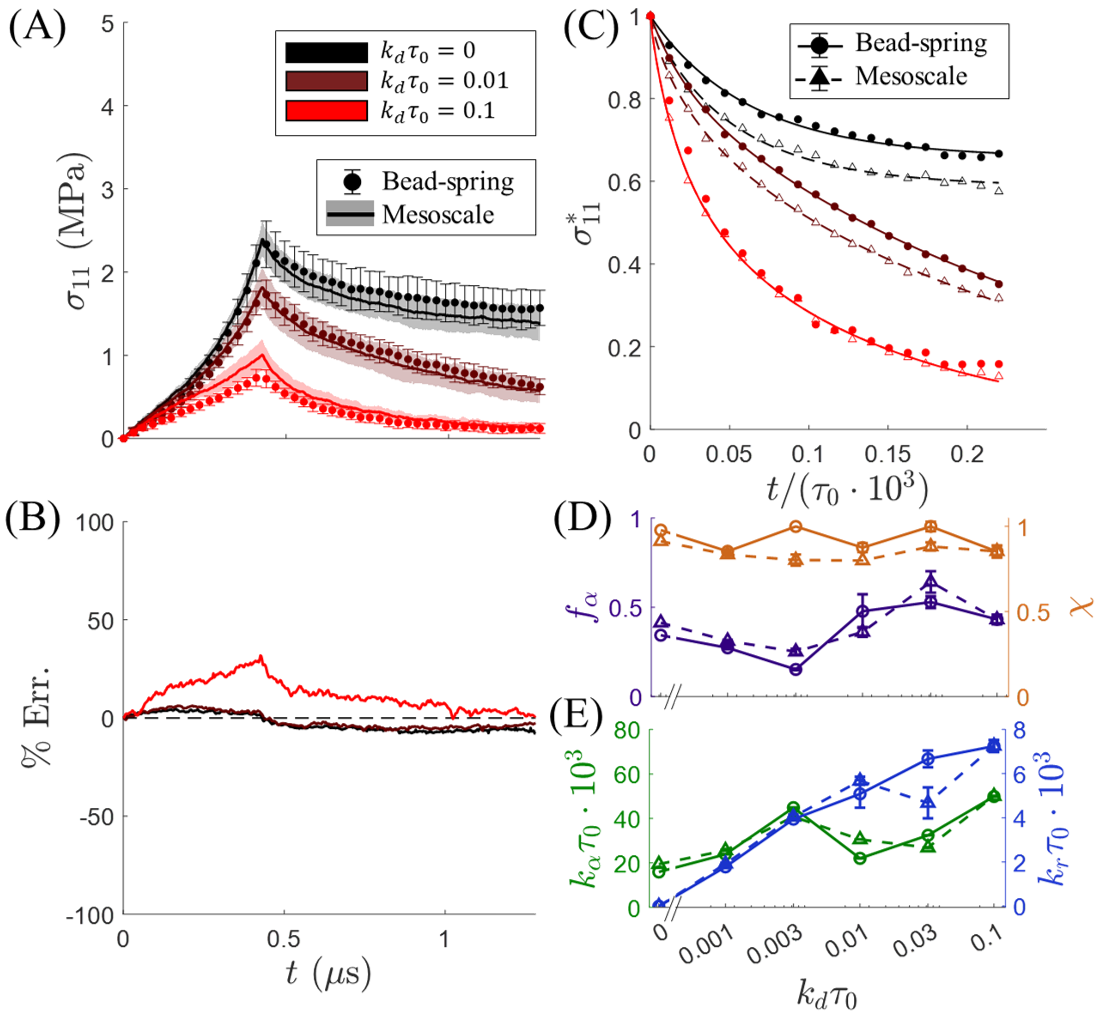}{\textbf{Effects of detachment rate. }}{\textbf{(A)} Normal Cauchy stress in the direction of loading and \textbf{(B)} percent relative error between the bead-spring and mesoscale models are plotted with respect to time when $\phi=0.2$, $N=12$, $\dot \varepsilon = (0.01) \tau_0^{-1}$ and $k_d =\{0,0.01,0.1\}\tau_0^{-1}$. \textbf{(C)} Cauchy stress normalized by peak stress during stress relaxation with respect to time. \textbf{(D-E)} Fit values of $f_\alpha$ (purple) and $\chi$ (orange), as well as \textbf{(E)} and $k_\alpha$ (green) and $k_r$ (blue), all with respect to $k_d$ for the bead-spring (circles) and mesoscale (triangles) models. \label{fig: Detachment rate sweep}}{0.8}


\section{Mesoscale access to larger spatiotemporal domains}
\label{sec: Spatiotemporal Extrapolation}

In \textbf{Section \ref{sec: Results of network-scale}}, we interrogated the effects of chain length, loading rate, and detachment rate at spatiotemporal scales readily accessible to the bead-spring model ($\sim 10$ nm and $\sim 1$ $\mu$s). Importantly, this restricts loading rates to fast regimes in which $\dot \varepsilon$ is on the order of 1\% to 10\% of $\tau_0^{-1}$ and frictional effects are dominant. By extension, this restricts worthwhile examination of the effects of detachment rates to regimes in which $k_d$ is roughly 10\% of $\tau_0^{-1}$, since crossover from predominantly elastic to viscous behavior tends to occur at loading rates on the order of $\dot \varepsilon \sim k_d$. Here we demonstrate how the mesoscale model's computational cost savings may be put towards modeling relatively large networks with dimensions on the order of 100 nm, over time domains on the order of 100 $\mu$s (i.e., $10^5 \tau_0$), thus lessening the gap between molecular theory and experimentally relevant scales. 

One feature observed in many dynamic polymers is the ability to undergo maximum global stretch ratios, $\lambda_{max}$, well over $10\times$ without failure \citep{zhang_superstretchable_2019,cai_highly_2022,xu_thermosensitive_2022} at relatively slow loading rates relative to $k_d$ (i.e., $\dot \varepsilon/k_d < 0.5$). However, to capture this effect in a Lagrangian RVE requires simulating slow loading rates ($\dot \varepsilon < k_d \ll \tau_0^{-1}$), at loading timescales up to $t_{load}= \ln (\lambda_{max})/\dot \varepsilon$. It also requires simulating RVEs with large enough initial dimensions, $L_0$, so that incompressible uniaxial extension does not result in domain widths less than the maximum allowable bond length (i.e., $L_0 \geq Nb \sqrt{\lambda_{max}}$). For instance, to model a network undergoing uniaxial stretch up to $\lambda_{max} = 20$, the initial RVE should have dimensions of no less than $4.6Nb$, which for the networks investigated here ($N=12$), corresponds to $L_0 \geq 34$ nm. Furthermore, supposing $\dot \varepsilon/k_d=0.5$, and $k_d\sim 10^{-3} \tau_0^{-1}$, then deformation would need to be applied for a duration of $t_{load}=\ln(20)/[0.5\times 10^{-3} \tau_0^{-1}]\approx 6\times 10^5\tau_0$. For the bead-spring model, whose time step is $\delta t=4\times 10^{-4}\tau_0$, this would require on the order of $10^9$ discrete iterations. While such simulations become impractical for the bead-spring approach, especially as $\lambda_{max}$ is increased or $\dot \varepsilon$ is decreased, the mesoscale model's reduced timestep and explicitly tracked number of particles renders modeling such deformations easily achievable.

To demonstrate the capabilities of the mesoscale model, we simulated networks with initial RVE dimensions of $L_0 \approx 70$ nm, undergoing incompressible, uniaxial extension to $\lambda_{max}=20$ per \textbf{Fig. \ref{fig: Large deformation}}. As in \textbf{Section \ref{sec: Results of network-scale}}, all simulations in this section were run using the ICX compute nodes of Stampede3 and LAMMPS MPI capability. Networks were initiated and equilibrated for $t_{eq}=220\tau_0$ per the procedure of \textbf{Appendix \ref{sec: Network Initiation}}. Based on the results of \textbf{Section \ref{sec: Results of network-scale}} the networks'  polymer chains are comprised of $N=12$ Kuhn lengths per chain segment, and the packing fraction was set to $\phi=0.2$ so that $n_p=5\times 10^3$ polymer chains are needed to attain the desired RVE size. In order to diminish the effects of frictional damping and meaningfully interrogate the effects of loading rate, we set $k_d=10^{-3} \tau_0^{-1}$ and swept the Weissenberg number over $\dot \varepsilon/k_d=\{ 0.125,0.25, 0.5, 1 \}$. Simulations were run until $\lambda_{max}=20$ was reached, or chains within the network approached stretches of $\sqrt N$ causing divergence in force and numerical instability.\footnote{Numerical instability may be circumvented by enforcing force-dependent bond dissociation \citep{puthur_theory_2002,shen_rate-dependent_2020,lamont_rate-dependent_2021,buche_chain_2021,wagner_network_2021,song_force-dependent_2021,buche_freely_2022,mulderrig_statistical_2023}. Here we focus on the rate-dependence originating from constant bond detachment rates and stopped simulations when chain lengths reached $r \approx Nb$ (at network stretches of $\lambda=10.5$ and $\lambda=4.5$ when $\dot \varepsilon/k_d = 0.5$ and $\dot \varepsilon/k_d=1$, respectively)}


\textbf{Fig. \ref{fig: Large deformation}B} displays the mechanical stress response of the simulated networks with respect to time for all four Weisenberg numbers. As expected, the mechanical response is highly rate-dependent. However, unlike the rate-dependence of \textbf{Section \ref{sec: Loading Rate Sweep}}, here we observe the onset of steady state stress for values of $\dot \varepsilon/k_d <0.5$, in accordance with TNT \citep{vernerey_statistically-based_2017}. This is representative of many dynamic polymers \citep{jeon_extremely_2016,cai_highly_2022} for which the rate of energy dissipation equilibrates with the rate of work input at sufficiently slow loading speeds. In contrast, when $\dot \varepsilon/k_d \geq 0.5$, an elastic-like response of monotonically increasing stress is observed. The true stress-strain responses for networks loaded at rates $\dot \varepsilon/k_d \geq 0.5$ assume the nonlinear shapes characteristic of rubbery, hyperelastic materials \citep{treloar_elasticity_1943}, and this effect is more pronounced when $\dot \varepsilon/k_d = 1$ than $\dot \varepsilon/k_d = 0.5$. \textbf{Fig. \ref{fig: Large deformation}C} display snapshots of sample networks at a stretch of $\lambda=4$ when $\dot \varepsilon/k_d=0.125$ and $\dot \varepsilon/k_d = 1$ to visibly convey differences in chain stretch for a network undergoing steady state creep (top) versus monotonically increasing stress (bottom). This study successfully reproduces a typical rate-dependent dynamic polymer response by probing the $10^{-7}$ m scale for over 90 $\mu$s in the longest cases, and does so in the span of just 7 wall-clock hours.

    \figuremacro{H}{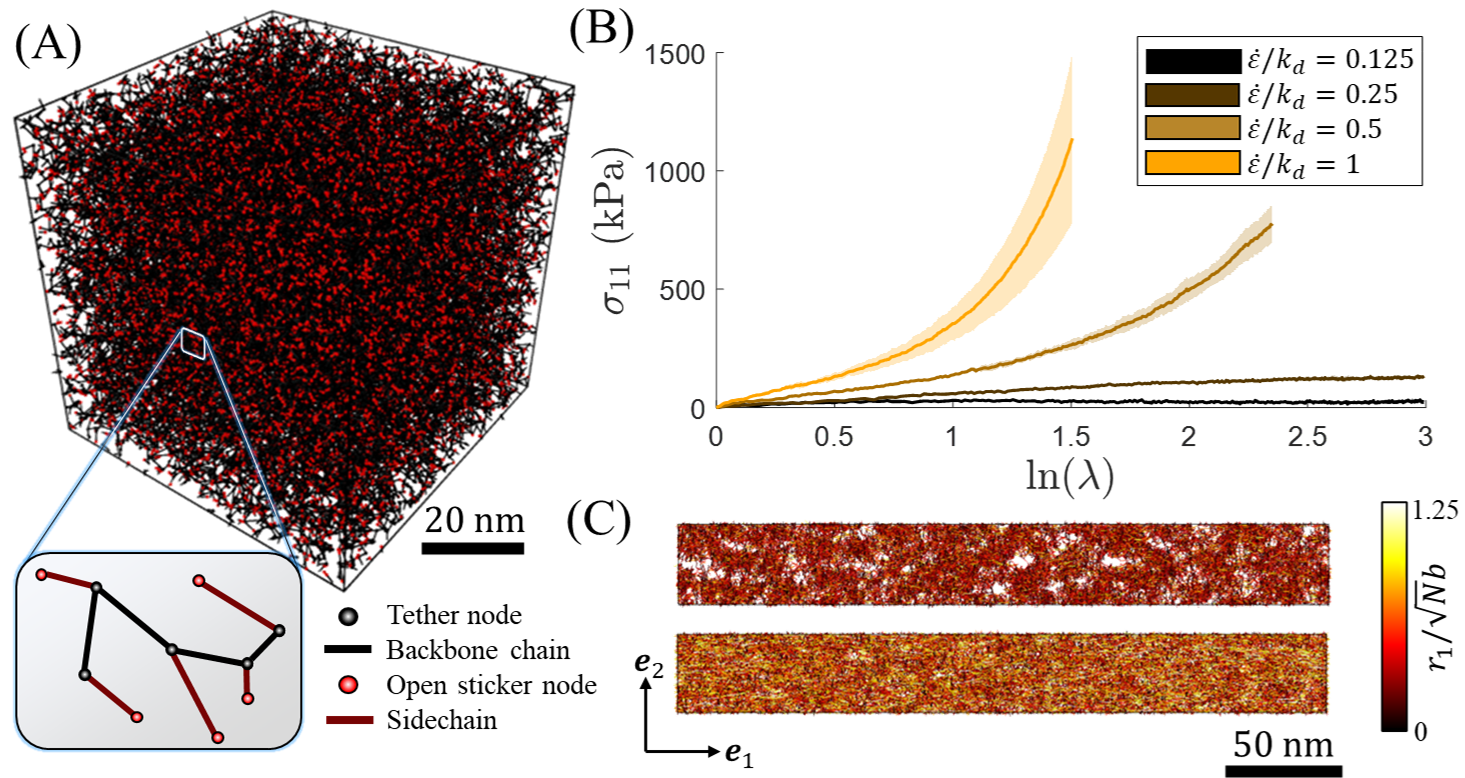}{Effects of loading rate during monotonic large deformation at constant true strain rate. }{\textbf{(A)} Initiated mesoscale model RVE comprised of $n_p=5 \times 10^3$ polymers, where red nodes indicate the positions of stickers. \textbf{(B)} Normal Cauchy stress in the direction of loading with respect to true strain, $\varepsilon = \ln(\lambda)$, when $\phi=0.2$, $N=12$, $k_d=10^{-3} \tau_0^{-1}$, and $\dot \varepsilon/k_d = \{ 0.125, 0.25, 0.5, 1 \}$. Data represents the average of $n=3$ samples and shaded regions represent S.E. of the mean. \textbf{(C)} Snapshots of two RVEs normal to the $\{ \bm e_i \}_{i=1,2}$ plane at a stretch of approximately $\lambda=4$ when $\dot \varepsilon/k_d = 0.125$ (top) and $\dot \varepsilon/k_d=1$ (bottom). Bond colors represents chain stretch in the direction of applied extension ($\bm e_1$). \label{fig: Large deformation}}{0.95}

Dynamic mechanical analysis (DMA) frequency sweeps are another common approach for probing strain rate dependence of polymers. We therefore simulated networks of $n_p=100$ polymers with $N=12$ and $\phi=0.2$ undergoing oscillatory shear. For elastomers and gels, pure shear loading conditions are typically applied via parallel plate rheology. To replicate pure shear within the orthonormal RVEs of the mesoscale model, we applied plane strain in directions $\bm e_1$ and $\bm e_2$ following $\epsilon_{11} = \epsilon_0 \sin(2\pi \omega t)$ and $\epsilon_{22}=-\epsilon_{11}$, respectively, where $\epsilon_0$ is the strain amplitude and $\omega$ is the angular frequency. For gels and other soft materials, $\epsilon_0$ is typically on the order of $0.01$ to maintain linearity; however, we set $\epsilon_0=0.05$ to enhance the signal-to-noise ratio. These loading conditions reasonably replicate pure shear for small strains so that shear strain and stress may be approximated using in-plane transformations as $\epsilon_{12} \approx - (\frac{\epsilon_{11}-\epsilon_{22}}{2}) \sin (2\theta)$ and $\sigma_{12} \approx - (\frac{\sigma_{11}-\sigma_{22}}{2}) \sin (2\theta)$, where $\theta \approx \pi/4$ is the orientation of maximum shear with respect to the the principle basis, $\{ \bm e_1,\bm e_2 \}$. Snapshots of an RVE undergoing oscillatory pure shear are shown in \textbf{Fig. \ref{fig: DMA}A}, while a sample of the corresponding applied strain and steady state stress response are depicted in \textbf{Fig. \ref{fig: DMA}B-C}. Storage modulus ($G^\prime = \sigma_{peak}/\epsilon_0 \cos(\delta)$) and loss modulus ($G^{\prime \prime} = \sigma_{peak}/\epsilon_0 \sin(\delta)$) are calculated from such stress and strain data, where $\sigma_{peak}$ is the peak value of the measured shear stress and $\delta$ is the phase shift between the applied strain and resulting stress.

    \figuremacro{H}{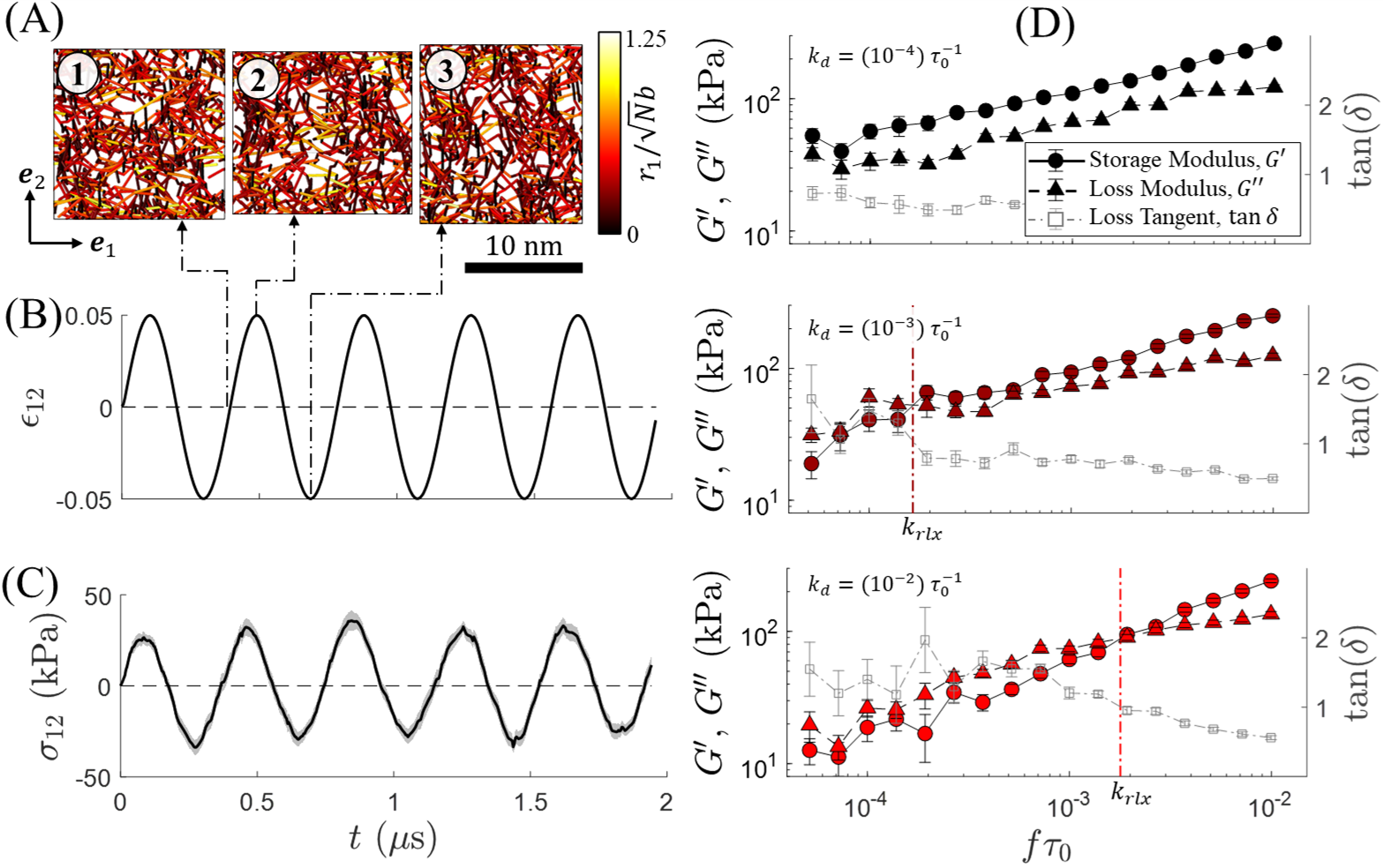}{Results of numerical DMA. }{\textbf{(A)} Snapshots of an RVE of $n_p=100$ polymers normal to the $\{ \bm e_i \}_{i=1,2}$ plane (1) in the undeformed state, (2) at $\epsilon_{11}=0.05$, and (3) at $\epsilon_{11}=-0.05$. \textbf{(B)} Sample loading history ($\epsilon_{12}$ versus time) over $2$ $\mu$s when $f=(10^{-2})\tau_0^{-1}$. \textbf{(C)} Cauchy stress, $\sigma_{12}$, resulting from the load history of \textbf{(B)}. \textbf{(D)} Storage and loss moduli with respect to normalized frequency, $f \tau_0$ for $k_d = (10^{-4})\tau_0^{-1}$ (\textbf{top}) $k_d = (10^{-3})\tau_0^{-1}$ (\textbf{center)}, and $k_d = (10^{-2})\tau_0^{-1}$ (\textbf{bottom}). Loss tangent, $\tan(\delta)$ is provided in faint grey. Vertical dotted-dashed lines denote $k_{rlx}$ (i.e., the frequency at which $\tan \delta = 1$). Data is ensemble averaged from $n=15$ samples at each frequency and detachment rate. Shaded region in \textbf{(C)} and error bars in \textbf{(D)} represent SE. \label{fig: DMA}}{1}
    
Storage and loss moduli, as well as corresponding loss tangents ($\tan \delta$), for the ensemble average of $n=15$ networks are plotted with respect to frequency ($f=2\pi \omega$) in \textbf{Fig. \ref{fig: DMA}D}, for detachment rates of $k_d=\{10^{-4},10^{-3},10^{-2} \}\tau_0^{-1}$. We choose to sweep detachment rate because $k_d$ is tunable in many material systems via crosslinker chemistry (e.g., cationic species and valency in metallopolymers) \citep{kloxin_covalent_2010,richardson_hydrazone_2019,vidavsky_tuning_2020,zhang_bridging_2020}. All $n=15$ networks were independently equilibrated and simulated at $17$ distinct frequencies. Cumulatively, each sample network was simulated for a duration in excess of $400$ $\mu$s over roughly 11 wall-clock hours, exhibiting this method's practicable access to large timescales. Evaluating the results, storage and loss moduli are consistently on the order of $10$ to $10^2$ kPa, which is typical of a gel \citep{okamoto_viscoelastic_2011} or soft elastomer \citep{chen_viscoelasticity_2015,shabbir_linear_2016,zhang_bridging_2020,peter_rheological_2021}. At all frequencies the loss tangent is relatively high ($\tan(\delta)>0.5$), signifying that the networks generally undergo a large degree of non-affine chain relaxation regardless of $k_d$.


Trends in $G^\prime$ and $G^{\prime \prime}$ are exemplary of a linear viscoelastic material in that high frequencies beget predominantly elastic response ($G^{\prime \prime}/G^\prime = \tan (\delta) < 1$), while low frequencies impart viscous behavior ($\tan (\delta) > 1$). The loading frequency at which this transition occurs (\textbf{Fig. \ref{fig: DMA}.D}, dotted-dashed lines) is interpretable as the relaxation rate, $k_{rlx}$ \citep{rubinstein_polymer_2003}. As the detachment rate increases (\textbf{Fig. \ref{fig: DMA}.D}, dashed lines), so too does $k_{rlx}$, and no transition is observed for the slowest detachment rate ($k_d= 10^{-4}\tau_0^{-1}$) for which $\tan(\delta)$ is always less than unity (indicative of rubbery response). The dependence of $k_{rlx}$ on $k_d$ confirm that this model is capturing the transition regime wherein relaxation is mediated by bond detachment, rather than $\alpha$-relaxation \citep{chen_viscoelasticity_2015,shabbir_linear_2016,zhang_bridging_2020} as in \textbf{Section \ref{sec: Loading Rate Sweep}}. For the two detachment rates with observable transitions ($k_d=\{10^{-3},10^{-2} \}\tau_0^{-1}$), $k_d/k_{rlx}\approx 6$ affirming that the relaxation rate is slower than $k_d$. This is attributed to bond lifetime renormalization \citep{stukalin_self-healing_2013} and is representative of many experimental dynamic polymers \citep{rubinstein_polymer_2003,chen_viscoelasticity_2015,shabbir_linear_2016,zhang_bridging_2020} whose rheological responses are best captured by the sticky Rouse model wherein chain relaxation is retarded by the presence of dynamic binding sites \citep{leibler_dynamics_1991}.\footnote{To measure partner exchange rates directly requires high frequency data output ($k_s\sim 0.1\tau_0^{-1}$) so that data curation becomes untenable for detailed bond exchange studies at these timescales.} The mesoscale model's capture of reconfigurational relaxation at small strains and slow loading rates, as well as its ability to model stretch-dependent relaxation spectra \citep{yu_effects_2014} at large strains and fast loading rates in \textbf{Section \ref{sec: Results of network-scale}}, spotlight its capacity for predicting physically realistic trends in polymers across many decades of time.

\section{Conclusion}

We have formulated an idealized mesoscale model for dynamic polymers at the single-chain and network scales and conducted detailed investigation on its validity. Unlike comparable prior studies, which implicitly modeled dynamically bonding stickers \citep{wagner_network_2021,wagner_coupled_2023}, the mesoscale model introduced here explicitly tracks sticker positions. Furthermore, whereas prior studies primarily compared mesoscale models to macroscale theory or experiments, we have taken a bottom-up approach in which the mesoscale model was compared to a Kremer-Grest coarse-grained MD approach \citep{kremer_dynamics_1990}, as well as to statistical mechanics-based molecular theory where possible \citep{rubinstein_polymer_2003,stukalin_self-healing_2013}. Our method revealed that the effective bond activation energy for telechelic association of polymer chains is penalized by the Helmholtz free energy of the end-to-end stretch they must assume for attachment. This relation is predicted by statistical mechanics \citep{buche_chain_2021}, has been discovered experimentally \citep{guo_association_2009}, and mirrors Bell's well-established model for force-dependent bond dissociation \citep{bell_models_1978}. While this finding may offer a route by which to implicitly model stickers under specific conditions (e.g., low sticker concentration), our bottom-up approach highlights the importance of explicitly tracking stickers. 

We found that seemingly small differences in bead-spring versus mesoscale sticker MSDs below the Rouse time culminate in distinct partner exchange and repeat attachment rates \citep{stukalin_self-healing_2013}, which in turn drive measurable structural differences for networks of long chains and low chain concentrations between models. These differences were reduced, but not eliminated, by setting the damping coefficient for a mesoscale node to $\gamma^\alpha=Nz\gamma_0/2$ following Rouse subdiffusion \citep{rubinstein_polymer_2003,stukalin_self-healing_2013}, or $\gamma^\alpha=N^{2/3}\gamma_0$ for chains tethered to a fixed point. Based on these findings, we recommend maintaining implicit chains with on the order of $N=12$ segments when employing these mesoscale methods, as seen in many polymers with high dynamic binding site concentrations \citep{colby_dynamics_1998,chen_linear_2013,zhang_bridging_2020,xu_thermosensitive_2022,xie_length_2024}. Future work could explore incorporating intermediate nodes along the lengths of mesoscale chains to simulate high molecular weight systems. We found that these models predicted mechanical network responses in excellent agreement with each other for networks comprised of short-chained ($N=12$), monodisperse polymers loaded at strain rates below 1\% of the monomer diffusion rate. Furthermore, they nicely predicted the chain length, loading rate, and detachment rate-dependent trends expected of dynamic elastomers and gels \citep{jeon_extremely_2016,zhang_superstretchable_2019,xu_thermosensitive_2022,cai_highly_2022}. 
The mesoscale model achieved these predictions with a 90\% reduction in computational time and data storage requirements compared to the bead-spring model. Via these savings, the mesoscale approach offers access to spatiotemporal scales not readily available using conventional MD approaches including large Lagrangian deformations and long, variable-timescale numerical experiments (e.g., frequency sweeps).

When applying this mesoscale method over longer durations, we emphasize the importance of accurately representing a material's structural evolution. Any discrepancies in the rates of bond restructuring and conformational changes are liable to cause error stack-up over timescales significantly longer than the renormalized bond lifetime. A major focus of future work for this mesoscale approach that may improve long-term structural accuracy is capturing crucial inter-chain and polymer-solvent interactions. Real materials such as gels and elastomers host innate homogenization mechanisms such as osmotic pressure \citep{flory_thermodynamics_1942,flory_molecular_1985} or excluded volume interactions between chains \citep{zimm_excluded_1953}. Both of these mechanisms culminate in a material-scale pressure ($\pi \bm I$ in Eq. \ref{eq: virial stress}) that resists clustering of chains \citep{mordvinkin_rheology_2021} and mediates bulk volumetric deformation in real materials and conventional MD approaches. In the case of gels, a recently introduced coarse-grained method for capturing osmotic pressure 
as a function of the $\chi$-parameter may be incorporated into this framework \citep{flory_thermodynamics_1942,doi_soft_2013,wagner_mesoscale_2022}. However, more work is required to develop mechanisms that encapsulate the effects of steric interactions between chains, especially as seen in denser polymers such as melts and elastomers, or highly entangled polymers with high molecular weights. Indeed, these same inter-chain interactions are responsible for the dissipative entanglements that partially cohere and greatly toughen many high molecular weight polymers \citep{sun_crossover_2006,ge_molecular_2013,schieber_entangled_2014,masubuchi_simulating_2014,ge_nanorheology_2018,kim_fracture_2021,steck_multiscale_2023,shi_highly_2023}. This mesoscale model may serve as a foundational framework into which researchers may incorporate novel reduced-order methods for such mechanisms and provide a powerful tool for the predictive design of dynamic polymers.

\appendix
\titleformat{\section}[block]{\normalfont\Large\bfseries}{\appendixname\ \thesection.}{1em}{}

\section{Model initiation procedures.}
\label{Appendix - initiation procedures}

\setcounter{equation}{0}    
\renewcommand{\theequation}{\thesection \arabic{equation}}
\setcounter{table}{0}    
\renewcommand{\thetable}{\thesection \arabic{table}}

Here we describe the numerical initiation procedures for all simulations of the various studies in the main text. Initiation procedures for the chains in tethered diffusion, bond kinetics, and bond exchange studies of \textbf{Section \ref{sec: Results of single-chain}} are described in \textbf{Section \ref{sec: Single Chain Initiation}}. The procedures used to generate RVEs for network-scale mechanics studies in \textbf{Sections \ref{sec: Results of network-scale}}-\textbf{\ref{sec: Spatiotemporal Extrapolation}} are described in \textbf{Section \ref{sec: Network Initiation}}. All initiation procedures were carried out using custom codes written in MATLAB2022a.



\subsection{Single-chain initiation procedures}
\label{sec: Single Chain Initiation}

To validate single-chain characteristics (e.g., end-to-end distributions) and bond kinetics (e.g., attachment, detachment, and exchange rates), various tethered chain studies were conducted in \textbf{Section \ref{sec: Results of single-chain}}, wherein an end of each chain was fixed while the rest of it diffused following Eq. (\ref{eq: Brownian Equation of Motion}). To initiate the chains in these studies, arrays of fixed, tethering nodes were first generated at the position set $\{ \bm x^1 _m\}$ where the index $m$ denotes the molecule number. The detailed configuration (e.g., number of nodes, their relative position to each other, etc.) of these initial tethering nodes are described in detail on a case-by-case basis in \textbf{Sections \ref{sec: Single Chain End-to-end Results}-\ref{sec: Bond Exchange Results}}. However, in every case, simulated polymer chains were randomly generated from these seeded tethering sites. 

For the bead-spring iterations of these studies, this was achieved using a 3D random walk approach whereby new beads were subsequently positioned at $\{ \bm x_m^\alpha \}$ based on the positions of the previous beads of their chain, $\{ \bm x_m^{\alpha-1} \}$, according to:
\begin{equation}
    \bm x_m^{\alpha+1} = \bm x_m^{\alpha} + b \bm{\hat r}^\alpha,
    \label{eq: Random Walk - bead-spring}
\end{equation}
until the desired number of Kuhn segments, denoted by index $\alpha$, was obtained (i.e., $\alpha\in[1,N]$). Here, the step size is that of a Kuhn length, $b$, and the directional vector, $\bm{\hat r}^\alpha=\bm{r}^\alpha / r^\alpha$, was determined according to:
\begin{equation}
    \bm{r}^\alpha = (\sin \varphi^\alpha \cos \theta^\alpha) \bm{\hat e}_1 + (\sin \varphi^\alpha \sin \theta^\alpha) \bm{\hat e}_2 + (\cos \varphi^\alpha) \bm{\hat e}_3
    \label{eq: Directional Vector}
\end{equation}
where $\{\bm e \} = \{\bm{\hat e}_1,\bm{\hat e}_2,\bm{\hat e}_3 \}$ denotes the orthonormal basis of the simulation, and the polar, $\theta$, and azimuthal, $\varphi$, angles were randomly sampled from the uniform distributions $\theta \in [0,2\pi)$ and $\varphi \in [0, \pi]$, respectively. 

For the mesoscale model, only one additional node was appended to each tethering site, which represents the distal end of the chain. This node was positioned according to:
\begin{equation}
    \bm x_m^2 = \bm x_m^1 + \bm r_m,  
    \label{eq: Initiating Mesoscale Chain}
\end{equation}
where the direction of vector $\bm r_m$ is again assigned per Eq. (\ref{eq: Directional Vector}), but the step-size, $|\bm r_m|$ is randomly sampled from the 3D joint PDF for finding a Gaussian chain with end-to-end vector, $\bm r$, given by \citep{rubinstein_polymer_2003}:
\begin{equation}
    P(\bm r) = \left( \frac{2}{3} Nb^2 \pi \right)^{-\frac{3}{2}} \exp \left( -\frac{3 \bm r^2}{2 N b^2} \right).
    \label{eq: Gaussian Joint PDF}
\end{equation}
Once initiated, all non-tethered nodes' positions were updated according to Eq. (\ref{eq: Brownian Equation of Motion}) and the details of \textbf{Section \ref{sec: Equation of Motion}}.   

\subsection{Network initiation}
\label{sec: Network Initiation}

To generate stable initial conditions for simulations in \textbf{Sections \ref{sec: Results of network-scale}-\ref{sec: Spatiotemporal Extrapolation}}, particle positions and bond configurations were initiated for both models using a custom code written in MATLAB2022a. First, a cubic RVE of dimensions $V = L^3$ was sized to achieve the desired polymer packing fraction, $\phi$, per the relation:
\begin{equation}
    L=\left\{ \phi^{-1} \left[ \frac{\pi}{6}b^3 n_p (2 n_t-1) N \right] \right\} ^{1/3},
\end{equation}
where the quantity $\frac{\pi}{6}b^3 n_p (2 n_t -1) N$ is the total occupied volume of polymer assuming that the volume of a single Kuhn segment is approximately $\pi b^3/6$. Additionally, $n_p$ (varied) is the number of polymers in the network, $n_t=5$ is the number of tethered side chains for each polymer, and $N$ is the number of Kuhn segments between crosslinks, so that $n_p (2 n_t - 1) N$ is the total number of Kuhn segments in the network based on the branched configuration of \textbf{Fig. \ref{fig: Network model summary}A}. The RVE was positioned with its center at Cartesian coordinates $\bm X_0=\left[ 0,0,0\right]$ and its initial boundaries spanning in each direction by $\pm L/2$. 

Once the RVE was positioned, $n_p n_t$ particles representing all tethering sights of side chains on the polymers' backbones were positioned using a Poisson growth process. The first particle was positioned at $\bm X_0$. Tether particles were then seeded in series at positions $\bm X_{i} = \bm X_{i-1} + \bm u_i$ where $i\in \left[1,n_p n_t\right]$ denotes the tether index and $\bm u_i$ is a randomly selected displacement vector. Direction vectors of $\bm u$ were assigned using the spherically uniform sampling procedure from Eq. \eqref{eq: Directional Vector}. The norms of $\bm u$ were randomly assigned from the uniform distribution $u \in \left[0.78,1.41\right] c_t^{-1/3}$, where $c_t=n_p n_t/V$ is the tether concentration so that $c_t^{-1/3}$ is their nominal separation. Particles were not permitted to be within a distance of less than $0.78 c_t^{-1/3}$ from nearby neighbors. If a particle was seeded outside of the boundaries of the RVE, a new seeding branch was begun by selecting an earlier seed particle at random. This process was carried out until the domain was occupied by $n_p n_t$ tether particles within the RVE boundaries. Any particles seeded outside of the RVE were removed. Once all tether crosslink positions were initiated, their radial distribution was homogenized by introducing an arbitrary soft, pairwise repulsive force of the form:
\begin{equation}
    \bm f_r = E \left(\frac{1}{\sigma_r} - \frac{\sigma_r^\alpha}{d^{\alpha+1}} \right) \hat{\bm d},
    \label{eq: Arbitrary soft repulsion}
\end{equation}
and then allowing their positions to equilibrate using an overdamped steepest descent algorithm and periodic boundary conditions according to \cite{wagner_network_2021}. Here, $E$ is an energy scale to modulate the magnitude of force, $\sigma_r$ is the characteristic length scale over which this force acts, $\alpha$ is a scaling parameter that controls the stiffness of the force, and $d$ is the distance between particles. Eq. \eqref{eq: Arbitrary soft repulsion} is phenomenological and merely used to erase any process-specific artifacts of the particle initiation procedure. To homogenize tether positions, we set $E=1.25 k_b T$, $\sigma_r=1.2 c_t^{-1/3}$, and $\alpha = 2$.

After tether homogenization, we polymerized the backbones of all $n_p$ polymers in series. For each polymer, an initial node was selected at random, and then a chain was added between it and its nearest unbound neighboring tether while observing RVE periodicity. This was carried out $n_t-1$ times for all $n_p$ polymers. To equilibrate polymer segments and mitigate stochastically initiated chains with lengths in excess of $Nb$, the system was then equilibrated using the repulsive forces of Eq. \eqref{eq: Arbitrary soft repulsion} and linear tensile forces between all attached crosslinks according to:
\begin{equation}
    \bm f_t = \frac{3k_b T}{Nb^2} \bm r,
    \label{eq: Arbitrary spring force}    
\end{equation}
where $\bm r$ is the end-to-end vector representing each chain segment. While Eq. \eqref{eq: Arbitrary spring force} is expressed in physical units, it is phenomenologically applied like Eq. \eqref{eq: Arbitrary soft repulsion} simply to erase initiation history. Again, equilibration was carried out per \cite{wagner_network_2021} while enforcing periodic boundaries. 

Next, side chains were grafted to each of the polymer backbones' $n_t$ tethering sites using the procedure described above via Eqs. \eqref{eq: Initiating Mesoscale Chain} and \eqref{eq: Gaussian Joint PDF}. Another step of phenomenological equilibration was then carried out per the procedure above (i.e., applying the forces of Eqs. \eqref{eq: Arbitrary soft repulsion} and \eqref{eq: Arbitrary spring force}) to homogenize all sticker and tether positions. Finally, $N$ Kuhn segments were placed at the linearly interpolated positions between the ends of every attached tether-tether and tether-sticker pair. To ensure stable positioning of each Kuhn segment with respect to its bonded neighbors (which occurs when they are approximately a distance of $b$ apart from one another), the soft pairwise repulsion of Eq. \eqref{eq: Arbitrary soft repulsion} was again applied but with its characteristic length scale set to $\sigma_r = b$. A final step of equilibration (without any pairwise tensile forces) was conducted to homogenize the Kuhn segments. This procedure was executed identically for both the bead-spring and mesoscale approaches so that initial crosslink distributions were statistically equivalent between models; however, the mesoscale models' intermediate Kuhn segments between stickers and tether crosslinks were retroactively removed and replaced by direct sticker-to-sticker or sticker-to-tether connections with implicit pairwise bond potentials through Eq. \eqref{eq: Langevin force}. Although the initiation procedure was the same between approaches (with the additional step of Kuhn segment removal for the mesoscale), distinct seeds were used for the mesoscale and bead-spring models to properly evaluate statistical agreement between random samples. With the initial chain positions established, the periodic boundaries were unwrapped in accordance with the requirements of LAMMPS and input files were automatically generated using MATLAB2022a to enact the loading criteria described in each section of \textbf{Sections \ref{sec: Results of network-scale}-\ref{sec: Spatiotemporal Extrapolation}}. 

\section{Unit conversion and parameter selection}
\label{sec: Parameters and Unit Normlization}

\setcounter{equation}{0}    
\renewcommand{\theequation}{\thesection \arabic{equation}}
\setcounter{table}{0}    
\renewcommand{\thetable}{\thesection \arabic{table}}

Conversions between the arbitrary units of the discrete model and SI units are provided in \textbf{Table \ref{Table 1}}. Conversions are prescribed directly for the fundamental units of temperature, time, and length. However, since the model is overdamped, particle masses were not prescribed and a conversion for the Boltzmann constant was prescribed instead. These four conversions are used to derive conversions for other pertinent units such as energy, force, and stress. While these conversions are provided for reference, results are generally provided in normalized units throughout the work unless specified otherwise.

    \tablemacro{H}{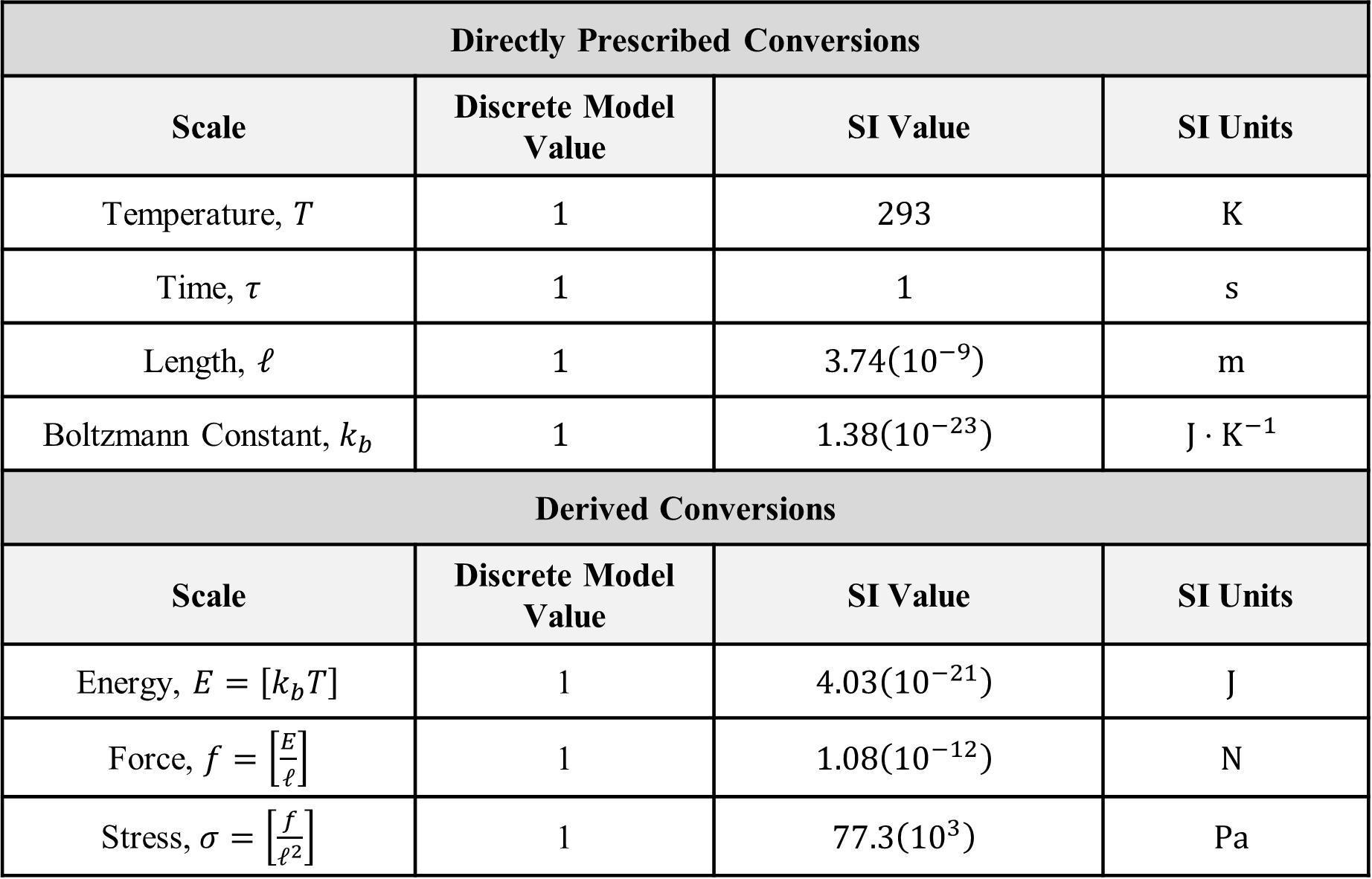}{}{Discrete model unit conversions. \label{Table 1}}{0.85}

Model parameters are listed in \textbf{Table \ref{Table 2}} in both SI and arbitrary model units. 

    \tablemacro{H}{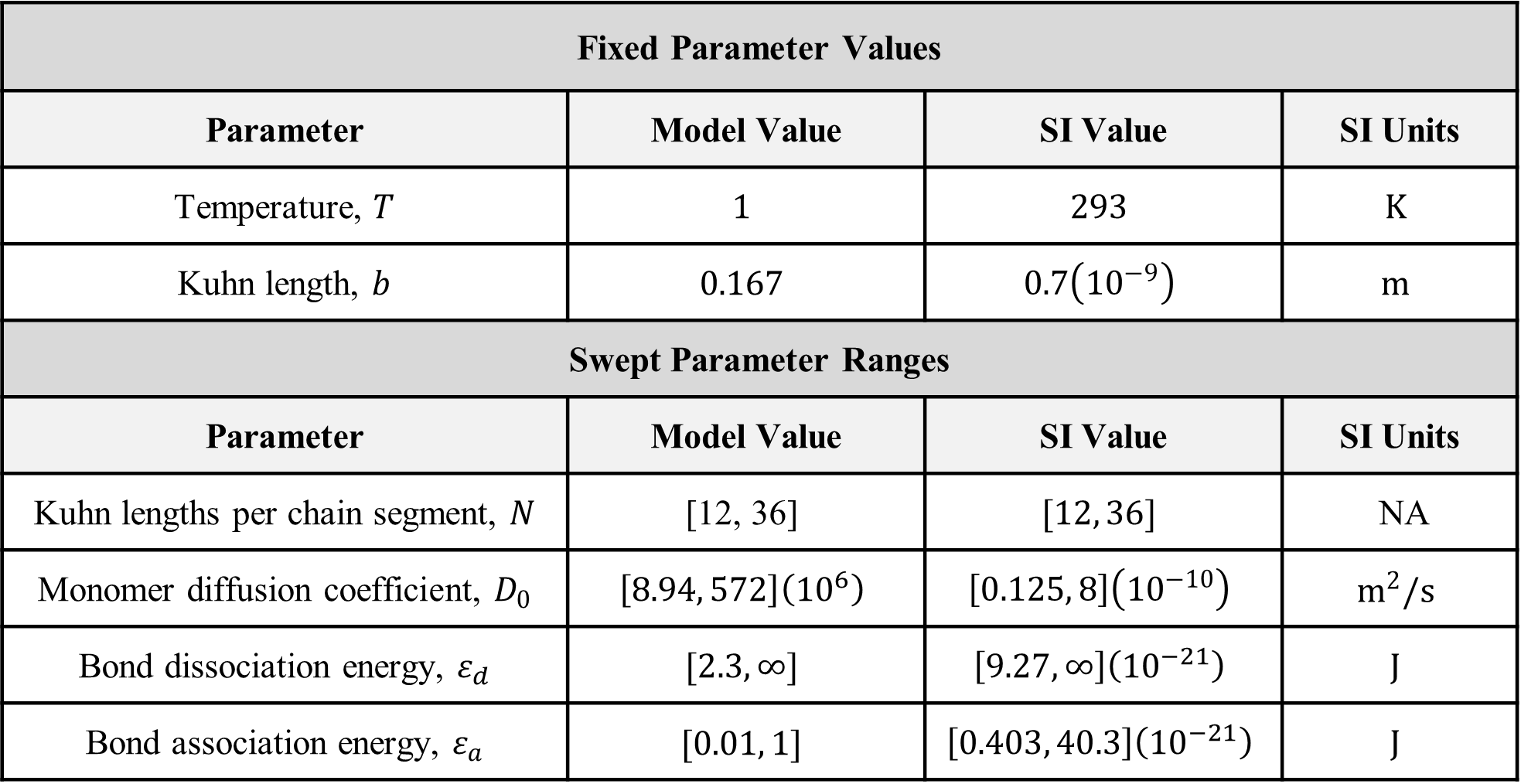}{}{Discrete model parameters. \label{Table 2}}{0.9}

Justifications for the parameter values of \textbf{Table \ref{Table 2}} are as follows:

\begin{itemize}
    \item \textbf{Temperature:} Temperature was set to 293 K based on typical ambient conditions. 

    \item \textbf{Kuhn length:} The Kuhn length was set to 0.7 nm based on the order of experimentally estimated values of Kuhn lengths for polymers such as poly(ethylene glycol) in near-theta solvent \citep{ahlawat_elasticity_2021,liese_hydration_2017,lee_molecular_2008}.

    \item \textbf{Number of Kuhn segments per polymer chain segment:} The number of Kuhn segments was swept over the range $N\in [12,36]$ (corresponding to chain lengths of $Nb \in [7.5,22.2]$ nm) per the justification of \textbf{Section \ref{sec: Single Chain End-to-end Results}}. 
    
    \item \textbf{Molecular weights:} Molecular weights, $M_w$, were considered for the branched polymers investigated in \textbf{Sections \ref{sec: Results of network-scale}-\ref{sec: Spatiotemporal Extrapolation}}, as related to the selected values of $N$. In full network-scale studies, in which $n_t-1$ chain segments of $N$ Kuhn segments are linked in series to form a polymer's backbone, with $n_t$ side-branches grafted to their sides (also of $N$ Kuhn segments), the total number of Kuhn segments per polymer is $N_p = (2 n_t - 1) N$. Given the estimate that each Kuhn segment is comprised of two mers \citep{liese_hydration_2017}, and the estimated molecular weight of ethylene glycol is $M_{eg} \approx 44.1$ kg mol$^{-1}$ \citep{wagner_mesoscale_2022}, then the molecular weight per polymer chain may be estimated as $M_w = 2 N_p M_{eg}$. Thus, setting $n_t=5$ (\textbf{Fig. \ref{fig: Network model summary}A}), then the molecular weights for polymers with $N=12$ and $N=36$ Kuhn segments are $M_w = 9.53$ kDa and $M_w=28.6$ kDa, respectively. These are reasonable values as compared to many low molecular weight polymers and gels. 

    \item \textbf{Diffusion rate:} The diffusion coefficient, $D_0$, was used to set the monomer damping coefficient, $\gamma_0= k_b T/D_0$, and monomer diffusion timescale, $\tau_0 = b^2/D_0$. It was swept over the range $D_0 \in \{ 0.125, 0.25, 0.5, 1, 2, 4, 8\}\times 10^{-10}$ m$^2$ s$^{-1}$, to study its effects on MSD (\textbf{Fig. \ref{fig: Extended MSD results - effect of damper}}). However, it had no meaningful effect on MSD characteristics and merely renormalized the models' timescales. Hence, thereafter it was fixed at $D_0 = 10^{-10}$ m$^2$ s$^{-1}$ per the justification of \textbf{Section \ref{sec: Single Chain End-to-end Results}}.

    \item \textbf{Bond detachment activation energies:} Bond detachment activation energy, $\varepsilon_d$, was set as specified within each study of \textbf{Sections \ref{sec: Results of single-chain}-\ref{sec: Spatiotemporal Extrapolation}}. However, it was broadly set as a multiple of $k_b T$ in the range $\varepsilon_d \in [2.3,\infty) k_b T$ such that $k_d \in [0,0.1]\tau_0^{-1}$ through Eq. \eqref{eq: Detachment rate}, depending on the need of each study. 
    
    \item \textbf{Bond attachment activation energies:} As with $\varepsilon_d$, bond attachment activation energy, $\varepsilon_a$, was set as specified within each study of \textbf{Sections \ref{sec: Results of single-chain}-\ref{sec: Spatiotemporal Extrapolation}} to multiples of $k_b T$. Besides when it was swept over $\varepsilon_a \in [0.01,1] k_b T$ in \textbf{Section \ref{sec: Results of single-chain}}, it was held at $\varepsilon_a=0.01k_b T$, begetting a fast intrinsic attachment rate of $k_a^{ap} \approx \tau_0^{-1}$ per Eq. \eqref{eq: Attachment rate} and resulting in percolated network structures.
\end{itemize}


\section{Calibrating the bead-spring potential}
\label{Appendix - bead-spring potential}

\setcounter{figure}{0}    
\renewcommand{\thefigure}{\thesection \arabic{figure}}

\setcounter{equation}{0}    
\renewcommand{\theequation}{\thesection \arabic{equation}}

To verify that the chains of the bead-spring model reproduced the force-extension relations of the ideal Langevin chains approximated by Eq. \eqref{eq: Langevin force}, we conducted a simple study in LAMMPS. Chains modeled as $N+1$ beads attached by springs per the description of \textbf{Section \ref{sec: Discrete modeling approaches}} were modeled at various end-to-end lengths, $r$, in the range $r\in [1,0.95N]b$, while their average end-to-end forces were measured. To achieve numerically stable chains with initial end-to-end separations of approximately $b$, chains were initiated using MATLAB 2022b. One end of each chain was fixed at Cartesian coordinates, $\bm x^0 = (0,0,0)$, and the other was initially positioned at $\bm x^N = (0.95 Nb,0,0)$. The positions of $N-1$ beads were then linearly interpolated between the endpoints, $\bm x^0$ and $\bm x^N$, and then adjacent beads were connected to generate $N$ Kuhn segments. These $N-1$ intermediate beads were then offset in the \textit{yz}-plane by some amount, $(0,v,w)$, to break axial symmetry, where $v$ and $w$ are random displacements in the range $\pm b\sqrt{0.05}$. The positions of the particles were then equilibrated using arbitrarily soft, pairwise repulsive potentials between all beads within distance $b$ of each other, and soft harmonic potentials between all connected beads. The soft pairwise repulsive potential is given by $\psi_r = 2\mu [b/\sigma-\sigma^2 b/ (\bm r^{\alpha})^3]$ \citep{wagner_network_2021}, where $\mu=7.5k_b T$ is an arbitrary energy scale and $\sigma=b$ is the separation length at which repulsive forces go to zero. The harmonic potential between attached beads is given by $\psi_s=K 3k_b T  \bm (\bm r^{\alpha})^2/b^2$, where $K=1/2$ is an arbitrary, dimensionless softening factor. The beads' positions were equilibrated using the overdamped steepest descent method of \cite{wagner_network_2021}. The phenomenological parameters $\mu$, $\sigma$, and $K$ were set so that stable convergence was always achieved for the initial conditions in LAMMPS. 

Once the initial bead positions were set, the chains were loaded into LAMMPS, and their beads' positions were stepped in time according to Eq. \eqref{eq: Brownian Equation of Motion} per the methods of \textbf{Section \ref{sec: Equation of Motion}}. To reach the initial conditions of the study, the $N^{th}$ beads of each chain were gradually stepped towards the fixed beads at position $(0,0,0)$ in increments of approximately $(-0.75,0,0)b$ until an end-to-end separation of $r=b$ was achieved. At each step, the chains were equilibrated for a duration of $400 \tau_0$. However, once the initial end-to-end separation of $r=b$ was reached, the chains with $N=12$, $N=18$, and $N=36$ Kuhn segments were equilibrated for variable durations of $16 \times 10^3\tau_0$, $32 \times 10^3 \tau_0$, and $68 \times 10^3 \tau_0$, respectively. This was done to erase any conformational memory of the initially stretched state. Once each chain was equilibrated with an end-to-end length of $b$, it was gradually extended by stepping the position of $N^{th}$ bead by increments of $(0.75,0,0)b$. At each step, the chains were held for $4 \times 10^3 \tau_0$ and the mean end-to-end force was calculated as the average force of each segment projected onto the chain's end-to-end axis, $\bm e_1$:
\begin{equation}
    \bar f =\langle  N^{-1} \sum_\alpha^N \bm f^{\alpha} \rangle_t \cdot \bm e_1,
    \label{eq: Average end-to-end force}
\end{equation}
where $\bm f^\alpha=-\partial \psi_b/\partial\bm r^\alpha$ is the force of Kuhn segment $\alpha$ with end-to-end vector $\bm r^\alpha$, $\psi_b$ is the bond potential from Eq. \eqref{eq: FENE}, and $\langle \Box \rangle_t$ denotes ensemble averaging over the $4 \times 10^3 \tau_0$ duration. This process was repeated for chains with $N=\{12,18,36\}$ Kuhn segments and when the characteristic energies of Eq. \eqref{eq: FENE} were set to $E=\{100,200,400,800\} k_b T$. The distribution of Kuhn lengths remains within $\pm 10\%$ of $b$ when $E=800k_b T$ (\textbf{Fig. \ref{fig: Appendix_bond_potential}.A}). Additionally, the force-extension relations are in good agreement with those predicted by differentiating Eq. \eqref{eq: Langevin force} with respect to $r$ when $E=800k_b T$ (\textbf{Fig. \ref{fig: Appendix_bond_potential}.B}). 

    \figuremacro{H}{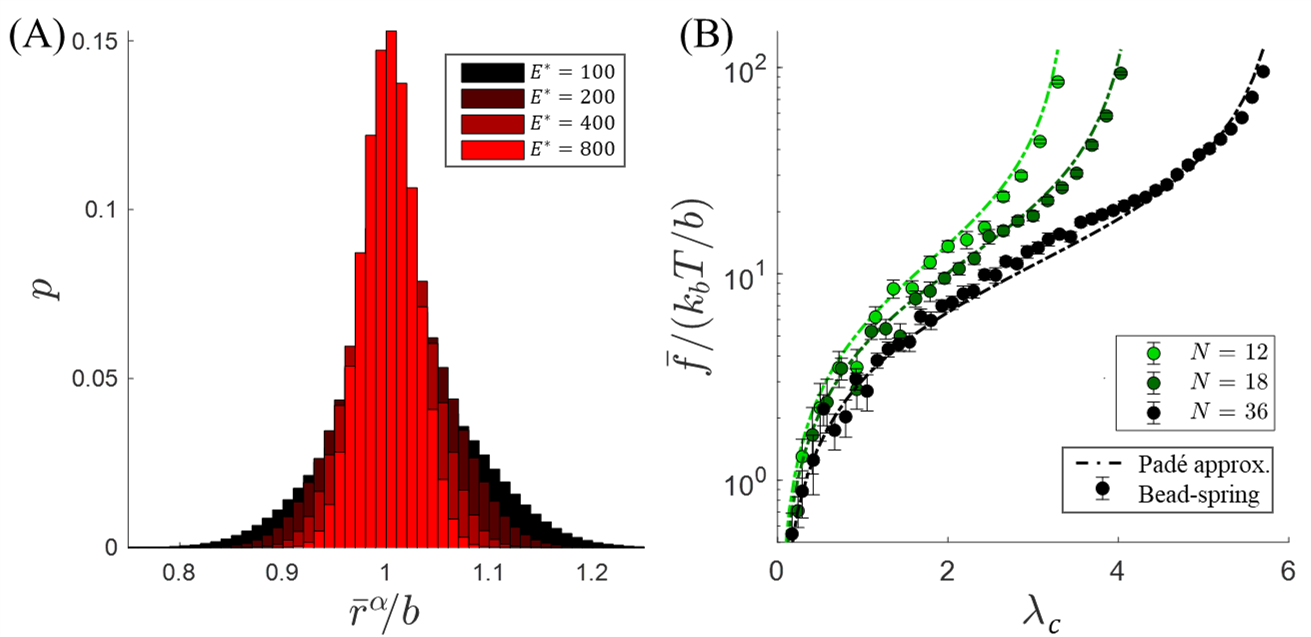}{Kuhn segment bond potential calibration: }{\textbf{(A)} End-to-end distributions of the Kuhn segment lengths of the bead-spring model for $E^*=E/k_b T=\{100,200,400,800\}$, $L=b$, and $N=36$. \textbf{(B)} Normalized force versus chain stretch for the Langevin potential of Eq. \eqref{eq: Langevin force} (dashed curves) and ensemble average force from the bead-spring model (discrete data) for various chain lengths ($N=\{12,24,36\}$ Kuhn segments) when $E=800 k_b T$ and $L=b$.\label{fig: Appendix_bond_potential}}{1}

\newpage

\section{Sampling frequency convergence study}
\label{Appendix - dt convergence}

Checking for dynamic bonding events incurs a computational cost, however bond kinetics must be sampled with an adequate frequency, $k_s$, to ensure that bonding opportunities are not missed as stickers oscillate to within distance $b$ of each other. To check for adequate sampling frequency, a convergence study was conducted using the single-chain attachment set-up of \textbf{Section \ref{sec: Pairwise attachment}}. Bond kinetic sampling frequency was swept over $k_s = \{10,15,20\} \tau_0^{-1}$ and bond kinetic rates and steady state attached/detached chain fractions were measured for both models. Results are displayed in \textbf{Fig. \ref{fig: Timescale Convergence Study on ka}}. Results do not change significantly as $k_s$ is increased from $15 \tau_0^{-1}$ to $20 \tau_0^{-1}$. This is consistent with the observation that the measured sticker diffusion timescale, $\tau_s$ is consistently $\tau_s \approx 0.07 \tau_0 \approx \tau_0/14$ per \textbf{Fig. \ref{fig: Extended MSD results - effect of damper}.H}. 

\setcounter{figure}{0}    
\renewcommand{\thefigure}{\thesection \arabic{figure}}

    \figuremacro{H} {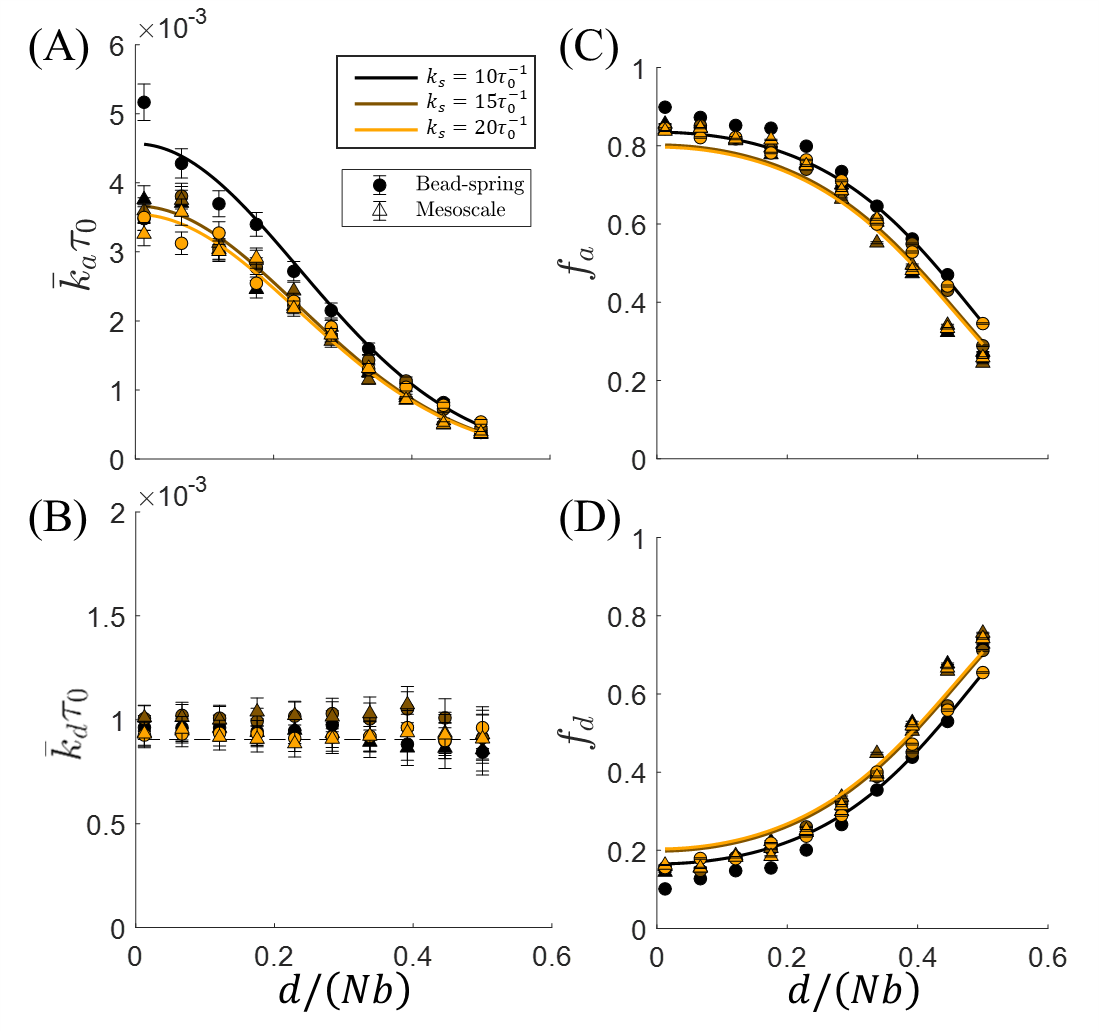}{Sampling frequency convergence of bond dynamics: }{\textbf{(A-B)} Average \textbf{(A)} attachment and \textbf{(B)} detachment rates with respect to tether-to-sticker separation distance, $d/(Nb)$ for bond kinetic sampling frequencies of $k_s=10 \tau_0^{-1}$ (black), $k_s=15 \tau_0^{-1}$ (brown), and $k_s=20 \tau_0^{-1}$ (orange). Results are provided for both the bead-spring (circles) and mesoscale (triangle) models. Solid curves represent the fitted scaling theory through Eq. \eqref{eq: Attachment rate scaling theory} with prefactor, $A$, treated as a fitting parameter. \textbf{(C-D)} Average fractions of \textbf{(C)} attached and \textbf{(D)} detached chains, plotted with respect to $d/(Nb)$, for the same sampling frequencies as \textbf{(A-B)}. Solid curves represent the predicted steady-state fractions of attached and detached chains, $f_a = k_a/(k_a + k_d)$, and $f_d = 1-f_a$, respectively \citep{vernerey_statistically-based_2017}. Here, $k_a$ is the distance-dependent attachment rate predicted by Eq. \eqref{eq: Attachment rate scaling theory} while $k_d$ is the detachment rate set \textit{a priori}.  \label{fig: Timescale Convergence Study on ka}}{0.78}

\section{Effects of diffusion coefficient}
\label{Appendix - extended single-chain diffusion}

\setcounter{figure}{0}    
\renewcommand{\thefigure}{\thesection \arabic{figure}}

\textbf{Fig. \ref{fig: Extended MSD results - effect of damper}} confirms that modifying the diffusion coefficient, $D_0$, has no effect on the results of either model in normalized time. This is true with respect to both the monomer diffusion timescale, $\tau_0 = b^2/D_0$, and the resulting Rouse time,  $\tau_r = \tau_s N^2$, where $\tau_s$ is the emergent timescale of distal sticker diffusion. \textbf{Fig. \ref{fig: Extended MSD results - effect of damper}H} confirms that the measured sticker diffusion timescale, $\tau_s = 0.07 \tau_0$, is relatively independent of both $D_0$ (i.e., $\tau_0$) and chain length (i.e., $N$).

    \figuremacro{H}{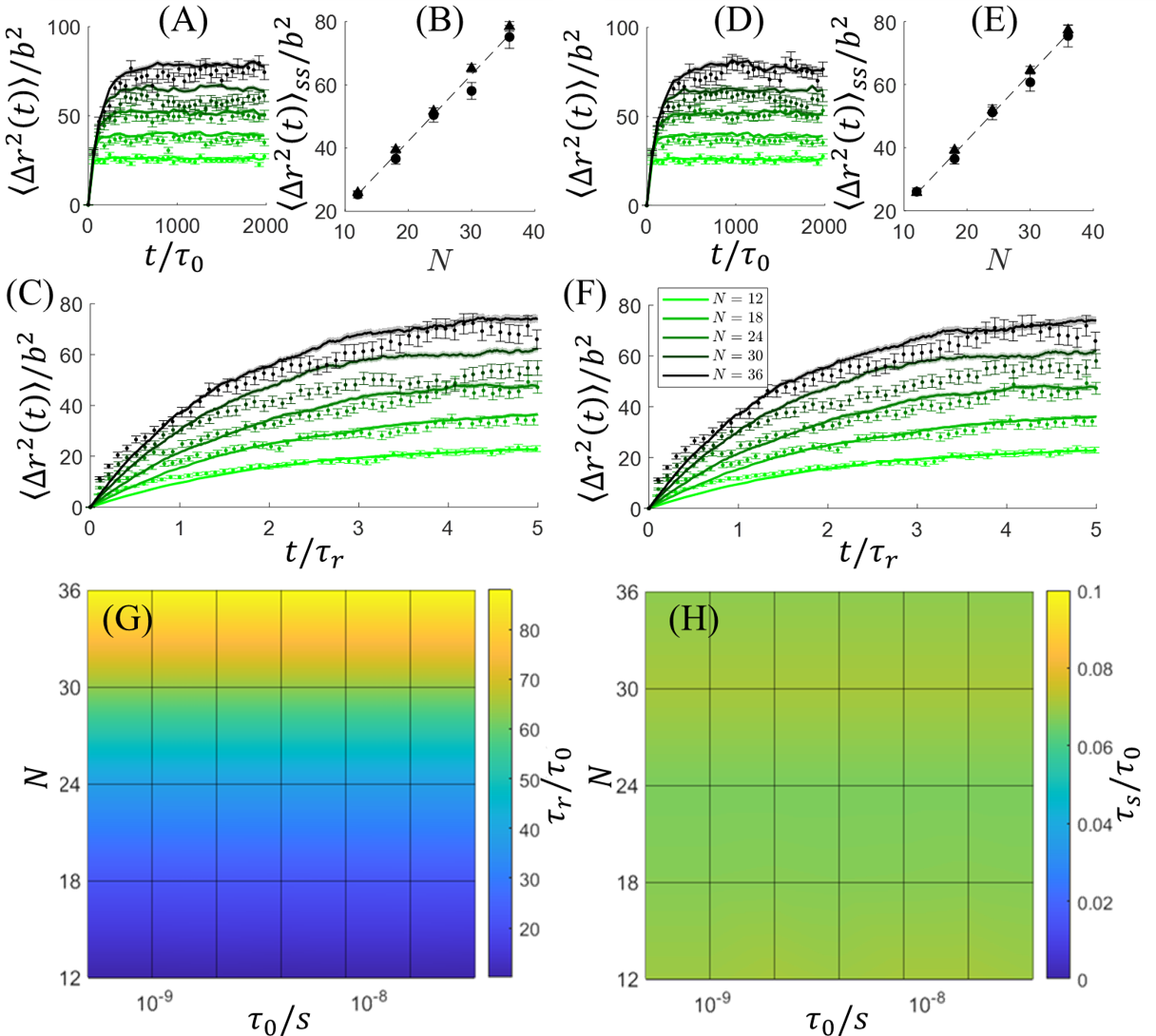}{Comparing diffusion coefficients: }{\textbf{(A-C)} MSD, steady state MSD with respect to $N$, and MSD at the Rouse timescale are presented for the case of $D_0=8 \times 10^{-10}$ m$^2$ s$^{-1}$, respectively. \textbf{(D-F)} The same plots are depicted for $D_0=0.125 \times 10^{-10}$ m$^2$ s$^{-1}$. The time-axes of \textbf{(A)} and \textbf{(D)} are normalized by the monomer diffusion timescale, $\tau_0 = b^2/D_0$, while the time-axes of \textbf{(C)} and \textbf{(F)} are normalized by the Rouse time, $\tau_r$. For both values of $D_0$, the steady-state MSD, $\langle \Delta r^2\rangle_{ss}$, scales linearly with respect to chain length (i.e., $N$). \textbf{(G)} Rouse time, $\tau_r$ and \textbf{(H)} emergent sticker diffusion time, $\tau_s$, both with respect to the prescribed diffusion timescale, $\tau_0$, and chain length via $N$. \label{fig: Extended MSD results - effect of damper}}{0.95}

\newpage
\noindent Characteristic sticker diffusion times, $\tau_s$, and Rouse times, $\tau_r$, were attained by identifying the time at which $\langle \Delta r^2 (t) \rangle \geq b^2$ (\textbf{Fig. \ref{fig: Extended MSD results - Rouse model}A}), and then fitting Eq. \eqref{eq: Rouse Model} to the data (\textbf{Fig. \ref{fig: Extended MSD results - Rouse model}B}).

    \figuremacro{H}{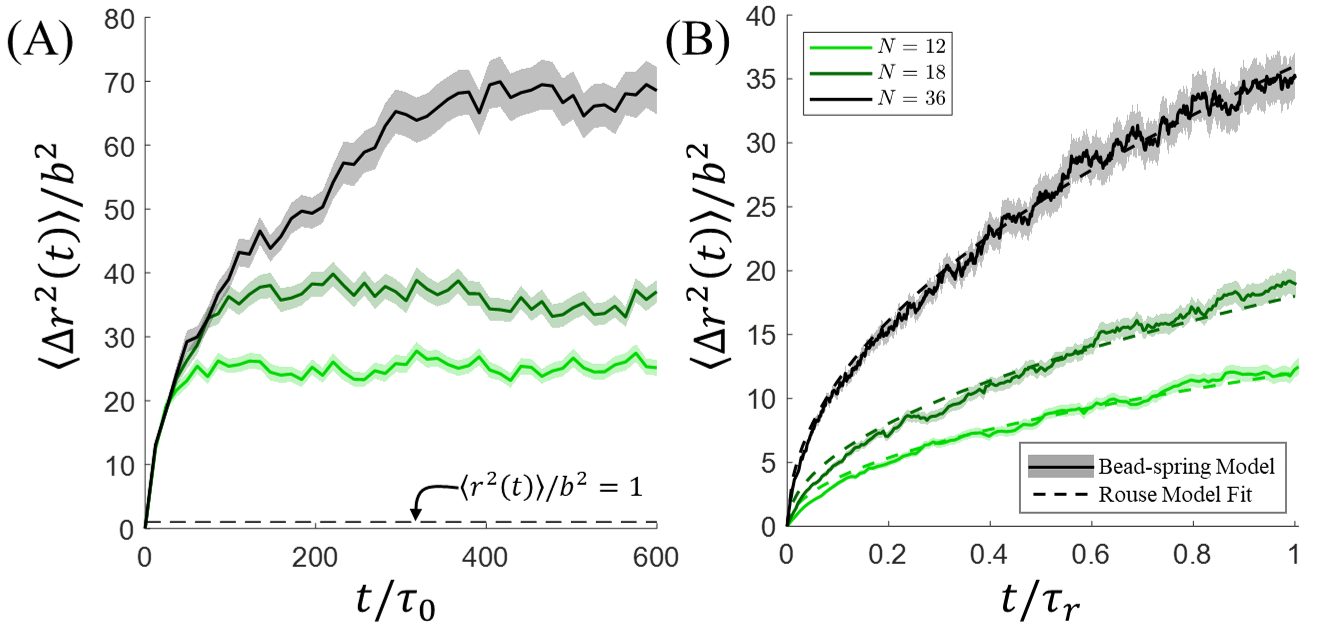}{The Rouse model for sticker diffusion: }{\textbf{(A)} Raw MSD data from the bead-spring model for $N\in \{ 12,18,24,30,36 \}$ is used to \textbf{(B)} compute the timescale, $\tau_s$, of distal sticker diffusion. \label{fig: Extended MSD results - Rouse model}}{0.8}

\newpage
\section{Validating sticker pairwise kinetic rates}
\label{Appendix - validating bond kinetics}

\setcounter{figure}{0}    
\renewcommand{\thefigure}{\thesection \arabic{figure}}
\setcounter{table}{0}    
\renewcommand{\thetable}{\thesection \arabic{table}}

Before examining the effects of chain exploration on bond association, we confirmed that the discrete model implementation of stochastic bond reactions reproduced the intrinsic detachment and attachment rates set \textit{a priori} through Eqs. (\ref{eq: Detachment rate}-\ref{eq: Poisson process}), without the extrinsic influence of diffusion kinetics. To do so, we simulated pairs of fixed stickers separated by distance $d_{ss}/b = 0.5$ that could bind and unbind to one another (see schematic of \textbf{Fig. \ref{fig: Kinetic Rates Validations}A}). For adequate sampling, arrays of $n_p=343$ pairs (separated from other pairs by distances $d_{ss}/b \gg 1$ so that no inter-pair interactions occur) were initialized and sampled concurrently for a duration of $t = 10^3 \tau_0$. \textbf{Fig. \ref{fig: Kinetic Rates Validations}B} confirms that the rates of detachment, $k_d$, and attachment, $k_a$, measured from the discrete model are in good agreement with those values set by Eqs. (\ref{eq: Detachment rate}) and (\ref{eq: Attachment rate}), respectively. Note that here, $\varepsilon_d$ was held constant at $\varepsilon_d = 0.1 k_b T$, while $\varepsilon_a$ was swept over the order of $\varepsilon_a \in [0.01,10]k_b T$. Maintaining a relatively high dissociation rate (via relatively low $\varepsilon_d$) ensured that ample numbers of dissociation events occurred within reasonable time domains for adequate statistical sampling of both event types.

    \figuremacro{H}{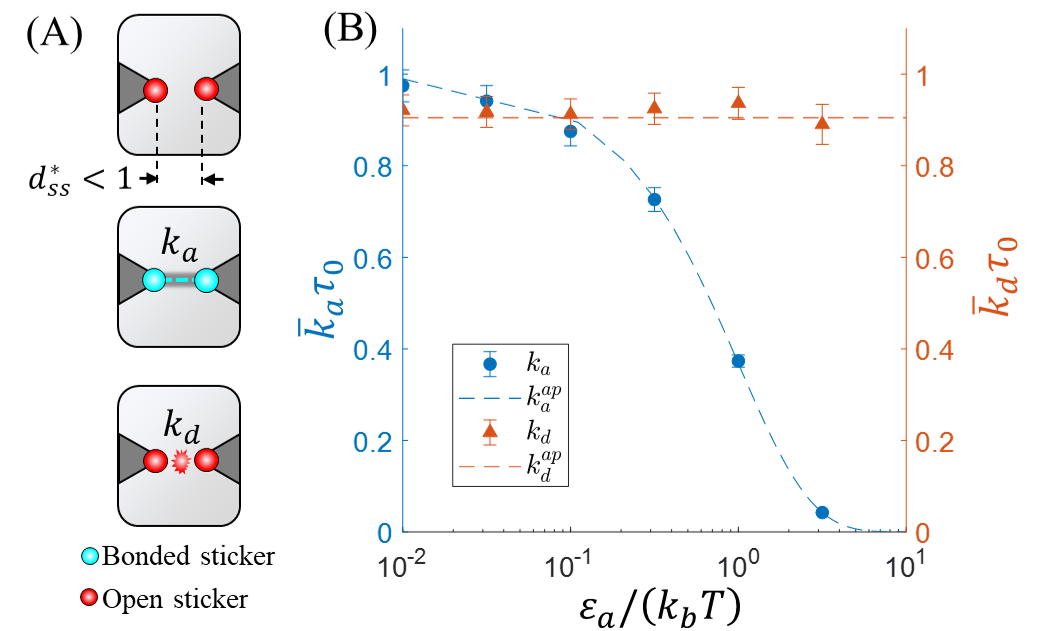}{Pairwise kinetics of adjacent stickers. }{\textbf{(A)} Illustration of a pair of fixed stickers separated by $d/b <1$ at their initial state \textbf{(top)}, immediately after an attachment event \textbf{(middle)}, and after the subsequent detachment event \textbf{(bottom)}. \textbf{(B)} Results comparing the discrete model predictions (discrete data points with error bars) of detachment and attachment rates to the \textit{a priori} values of $\varepsilon_a$ set through Eq. \eqref{eq: Attachment rate}. Note that $\varepsilon_d$ was held constant over all six sets of simulations, hence the lack of variation in $\bar k_d$. Error bars represent SE. \label{fig: Kinetic Rates Validations}}{0.75}

\newpage
\section{Extended pairwise kinetics results}
\label{Appendix - extended pairwise kinetics}

\setcounter{figure}{0}    
\renewcommand{\thefigure}{\thesection \arabic{figure}}
\setcounter{table}{0}    
\renewcommand{\thetable}{\thesection \arabic{table}}

To check whether the prefactor, $A$, required for fitting Eq. \eqref{eq: Attachment rate scaling theory} results from some effect of repeat attachments, we compare the overall average attachment rates of \textbf{Fig. \ref{fig: Single chain bond kinetics}D-E} (which includes repeat attachment events when computing the mean) to the average attachment rates computed using only first-time attachment events, $\bar k_{a,1}$, in \textbf{Fig. \ref{fig: ka_vs_ka1}}. While eliminating repeat attachment events from the computation of $\bar k_a$ does slightly reduce the attachment rate in all cases, it does not alter the magnitudes of $\bar k_a$ sufficiently to influence the necessity of prefactor $A$ or explain its trends with respect to $N$ and $\varepsilon_a$.

     \figuremacro{H}{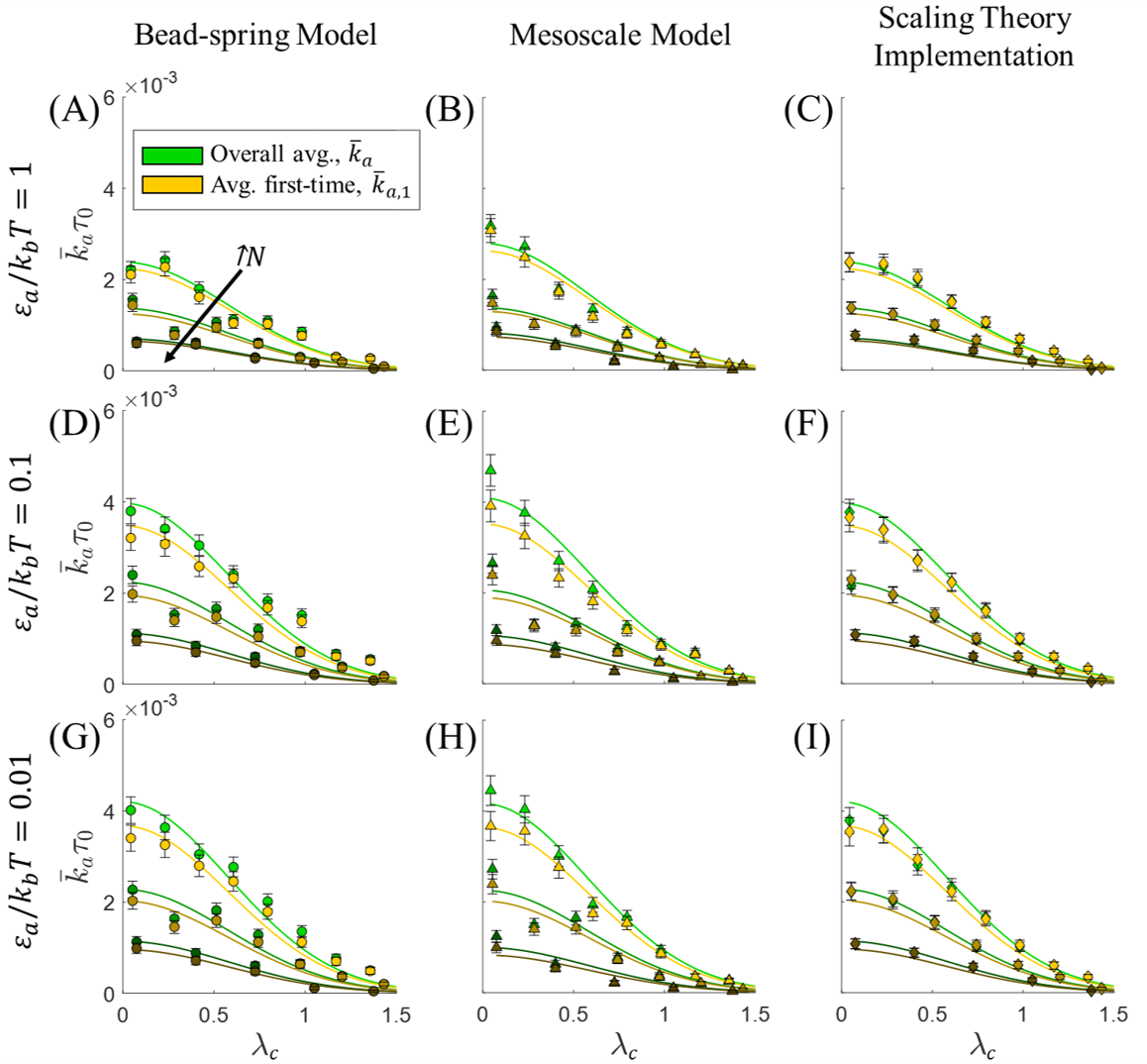}{First-time versus overall average attachment rates of a single bond. }{Average attachment rates, $\bar k_a$ (green) and first-time attachment rates, $\bar k_{a,1}$, (yellow) with respect to normalized separation distance, $\lambda_c = d_{ts}/(\sqrt N b)$. \textbf{(A-C)}, \textbf{(D-F)}, and \textbf{(G-I)} provide data when $\varepsilon_a = \{1,0.1,0.01 \} k_b T$, respectively. The left column \textbf{(A,D,G)}, center column \textbf{(B,E,H)}, and right column \textbf{(C,F,I)} display results for the bead-spring model, mesoscale model, and LAMMPS implementation of Eq. \eqref{eq: Attachment rate scaling theory}, respectively. Data are provided for $N=\{12, 18,36\}$. \label{fig: ka_vs_ka1}}{0.95}
     
Besides interrogating the associative kinetics of single chains to fixed stickers, we also evaluate the bead-spring and mesoscale models' predicted attachment rates for sets of two chains undergoing pairwise bonding (\textbf{Fig. \ref{fig: Pairwise kinetics analysis}A-C}). Average attachment rates are plotted with respect to tether-to-tether separation distance, $d_{tt}$, for $N=\{12,18,36\}$ and  $\varepsilon_a=0.01 k_b T$ in \textbf{Fig. \ref{fig: Pairwise kinetics analysis}D}. Prefactor $A$ remains necessary (\textbf{Fig. \ref{fig: Pairwise kinetics analysis}E-F}), although its value for the two-chain system generally exceeds that of the single-chain system, for reasons requiring further investigation in future work.

     \figuremacro{H}{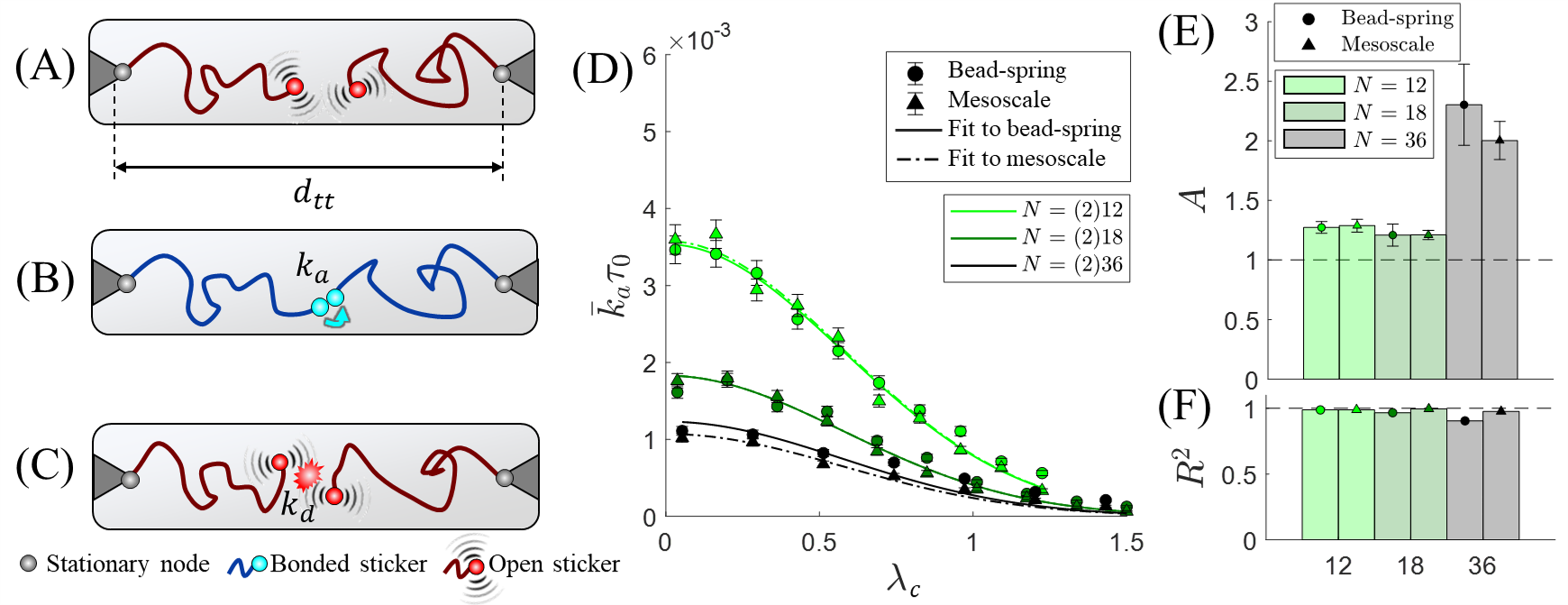}{Pairwise bond kinetics between two chains. }{\textbf{(A-C)} Illustration of two adjacent tethered chains with fixed ends (grey) separated by distance, $d_{tt}$, and free ends (red/blue stickers) that may bind/unbind to one another. \textbf{(D)} Average normalized attachment rates, $\bar k_a \tau_0$, with respect to normalized separation distance, $\lambda_c = d_{tt}/(\sqrt{2 N} b)$, when $\varepsilon_a = 0.01k_b T$. Error bars represent SE. Best fits of Eq. \eqref{eq: Attachment rate scaling theory} for the bead-spring (solid curves) and mesoscale (dashed curves) data are displayed (where $A$ is a fitting parameter). \textbf{(E)} Prefactor, $A$, for all chain lengths. Light green, dark green, and grey bars/markers correspond to $N=12$, $N=18$, and $N=36$, respectively. Error bars represent the 95$\%$ confidence interval. \textbf{(F)} Goodness of fit between the discrete models and scaling theory, characterized by $R^2$. \label{fig: Pairwise kinetics analysis}}{1}

    \tablemacro{H}{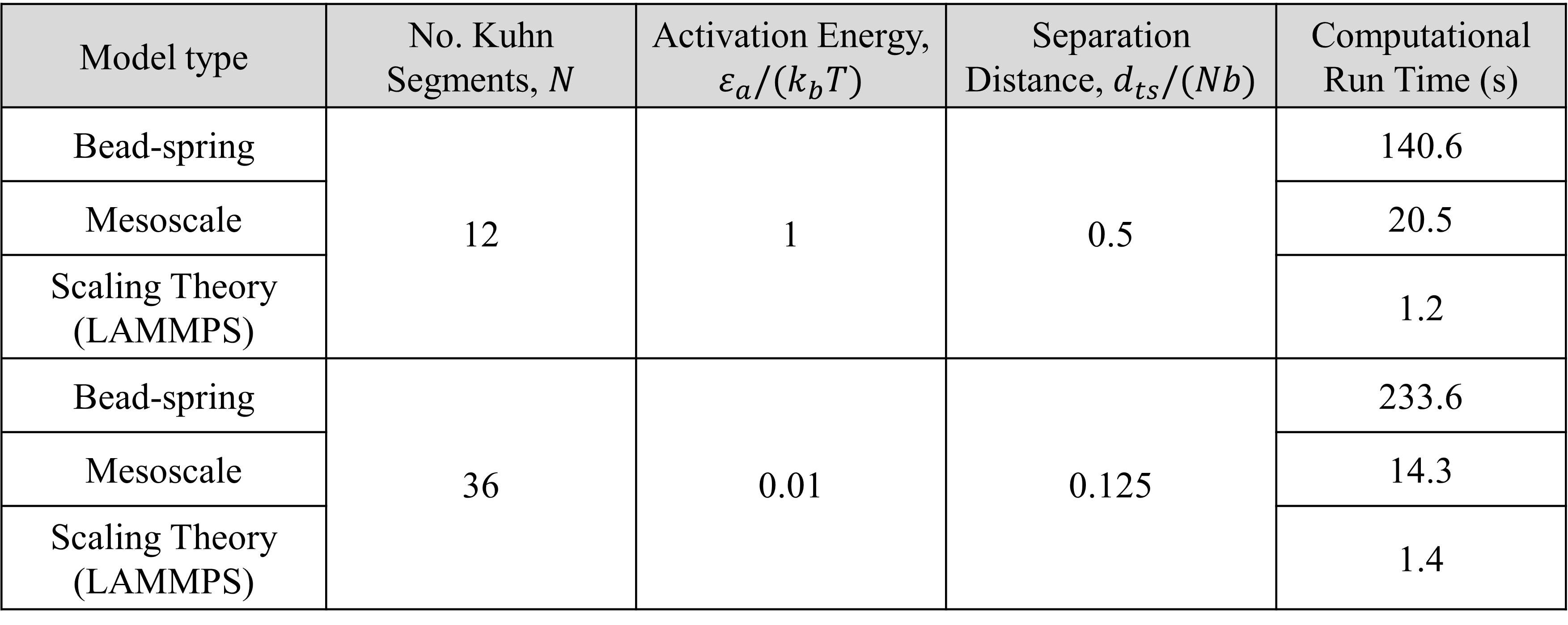}{}{Computational run times for each model's pairwise bonding study of \textbf{Section \ref{sec: Pairwise attachment}} for the parameter combinations with the lowest ($N=12$, $\varepsilon_a=k_b T$, $d_{ts} = 0.5 Nb$) and highest ($N=36$, $\varepsilon_a=0.01 k_b T$, $d_{ts} = 0.125 Nb$) computational costs. \label{Table 3}}{0.9}


\section{On prescribing polymer packing fractions}
\label{Appendix - polymer packing}

Historically, exact packing fractions -- or the more frequently reported free volume fractions ($1-\phi$) -- of polymers have not been consistently defined within the literature \citep{consolati_probing_2023}. Moreover, they have proved difficult to estimate experimentally due to the characteristic size of free volume features (e.g., pores and voids) at the molecular scale. Consequently, $\phi$ is often characterized by the differential volume fraction, $\phi_0 - \phi=\alpha_f (T-T_\infty)$, between a polymer at its current and Vogel (or reference) temperatures ($T$ and $T_\infty$, respectively) \citep{rubinstein_polymer_2003} where $\alpha_f$ is the coefficient of thermal expansion of the free volume, which is on the order of $10^{-4}$ to $10^{-3}$ K$^{-1}$ for highly crosslinked, rubbery polymers \citep{marzocca_physical_2013}. The Vogel temperature is commonly taken as 50 K below glass transition temperature, $T_g$, and is typically ascribed as the temperature at which free volume approximates zero (i.e., $\phi_0=1$), which -- while an idealization -- provides a pragmatic, relative estimate of empirical free volume. Experimental efforts that utilize a combination of dynamic mechanical analysis (DMA), differential scanning calorimetry (DSC), and positron annihilation lifetime spectroscopy (PALS) have indeed estimated that the reference, free volume fraction around $T_g$ is close to the order of 3-5$\%$ depending on the polymer's crosslink density \citep{marzocca_physical_2013}. Based on these values, one would approximate that the polymer packing fraction remains $\sim90$-$95\%$ even $10$ to $100$ K above $T_g$ in the elastomeric regime. 

However, recent studies conducted using all-atom MD and subsequently extrapolated machine learning predictions have estimated that the packing fraction of homopolymers and polyamides at ambient conditions (300 K and 0.1 MPa) are on the order of 55-70$\%$ \citep{tao_machine_2023}. Meanwhile, those of more broadly defined microporous polymers (with highly variable functional side group chemistries) at the same ambient conditions range from 55-80$\%$ \citep{tao_machine_2023}. While the work of \cite{tao_machine_2023} does not distinguish which polymers are in the glassy versus rubbery state when their free volume fractions are measured, these estimates provide a reasonable upper limit of polymer packing fraction on the order of 60-70$\%$. Based on these estimates and the realization that the freely jointed, ideal chain assumption has diminishing validity at higher packing fractions, we here limit packing fractions to a maximum value of $\phi \sim 0.5$. Lower limits of the packing fraction in the models are constrained only by the attainment of percolated, gel-like networks, which here occurred around $\phi \sim 0.2$.

\newpage    
\section{Extended ensemble bond dynamics}
\label{Appendix - extended exchange reaction studies}

\setcounter{figure}{0}    
\renewcommand{\thefigure}{\thesection \arabic{figure}}
\setcounter{table}{0}    
\renewcommand{\thetable}{\thesection \arabic{table}}

Average bond detachment rates, $\bar k_d$, and steady state detached and attached bond fractions ($\bar f_d$ and $\bar f_d$, respectively) from the study of \textbf{Section \ref{sec: Bond Exchange Results}} are plotted against nominal chain separation, $\bar d$, in \textbf{Fig. \ref{fig: Extended bond exchange results 1}}. 

   \figuremacro{H}{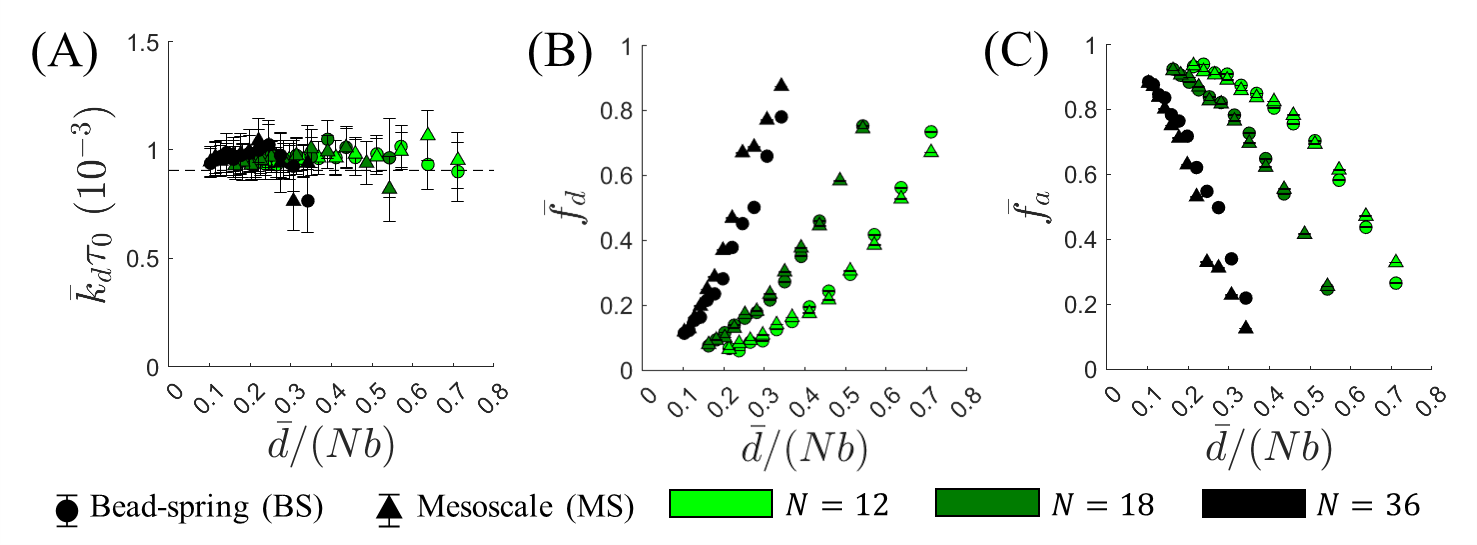}{Dissociative ensemble bond kinetics and steady state bond fractions. }{\textbf{(A)} bond detachment rates, $\bar k_d$, with respect to normalized chain separation, $d/(Nb)$. Dashed line indicates the value of $k_d$ prescribed \textit{a priori}. \textbf{(B-C)} Fractions of \textbf{(B)} detached and \textbf{(C)} attached chains with respect to $d/(Nb)$. Error bars represent standard error of the mean. \label{fig: Extended bond exchange results 1}}{1}
   
\noindent The average attached bond lifetimes, $\tau_a$, and renormalized bond lifetimes, $\bar \tau_{rnm}$, from the study of \textbf{Section \ref{sec: Bond Exchange Results}}  are plotted with respect to $\bar d$ in \textbf{Fig. \ref{fig: Extended bond exchange results 2}}. Renormalized bond lifetimes are roughly twice that of conventionally measured attached bond lifetimes, and the inverse renormalized bond lifetime is on the order of $\tau_{rnm}^{-1}\sim 10^{-3} \tau_0^{-1}$, which his consistent with the order of average partner exchange rates ($k_{exc}\sim  10^{-3} \tau_0^{-1}$ from \textbf{Fig. \ref{fig: Associative kinetics}B}), as expected.

    \figuremacro{H}{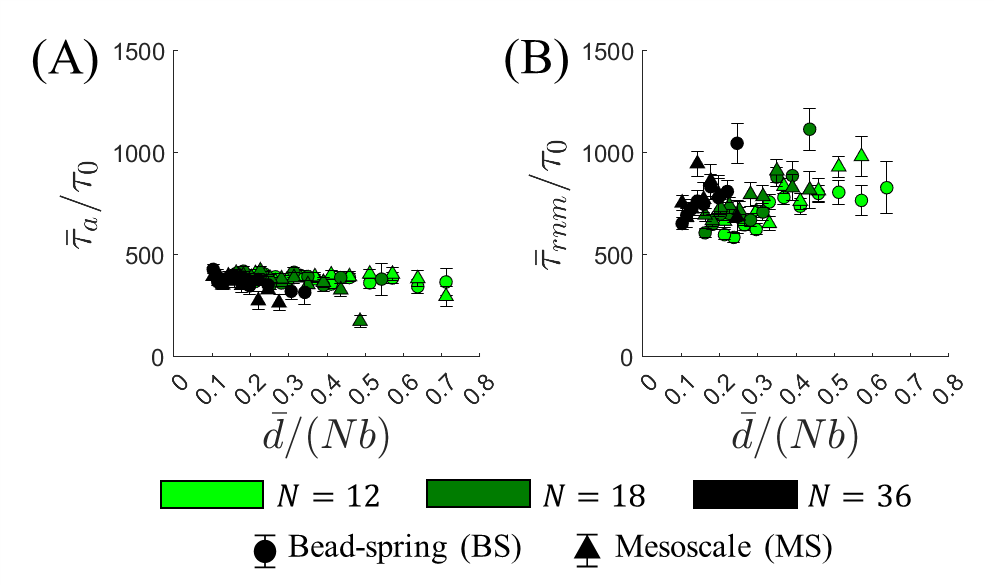}{Attached ensemble bond lifetimes. }{Average \textbf{(A)} attached bond lifetime, $\bar \tau_a$, and \textbf{(B)} renormalized bond lifetime, $\bar \tau_{rnm}$, with respect to normalized chain separation, $d/(Nb)$. Error bars represent standard error of the mean.  \label{fig: Extended bond exchange results 2}}{0.65}
   
\noindent The data from \textbf{Fig. \ref{fig: Associative kinetics}} are plotted with respect to chain (and therefore sticker) concentration, $c$, in \textbf{Fig. \ref{fig: Plot timescales wrt conc}} to illustrate the collapse of kinetic attachment rates and pertinent timescales in concentration-space. 

    \figuremacro{H}{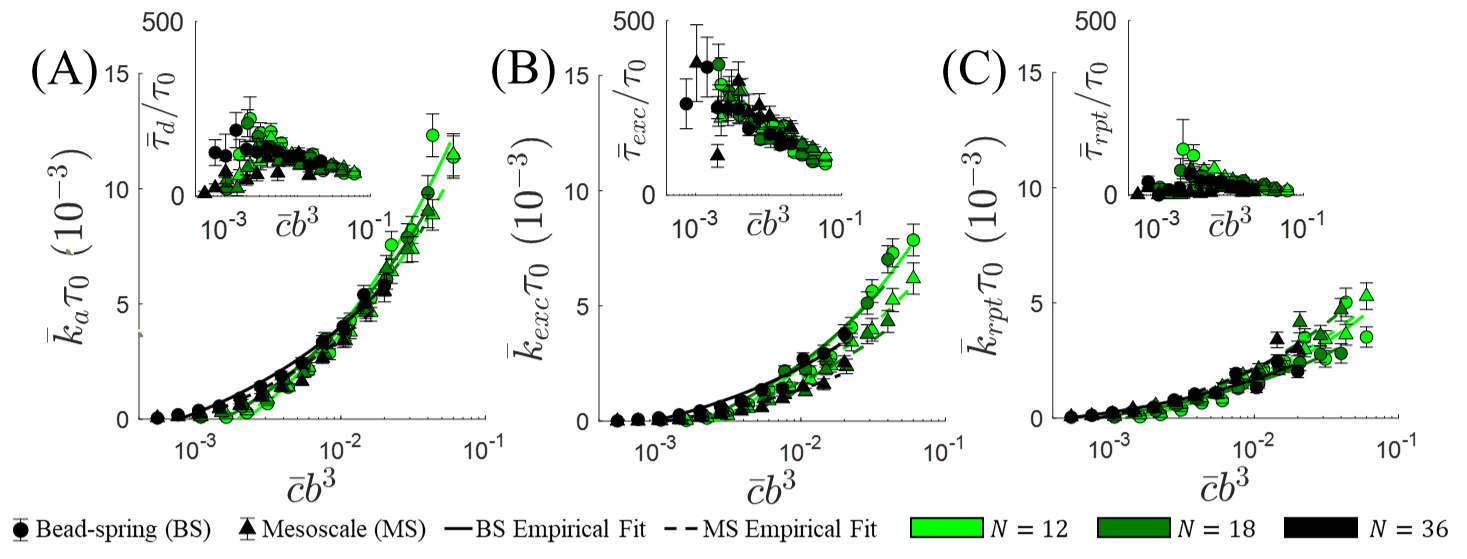}{}{\textbf{Associative kinetic rates versus sticker concentration.} \textbf{(A)} Bond attachment, \textbf{(B)} partner exchange, \textbf{(C)} and repeat attachment rates with respect to $cb^3 = (b/d)^3$ for both models when $N=\{12,18,36\}$. Error bars represent S.E. Dashed curves in \textbf{(A-C)} represent empirical fits according to Eq. (\ref{eq: empirical ensemble attachment}), where ${c}^{-1/3}$ is substituted for $d$. \label{fig: Plot timescales wrt conc}}{0.95}

\noindent Empirically fit parameters of Eq. \eqref{eq: empirical ensemble attachment} are plotted against chain length in \textbf{Fig. \ref{fig: emperical kinetics fits}}. 

   \figuremacro{H}{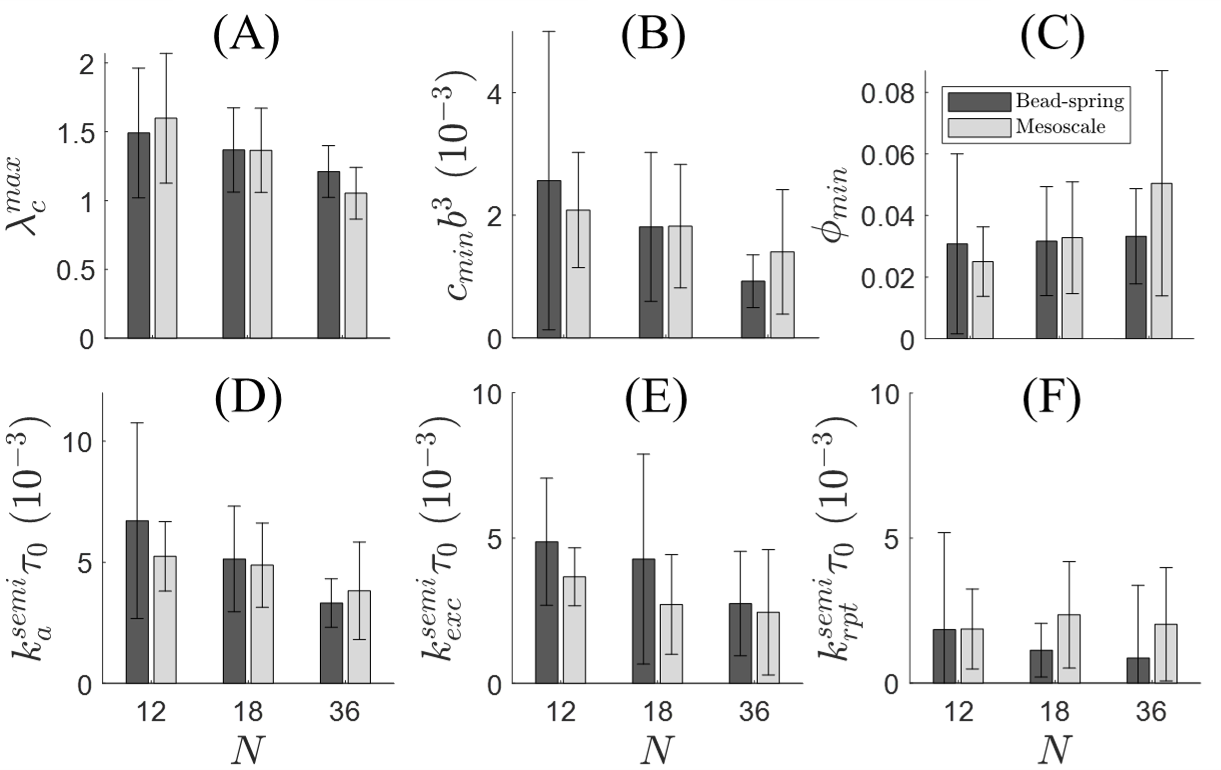}{\textbf{Average, empirically fit parameters. }}{\textbf{(A)} Maximum chain stretches, $\lambda_c^{max}=d_{max} \sqrt{N/2}$, \textbf{(B)} minimum chain concentrations, $c_{min}=d_{max}^{-1/3}$, \textbf{(C)} and minimum polymer packing fractions, $\phi_{min}=\pi N b^3/(6 d_{max}^3)$, as well as  characteristic \textbf{(D)} bond attachment rates, $k_a^{semi}$, \textbf{(E)} partner exchange rates, $k_{exc}^{semi}$, and \textbf{(F)} repeat attachment rates, $k_{rpt}^{semi}$ against $N$ for both models. Error bars represent the $95\%$ confidence interval from nonlinear, least-squares regression analysis. \label{fig: emperical kinetics fits}}{0.75}


\section{Extended network simulation results}
\label{Appendix - extended network results}

\setcounter{figure}{0}    
\renewcommand{\thefigure}{\thesection \arabic{figure}}
\setcounter{table}{0}    
\renewcommand{\thetable}{\thesection \arabic{table}}

Comparisons of the bead-spring virial stress evolution as computed using Eq. \eqref{eq: virial stress} when the sum is taken over all Kuhn segments versus when it is taken at the polymer chain end-to-end scale is provided in \textbf{Fig. \ref{fig: Virial Lammps vs mesoscale}} for all four sweeping parameters. Stresses are generally in reasonable agreement, but with considerably more noise for the Kuhn-scale computed virial stresses than those computed at the mesh-scale.
    
    \figuremacro{H}{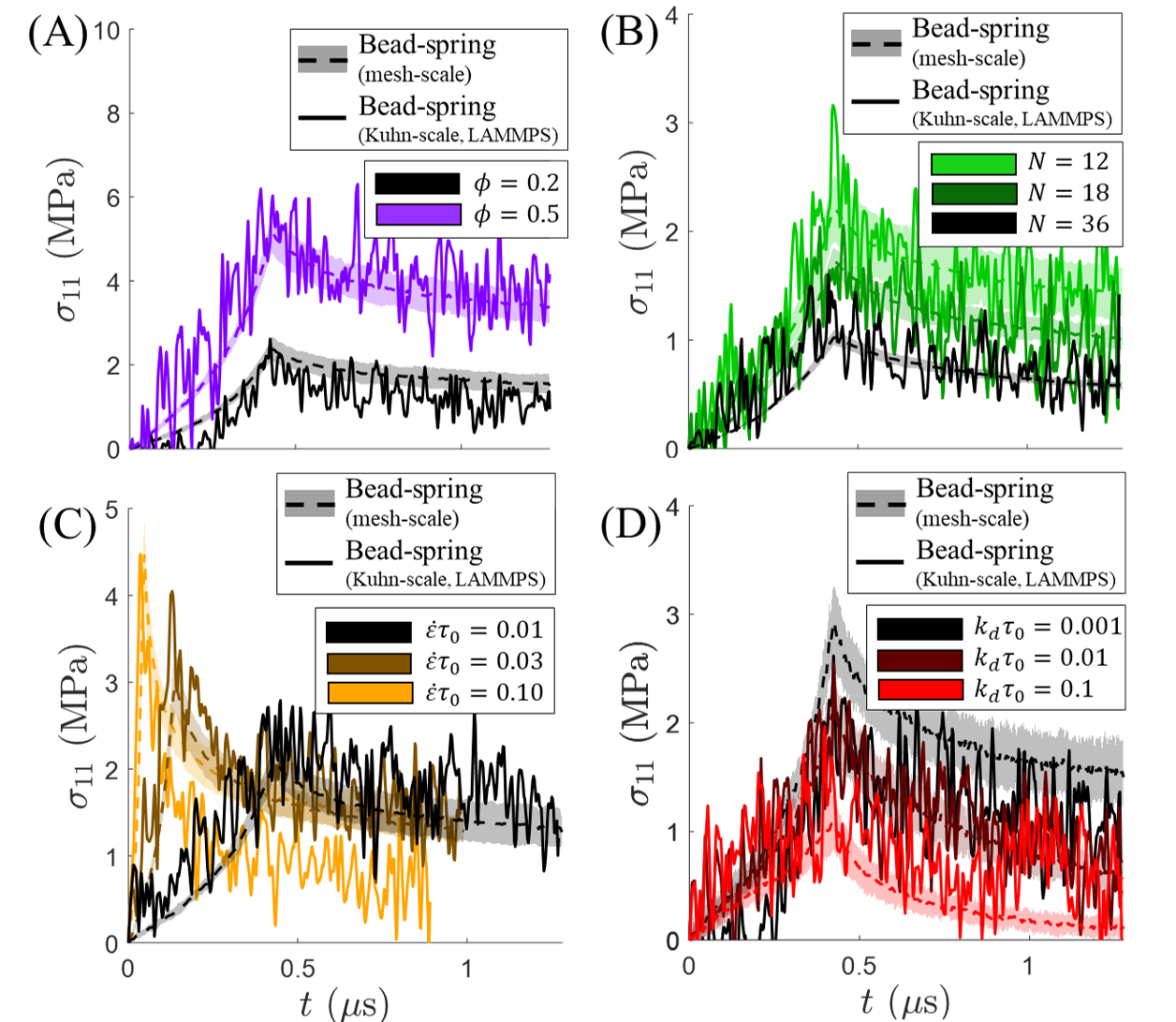}{\textbf{Kuhn versus mesh-scale virial stress comparison for bead-spring model. }}{Virial stress from Eq. \eqref{eq: virial stress} versus time for the bead-spring when computed using all Kuhn segments versus the polymer end-to-end vectors and the force-extension relation from \textbf{Fig. \ref{fig: Appendix_bond_potential}}. Parameters were set as \textbf{(A)} $\phi=\{0.2,0.5\}$, $N=12$, $\dot \varepsilon = 10^{-2}\tau_0^{-1}$, and $k_d=0$; \textbf{(B)} $\phi=0.2$, $N=\{12,18,36\}$, $\dot \varepsilon = 10^{-2}\tau_0^{-1}$, and $k_d=0$; \textbf{(C)} $\phi=0.2$, $N=12$, $\dot \varepsilon = \{1,3,10\}\times 10^{-2}\tau_0^{-1}$, and $k_d=0.1 \tau_0^{-1}$; and \textbf{(D)} $\phi=0.2$, $N=12$, $\dot \varepsilon = 10^{-2}\tau_0^{-1}$, and $k_d=\{10^{-3},10^{-2},10^{-1}\}\tau_0^{-1}$. \label{fig: Virial Lammps vs mesoscale}}{0.85}

\newpage
\noindent CPU run times and required storage per additional step of stored data of the full network-scale models from \textbf{Section \ref{sec: Results of network-scale}} are provided in \textbf{Fig. \ref{fig: Computational cost}}.

    \figuremacro{H}{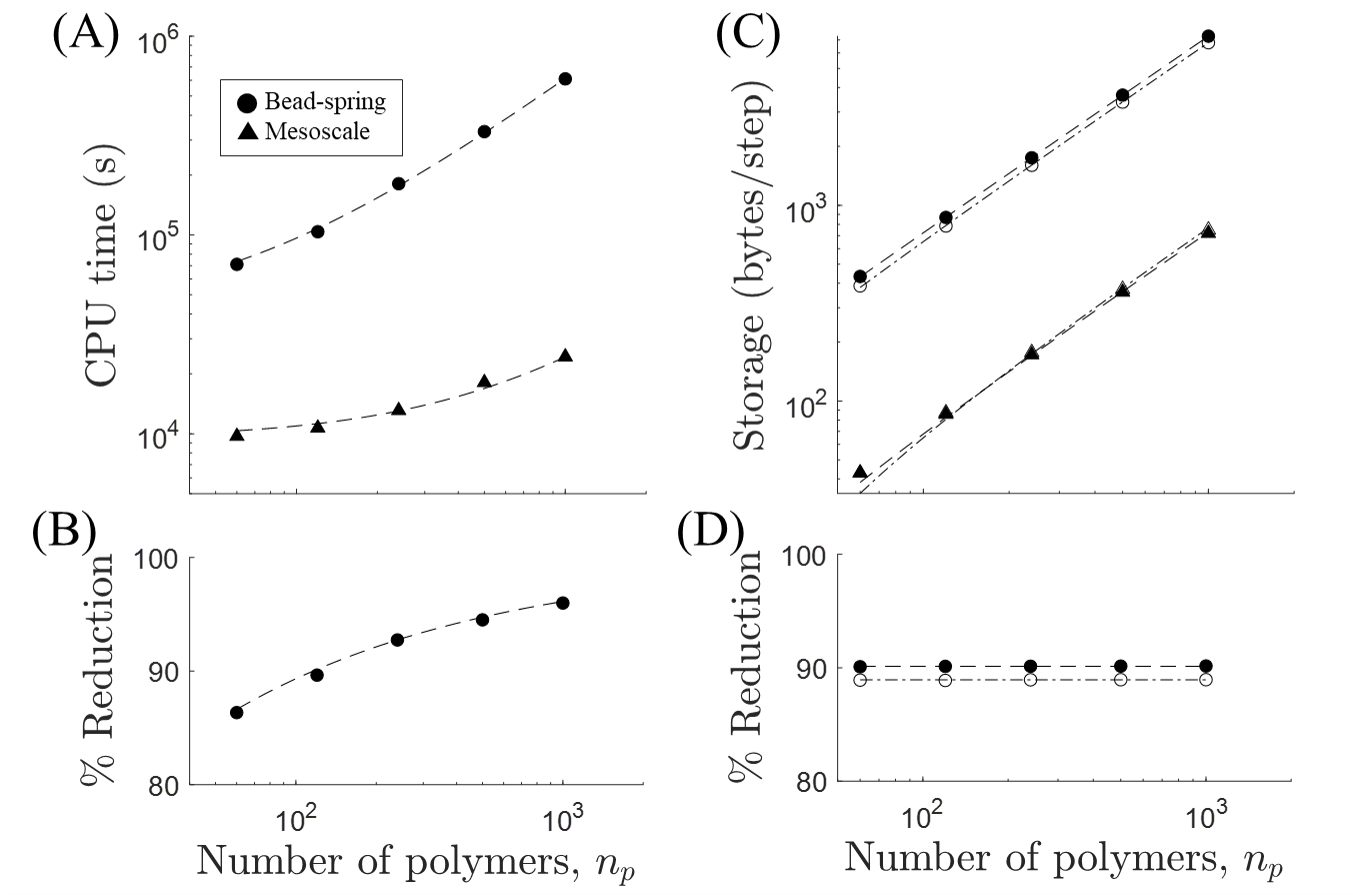}{\textbf{Computational run times and data storage requirements. }}{ \textbf{(A)} CPU time and \textbf{(B)} percent reduction in CPU time (from bead-spring to mesoscale models) with respect to simulated number of polymer chains, $n_p$. CPU time for the bead-spring and mesoscale models scale as $t_{cpu} \propto [547 $s/polymer$] n_p$  ($R^2=1.00$), and $t_{cpu} \propto [15 $s/polymer$] n_p$  ($R^2=1.00$), respectively. Percent reduction in CPU time scales as $\%\Delta t_{cpu} = 1-(0.81) n_p^{-0.44}$ ($R^2=1.00$). All simulations were run on  80 cores of a single Ice Lake (ICX) node using Stampede3. \textbf{(C)} Data storage required per additional step of data output using LAMMPS's atom.dump (filled) and bond.dump (open) files. Storage requirements for the bead-spring and mesoscale models scale proportionately to $[7.3$ Kb/step/polymer$] n_p$ ($R^2=1.00)$ and $[0.7$ Kb/step/polymer$] n_p$ ($R^2=1.00)$, respectively. $N=12$, $\phi=0.5$, $k_d = 0.01 \tau_0^{-1}$, and $\dot \varepsilon = 0.01 \tau_0^{-1}$ for all simulations. Percent reduction in storage requirements is constant and on the order of $90\%$ for both file types.  \label{fig: Computational cost}}{0.9}

\newpage
\noindent Analogous data to that of \textbf{Fig. \ref{fig: N12}} in \textbf{Section \ref{sec: Results of network-scale}} are provided in \textbf{Fig. \ref{fig: N36}} when $N=36$.

    \figuremacro{H}{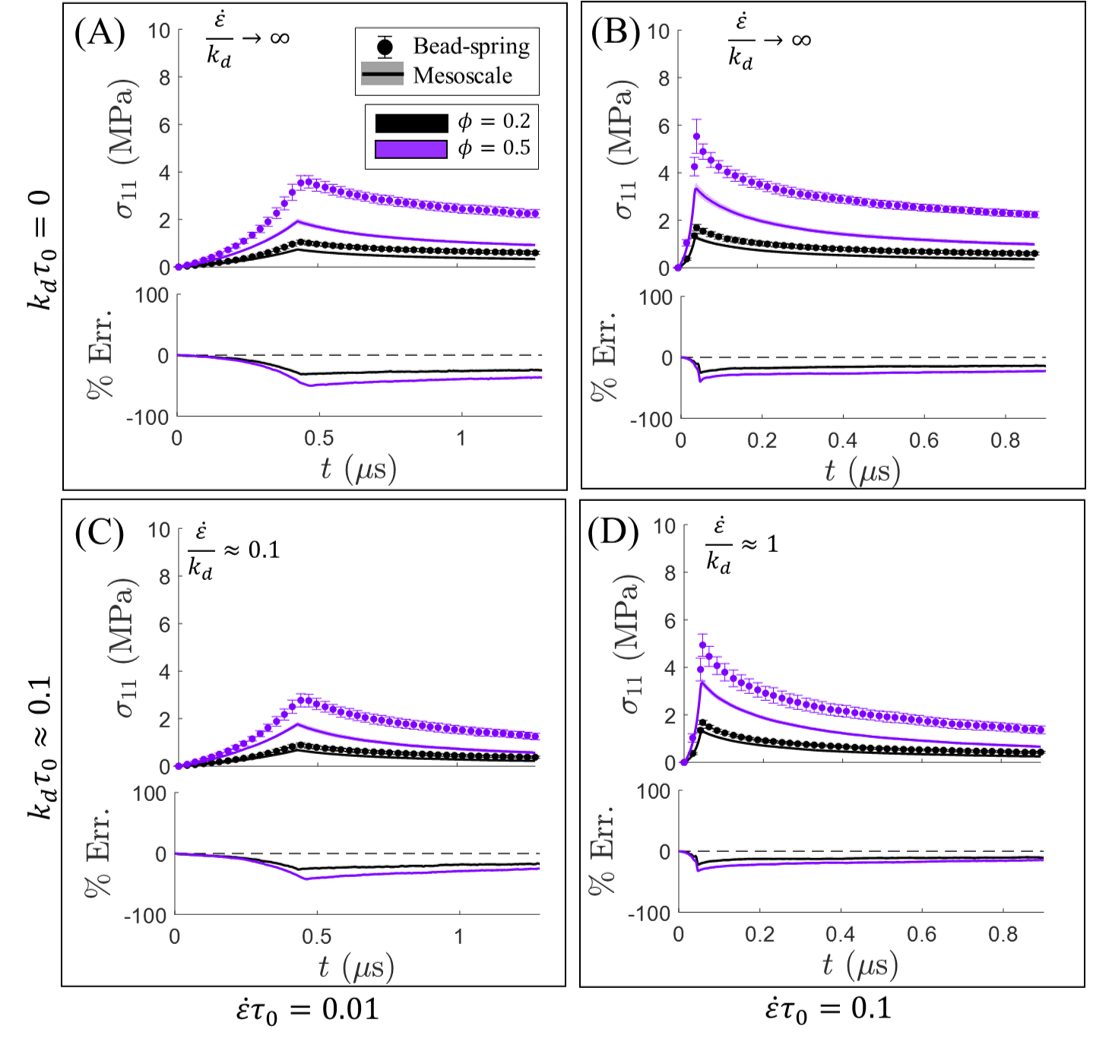}{Mechanical response at the parameter extremes when $\bm{N=36}$. }{Normal Cauchy stress in the direction of loading for the bead-spring and mesoscale models, as well as relative error are plotted with respect to time at both polymer packing fractions ($\phi = \{0.2,0.5\}$), loading rates ($\dot \varepsilon = \{0.01,0.1\}\tau_0^{-1}$), and bond detachment rates ($k_d=\{0,0.1\}\tau_0^{-1}$) when $N=36$. Horizontal dashed lines on error plots denote zero error.\label{fig: N36}}{0.95}
    

\newpage
\noindent Fractions of attached and detached chains, as well as the average number of bonds between two side chains of the same molecule are plotted in \textbf{Fig. \ref{fig: Chain length attached fracs}} with respect to time (during initial network equilibration), as measured from the simulations of \textbf{Section \ref{sec: Chain Length Sweep}}. These demonstrate how, despite similar values of $\bar f_a$ and $\bar f_d$, the two modeling approaches predict disparate degrees of self-attachment for networks of longer chains ($N=36$).

    \figuremacro{H}{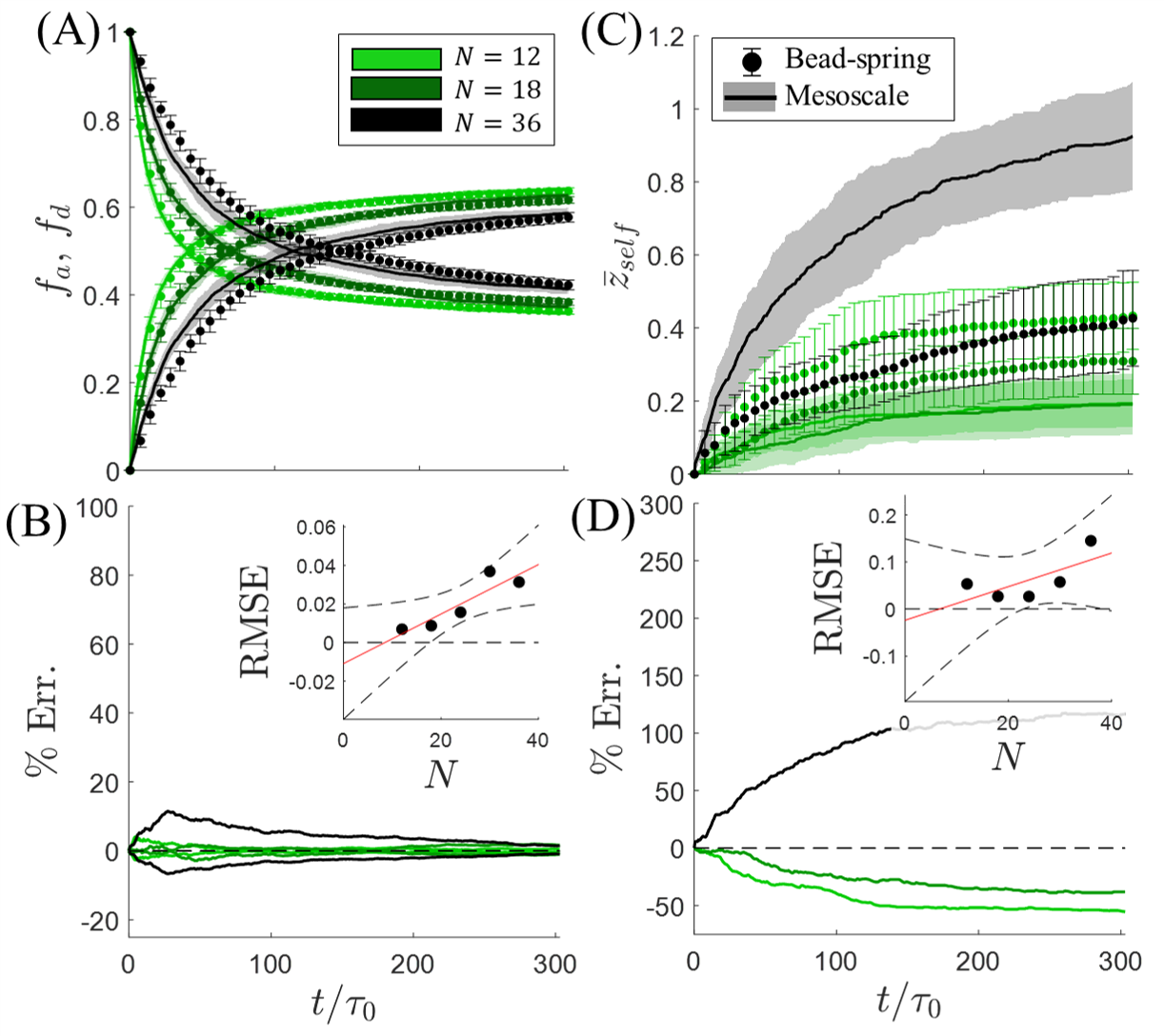}{\textbf{Effects of chain length on network connectivity. }}{\textbf{(A)} Attached/detached chain fractions and \textbf{(B)} relative error between the models (bead-spring model as reference) with respect to time during initial network equilibration. \textbf{(C)} Average degree of intra-molecular dynamic bonds and \textbf{(D)} relative error between the models (bead-spring model as reference), also with respect to time during initial network equilibration. Insets of \textbf{(B,D)} display root mean-square error (RMSE) of \textbf{(A,C)}, respectively, with respect to $N$. The red lines represents the best fit from linear regression analysis with a 95\% confidence interval ($p=0.04$ for \textbf{(B)} and $p=0.19$ for \textbf{(D)}). Although the degree of intra-molecular attachment appears biphasic with respect to $N$, RMSE is considerably higher for $N=36$ (RMSE $=15\%$) than the next closest value (RMSE $=6\%$ for $N=30$). \label{fig: Chain length attached fracs}}{0.8}

\newpage
\noindent MSD data and relative error between the bead-spring and mesoscale models from the simulations of \textbf{Section \ref{sec: Chain Length Sweep}} are plotted with respect to time (during initial network equilibration) in \textbf{Fig. \ref{fig: Chain length MSDs}}. These demonstrate how the mesoscale model captures slower diffusion than that of the bead-spring model for both sticker and tether nodes.

    \figuremacro{H}{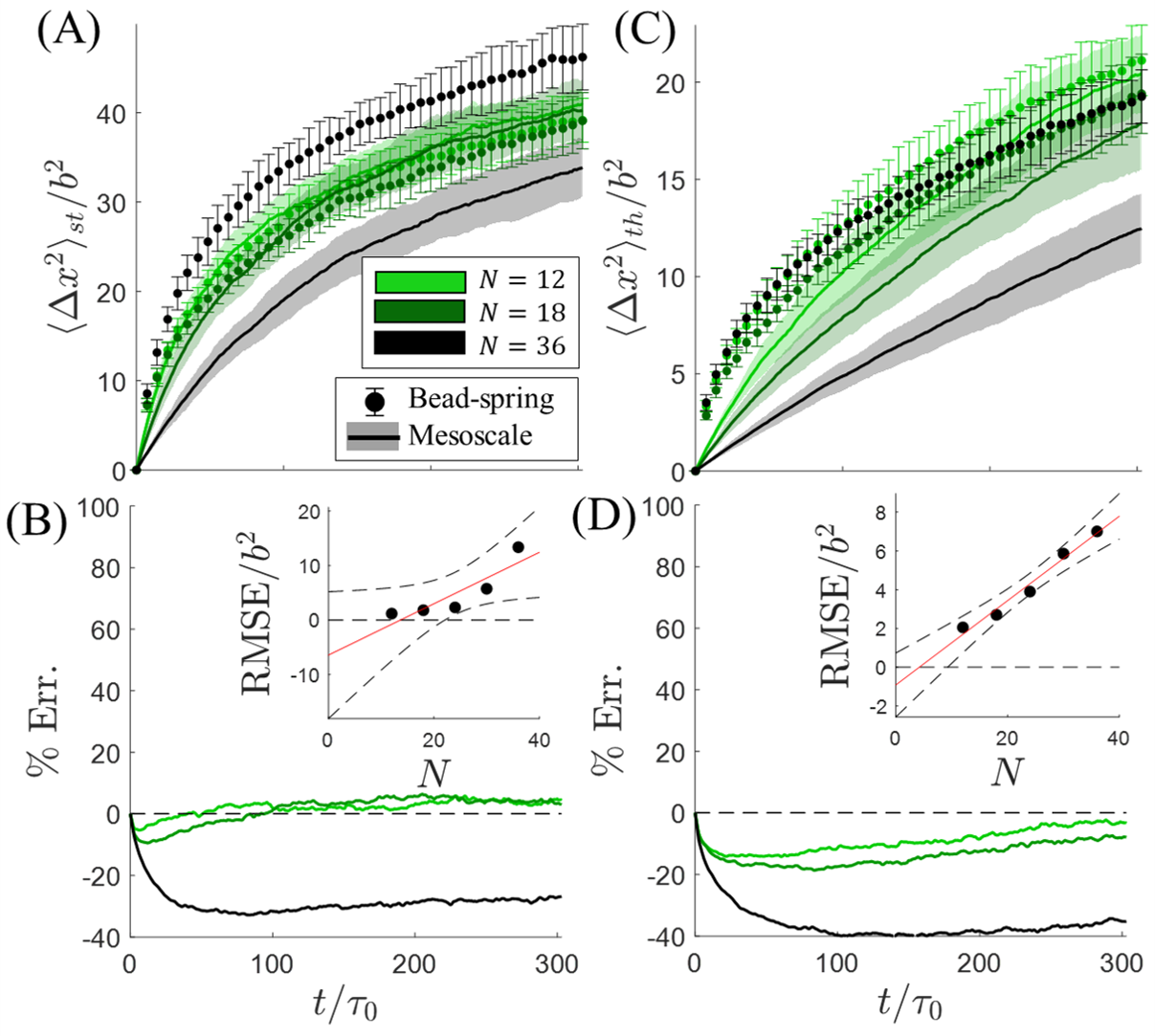}{\textbf{Effects of chain length on sticker and tether diffusion. }}{\textbf{(A)} MSD and \textbf{(B)} relative error between the models (bead-spring model as reference) with respect to time during initial network equilibration for the sticker nodes. \textbf{(C-D)} The same data as \textbf{(A-B)}, except it is provided for the tether nodes. Insets of \textbf{(B,D)} display RMSE of \textbf{(A,C)}, respectively, against $N$. The red lines represent the best fits from linear regression analysis with a 95\% confidence interval ($p=0.04$ for \textbf{(B)} and $p<0.01$ for \textbf{(D)}), illustrating the correlation between $N$ and error between the MSD values of both models.  \label{fig: Chain length MSDs}}{0.8}

\titleformat{\section}[block]{\normalfont\Large\bfseries}{}{0pt}{}
\section{CRediT authorship contribution statement}

\textbf{RJW:} Conceptualization, Methodology, Validation, Investigation, Formal analysis, Visualization, Software, Data curation, Resources, Writing – original draft, Writing – review \& editing. \textbf{MNS:} Conceptualization, Funding acquisition, Project administration, Resources, Supervision, Writing – original draft, Writing – review \& editing.

\section{Declaration of competing interest}

The authors declare that they have no known competing financial interests or personal relationships that could have appeared to influence the work reported in this paper.

\section{Acknowledgements}

This work was supported by the United States National Air Force Office of Scientific Research (AFOSR) under Contract No. FA9550-22-1-0030 and the United States National Science Foundation (NSF) under Grant No. CAREER-1653059. This work used Stampede3 at the Texas Advanced Computing Center (TACC) at The University of Texas at Austin through allocation MCH230053 from the Advanced Cyberinfrastructure Coordination Ecosystem: Services \& Support (ACCESS) program, which is supported by National Science Foundation grants \#2138259, \#2138286, \#2138307, \#2137603, and \#2138296 \citep{boerner_access_2023}. This content is solely the responsibility of the authors and does not necessarily represent the official views of AFOSR, NSF, TACC, or ACCESS. The authors would like to thank Dr. S. Lamont and Professor F. Vernerey (CU Boulder) for orchestrating the implementation of the Transient Network Theory (TNT) package into LAMMPS based on the work of \cite{wagner_network_2021}.


\bibliography{Mesoscale} 

\begin{thebibliography}{135}
\expandafter\ifx\csname natexlab\endcsname\relax\def\natexlab#1{#1}\fi
\providecommand{\url}[1]{\texttt{#1}}
\providecommand{\href}[2]{#2}
\providecommand{\path}[1]{#1}
\providecommand{\DOIprefix}{doi:}
\providecommand{\ArXivprefix}{arXiv:}
\providecommand{\URLprefix}{URL: }
\providecommand{\Pubmedprefix}{pmid:}
\providecommand{\doi}[1]{\href{http://dx.doi.org/#1}{\path{#1}}}
\providecommand{\Pubmed}[1]{\href{pmid:#1}{\path{#1}}}
\providecommand{\bibinfo}[2]{#2}
\ifx\xfnm\relax \def\xfnm[#1]{\unskip,\space#1}\fi
\bibitem[{Chung(2019)}]{chung_review_2019}
\bibinfo{author}{Chung D.~D.~L.},
\newblock \bibinfo{title}{A review of multifunctional polymer-matrix structural composites},
\newblock \bibinfo{journal}{Composites Part B: Engineering} \bibinfo{volume}{160} (\bibinfo{year}{2019}) \bibinfo{pages}{644--660}. \DOIprefix\doi{10.1016/j.compositesb.2018.12.117}.
\bibitem[{Rydz et~al.(2021)Rydz, Opálková~Šišková, Zawidlak-Wegrzyńska, and Duale}]{rydz_chapter_2021}
\bibinfo{author}{Rydz J.}, \bibinfo{author}{Opálková~Šišková A.}, \bibinfo{author}{Zawidlak-Wegrzyńska B.}, \bibinfo{author}{Duale K.},
\newblock \bibinfo{title}{Chapter 1 - {High}-performance polymer applications for renewable energy},
\newblock in: \bibinfo{editor}{Devasahayam S.}, \bibinfo{editor}{Hussain C.~M.} (Eds.), \bibinfo{booktitle}{Nano {Tools} and {Devices} for {Enhanced} {Renewable} {Energy}}, Micro and {Nano} {Technologies}, \bibinfo{publisher}{Elsevier}, \bibinfo{year}{2021}, pp. \bibinfo{pages}{3--26}.
\bibitem[{Vallejos et~al.(2022)Vallejos, Trigo-L{\'o}pez, Arnaiz, Miguel, Mu{\~n}oz, Mend{\'\i}a, and Garc{\'\i}a}]{vallejos_classical_2022}
\bibinfo{author}{Vallejos S.}, \bibinfo{author}{Trigo-L{\'o}pez M.}, \bibinfo{author}{Arnaiz A.}, \bibinfo{author}{Miguel {\'A}.}, \bibinfo{author}{Mu{\~n}oz A.}, \bibinfo{author}{Mend{\'\i}a A.}, \bibinfo{author}{Garc{\'\i}a J.~M.},
\newblock \bibinfo{title}{From classical to advanced use of polymers in food and beverage applications},
\newblock \bibinfo{journal}{Polymers} \bibinfo{volume}{14} (\bibinfo{year}{2022}) \bibinfo{pages}{4954}. \DOIprefix\doi{10.3390/polym14224954}.
\bibitem[{Ibrahim et~al.(2022)Ibrahim, Hamam, Sadiku, Ndambuki, Kupolati, Jamiru, Eze, and Snyman}]{ibrahim_need_2022}
\bibinfo{author}{Ibrahim I.~D.}, \bibinfo{author}{Hamam Y.}, \bibinfo{author}{Sadiku E.~R.}, \bibinfo{author}{Ndambuki J.~M.}, \bibinfo{author}{Kupolati W.~K.}, \bibinfo{author}{Jamiru T.}, \bibinfo{author}{Eze A.~A.}, \bibinfo{author}{Snyman J.},
\newblock \bibinfo{title}{Need for {Sustainable} {Packaging}: {An} {Overview}},
\newblock \bibinfo{journal}{Polymers} \bibinfo{volume}{14} (\bibinfo{year}{2022}) \bibinfo{pages}{4430}. \DOIprefix\doi{10.3390/polym14204430}.
\bibitem[{Heinzmann et~al.(2016)Heinzmann, Weder, and Espinosa}]{heinzmann_supramolecular_2016}
\bibinfo{author}{Heinzmann C.}, \bibinfo{author}{Weder C.}, \bibinfo{author}{Espinosa L.~M.~d.},
\newblock \bibinfo{title}{Supramolecular polymer adhesives: advanced materials inspired by nature},
\newblock \bibinfo{journal}{Chemical Society Reviews} \bibinfo{volume}{45} (\bibinfo{year}{2016}) \bibinfo{pages}{342--358}. \DOIprefix\doi{10.1039/C5CS00477B}.
\bibitem[{Chen et~al.(2022)Chen, Chen, Chen, Cui, Duan, Kang, Liu, Liu, Muhammad, Shao, Tang, Wang, Wang, Xiong, Yin, Zhang, Zhang, Zhen, Feng, Gao, Gu, He, Ji, Jiang, Liu, Liu, Peng, Shen, Shi, Sun, Wang, Wang, Xiao, Xu, Zhong, Zhang, and Chen}]{chen_biomedical_2022}
\bibinfo{author}{Chen W.-H.}, \bibinfo{author}{Chen Q.-W.}, \bibinfo{author}{Chen Q.}, \bibinfo{author}{Cui C.}, \bibinfo{author}{Duan S.}, \bibinfo{author}{Kang Y.}, \bibinfo{author}{Liu Y.}, \bibinfo{author}{Liu Y.}, \bibinfo{author}{Muhammad W.}, \bibinfo{author}{Shao S.}, \bibinfo{author}{Tang C.}, \bibinfo{author}{Wang J.}, \bibinfo{author}{Wang L.}, \bibinfo{author}{Xiong M.-H.}, \bibinfo{author}{Yin L.}, \bibinfo{author}{Zhang K.}, \bibinfo{author}{Zhang Z.}, \bibinfo{author}{Zhen X.}, \bibinfo{author}{Feng J.}, \bibinfo{author}{Gao C.}, \bibinfo{author}{Gu~Z.}, \bibinfo{author}{He~C.}, \bibinfo{author}{Ji~J.}, \bibinfo{author}{Jiang X.}, \bibinfo{author}{Liu W.}, \bibinfo{author}{Liu Z.}, \bibinfo{author}{Peng H.}, \bibinfo{author}{Shen Y.}, \bibinfo{author}{Shi L.}, \bibinfo{author}{Sun X.}, \bibinfo{author}{Wang H.}, \bibinfo{author}{Wang J.}, \bibinfo{author}{Xiao H.}, \bibinfo{author}{Xu~F.-J.}, \bibinfo{author}{Zhong Z.}, \bibinfo{author}{Zhang X.-Z.}, \bibinfo{author}{Chen X.},
\newblock \bibinfo{title}{Biomedical polymers: synthesis, properties, and applications},
\newblock \bibinfo{journal}{Science China Chemistry} \bibinfo{volume}{65} (\bibinfo{year}{2022}) \bibinfo{pages}{1010--1075}. \DOIprefix\doi{10.1007/s11426-022-1243-5}.
\bibitem[{Kloxin et~al.(2010)Kloxin, Scott, Adzima, and Bowman}]{kloxin_covalent_2010}
\bibinfo{author}{Kloxin C.~J.}, \bibinfo{author}{Scott T.~F.}, \bibinfo{author}{Adzima B.~J.}, \bibinfo{author}{Bowman C.~N.},
\newblock \bibinfo{title}{Covalent {Adaptable} {Networks} ({CANs}): {A} {Unique} {Paradigm} in {Cross}-{Linked} {Polymers}},
\newblock \bibinfo{journal}{Macromolecules} \bibinfo{volume}{43} (\bibinfo{year}{2010}) \bibinfo{pages}{2643--2653}. \DOIprefix\doi{10.1021/ma902596s}.
\bibitem[{Yue et~al.(2020)Yue, Guo, Kennedy, Patel, Gong, Ju, Gray, and Manas-Zloczower}]{yue_vitrimerization_2020}
\bibinfo{author}{Yue L.}, \bibinfo{author}{Guo H.}, \bibinfo{author}{Kennedy A.}, \bibinfo{author}{Patel A.}, \bibinfo{author}{Gong X.}, \bibinfo{author}{Ju~T.}, \bibinfo{author}{Gray T.}, \bibinfo{author}{Manas-Zloczower I.},
\newblock \bibinfo{title}{Vitrimerization: {Converting} {Thermoset} {Polymers} into {Vitrimers}},
\newblock \bibinfo{journal}{ACS Macro Letters} \bibinfo{volume}{9} (\bibinfo{year}{2020}) \bibinfo{pages}{836--842}. \DOIprefix\doi{10.1021/acsmacrolett.0c00299}.
\bibitem[{Shen et~al.(2021)Shen, Song, Cai, and Vernerey}]{shen_nonsteady_2021}
\bibinfo{author}{Shen T.}, \bibinfo{author}{Song Z.}, \bibinfo{author}{Cai S.}, \bibinfo{author}{Vernerey F.~J.},
\newblock \bibinfo{title}{Nonsteady fracture of transient networks: {The} case of vitrimer},
\newblock \bibinfo{journal}{Proceedings of the National Academy of Sciences} \bibinfo{volume}{118} (\bibinfo{year}{2021}) \bibinfo{pages}{e2105974118}. \DOIprefix\doi{10.1073/pnas.2105974118}.
\bibitem[{Hubbard et~al.(2021)Hubbard, Ren, Konkolewicz, Sarvestani, Picu, Kedziora, Roy, Varshney, and Nepal}]{hubbard_vitrimer_2021}
\bibinfo{author}{Hubbard A.~M.}, \bibinfo{author}{Ren Y.}, \bibinfo{author}{Konkolewicz D.}, \bibinfo{author}{Sarvestani A.}, \bibinfo{author}{Picu C.~R.}, \bibinfo{author}{Kedziora G.~S.}, \bibinfo{author}{Roy A.}, \bibinfo{author}{Varshney V.}, \bibinfo{author}{Nepal D.},
\newblock \bibinfo{title}{Vitrimer {Transition} {Temperature} {Identification}: {Coupling} {Various} {Thermomechanical} {Methodologies}},
\newblock \bibinfo{journal}{ACS Applied Polymer Materials} \bibinfo{volume}{3} (\bibinfo{year}{2021}) \bibinfo{pages}{1756--1766}. \DOIprefix\doi{10.1021/acsapm.0c01290}.
\bibitem[{Hubbard et~al.(2022)Hubbard, Ren, Picu, Sarvestani, Konkolewicz, Roy, Varshney, and Nepal}]{hubbard_creep_2022}
\bibinfo{author}{Hubbard A.~M.}, \bibinfo{author}{Ren Y.}, \bibinfo{author}{Picu C.~R.}, \bibinfo{author}{Sarvestani A.}, \bibinfo{author}{Konkolewicz D.}, \bibinfo{author}{Roy A.~K.}, \bibinfo{author}{Varshney V.}, \bibinfo{author}{Nepal D.},
\newblock \bibinfo{title}{Creep {Mechanics} of {Epoxy} {Vitrimer} {Materials}},
\newblock \bibinfo{journal}{ACS Applied Polymer Materials} \bibinfo{volume}{4} (\bibinfo{year}{2022}) \bibinfo{pages}{4254--4263}. \DOIprefix\doi{10.1021/acsapm.2c00230}.
\bibitem[{Brunsveld et~al.(2001)Brunsveld, Folmer, Meijer, and Sijbesma}]{brunsveld_supramolecular_2001}
\bibinfo{author}{Brunsveld L.}, \bibinfo{author}{Folmer B.~J.~B.}, \bibinfo{author}{Meijer E.~W.}, \bibinfo{author}{Sijbesma R.~P.},
\newblock \bibinfo{title}{Supramolecular {Polymers}},
\newblock \bibinfo{journal}{Chemical Reviews} \bibinfo{volume}{101} (\bibinfo{year}{2001}) \bibinfo{pages}{4071--4098}. \DOIprefix\doi{10.1021/cr990125q}.
\bibitem[{Yount et~al.(2005)Yount, Loveless, and Craig}]{yount_strong_2005}
\bibinfo{author}{Yount W.~C.}, \bibinfo{author}{Loveless D.~M.}, \bibinfo{author}{Craig S.~L.},
\newblock \bibinfo{title}{Strong {Means} {Slow}: {Dynamic} {Contributions} to the {Bulk} {Mechanical} {Properties} of {Supramolecular} {Networks}},
\newblock \bibinfo{journal}{Angewandte Chemie International Edition} \bibinfo{volume}{44} (\bibinfo{year}{2005}) \bibinfo{pages}{2746--2748}. \DOIprefix\doi{10.1002/anie.200500026}.
\bibitem[{Vidavsky et~al.(2020)Vidavsky, Buche, Sparrow, Zhang, Yang, DiStasio, and Silberstein}]{vidavsky_tuning_2020}
\bibinfo{author}{Vidavsky Y.}, \bibinfo{author}{Buche M.~R.}, \bibinfo{author}{Sparrow Z.~M.}, \bibinfo{author}{Zhang X.}, \bibinfo{author}{Yang S.~J.}, \bibinfo{author}{DiStasio R.~A.~J.}, \bibinfo{author}{Silberstein M.~N.},
\newblock \bibinfo{title}{Tuning the {Mechanical} {Properties} of {Metallopolymers} via {Ligand} {Interactions}: {A} {Combined} {Experimental} and {Theoretical} {Study}},
\newblock \bibinfo{journal}{Macromolecules} \bibinfo{volume}{53} (\bibinfo{year}{2020}) \bibinfo{pages}{2021--2030}. \DOIprefix\doi{10.1021/acs.macromol.9b02756}.
\bibitem[{Mordvinkin et~al.(2021)Mordvinkin, Döhler, Binder, Colby, and Saalwächter}]{mordvinkin_rheology_2021}
\bibinfo{author}{Mordvinkin A.}, \bibinfo{author}{Döhler D.}, \bibinfo{author}{Binder W.~H.}, \bibinfo{author}{Colby R.~H.}, \bibinfo{author}{Saalwächter K.},
\newblock \bibinfo{title}{Rheology, {Sticky} {Chain}, and {Sticker} {Dynamics} of {Supramolecular} {Elastomers} {Based} on {Cluster}-{Forming} {Telechelic} {Linear} and {Star} {Polymers}},
\newblock \bibinfo{journal}{Macromolecules} \bibinfo{volume}{54} (\bibinfo{year}{2021}) \bibinfo{pages}{5065--5076}. \DOIprefix\doi{10.1021/acs.macromol.1c00655}.
\bibitem[{Xu et~al.(2022)Xu, Fu, Wagner, Zou, He, Li, Pan, Ding, and Vernerey}]{xu_thermosensitive_2022}
\bibinfo{author}{Xu~L.}, \bibinfo{author}{Fu~Y.}, \bibinfo{author}{Wagner R.~J.}, \bibinfo{author}{Zou X.}, \bibinfo{author}{He~Q.}, \bibinfo{author}{Li~T.}, \bibinfo{author}{Pan W.}, \bibinfo{author}{Ding J.}, \bibinfo{author}{Vernerey F.~J.},
\newblock \bibinfo{title}{Thermosensitive {P}({AAc}-co-{NIPAm}) {Hydrogels} {Display} {Enhanced} {Toughness} and {Self}-{Healing} via {Ion}–{Ligand} {Interactions}},
\newblock \bibinfo{journal}{Macromolecular Rapid Communications} \bibinfo{volume}{43} (\bibinfo{year}{2022}) \bibinfo{pages}{2200320}. \DOIprefix\doi{10.1002/marc.202200320}.
\bibitem[{Haque et~al.(2011)Haque, Kurokawa, Kamita, and Gong}]{haque_lamellar_2011}
\bibinfo{author}{Haque M.~A.}, \bibinfo{author}{Kurokawa T.}, \bibinfo{author}{Kamita G.}, \bibinfo{author}{Gong J.~P.},
\newblock \bibinfo{title}{Lamellar {Bilayers} as {Reversible} {Sacrificial} {Bonds} {To} {Toughen} {Hydrogel}: {Hysteresis}, {Self}-{Recovery}, {Fatigue} {Resistance}, and {Crack} {Blunting}},
\newblock \bibinfo{journal}{Macromolecules} \bibinfo{volume}{44} (\bibinfo{year}{2011}) \bibinfo{pages}{8916--8924}. \DOIprefix\doi{10.1021/ma201653t}.
\bibitem[{Gong et~al.(2016)Gong, Zhang, Zeng, Li, Li, Huang, Sun, and Wong}]{gong_high-strength_2016}
\bibinfo{author}{Gong Z.}, \bibinfo{author}{Zhang G.}, \bibinfo{author}{Zeng X.}, \bibinfo{author}{Li~J.}, \bibinfo{author}{Li~G.}, \bibinfo{author}{Huang W.}, \bibinfo{author}{Sun R.}, \bibinfo{author}{Wong C.},
\newblock \bibinfo{title}{High-{Strength}, {Tough}, {Fatigue} {Resistant}, and {Self}-{Healing} {Hydrogel} {Based} on {Dual} {Physically} {Cross}-{Linked} {Network}},
\newblock \bibinfo{journal}{ACS Applied Materials \& Interfaces} \bibinfo{volume}{8} (\bibinfo{year}{2016}) \bibinfo{pages}{24030--24037}. \DOIprefix\doi{10.1021/acsami.6b05627}.
\bibitem[{Bai et~al.(2018)Bai, Yang, Morelle, Yang, and Suo}]{bai_fatigue_2018}
\bibinfo{author}{Bai R.}, \bibinfo{author}{Yang J.}, \bibinfo{author}{Morelle X.~P.}, \bibinfo{author}{Yang C.}, \bibinfo{author}{Suo Z.},
\newblock \bibinfo{title}{Fatigue {Fracture} of {Self}-{Recovery} {Hydrogels}},
\newblock \bibinfo{journal}{ACS Macro Letters} \bibinfo{volume}{7} (\bibinfo{year}{2018}) \bibinfo{pages}{312--317}. \DOIprefix\doi{10.1021/acsmacrolett.8b00045}.
\bibitem[{Li and Gong(2022)}]{li_role_2022}
\bibinfo{author}{Li~X.}, \bibinfo{author}{Gong J.~P.},
\newblock \bibinfo{title}{Role of dynamic bonds on fatigue threshold of tough hydrogels},
\newblock \bibinfo{journal}{Proceedings of the National Academy of Sciences} \bibinfo{volume}{119} (\bibinfo{year}{2022}) \bibinfo{pages}{e2200678119}. \DOIprefix\doi{10.1073/pnas.2200678119}.
\bibitem[{Tuncaboylu et~al.(2011)Tuncaboylu, Sari, Oppermann, and Okay}]{tuncaboylu_tough_2011}
\bibinfo{author}{Tuncaboylu D.~C.}, \bibinfo{author}{Sari M.}, \bibinfo{author}{Oppermann W.}, \bibinfo{author}{Okay O.},
\newblock \bibinfo{title}{Tough and {Self}-{Healing} {Hydrogels} {Formed} via {Hydrophobic} {Interactions}},
\newblock \bibinfo{journal}{Macromolecules} \bibinfo{volume}{44} (\bibinfo{year}{2011}) \bibinfo{pages}{4997--5005}. \DOIprefix\doi{10.1021/ma200579v}.
\bibitem[{Jeon et~al.(2016)Jeon, Cui, Illeperuma, Aizenberg, and Vlassak}]{jeon_extremely_2016}
\bibinfo{author}{Jeon I.}, \bibinfo{author}{Cui J.}, \bibinfo{author}{Illeperuma W.~R.~K.}, \bibinfo{author}{Aizenberg J.}, \bibinfo{author}{Vlassak J.~J.},
\newblock \bibinfo{title}{Extremely {Stretchable} and {Fast} {Self}-{Healing} {Hydrogels}},
\newblock \bibinfo{journal}{Advanced Materials} \bibinfo{volume}{28} (\bibinfo{year}{2016}) \bibinfo{pages}{4678--4683}. \DOIprefix\doi{10.1002/adma.201600480}.
\bibitem[{Zhang et~al.(2019)Zhang, Wu, Yang, Wang, Yu, Lai, Shi, Wang, Cui, Xiang, Zhao, and Xu}]{zhang_superstretchable_2019}
\bibinfo{author}{Zhang H.}, \bibinfo{author}{Wu~Y.}, \bibinfo{author}{Yang J.}, \bibinfo{author}{Wang D.}, \bibinfo{author}{Yu~P.}, \bibinfo{author}{Lai C.~T.}, \bibinfo{author}{Shi A.-C.}, \bibinfo{author}{Wang J.}, \bibinfo{author}{Cui S.}, \bibinfo{author}{Xiang J.}, \bibinfo{author}{Zhao N.}, \bibinfo{author}{Xu~J.},
\newblock \bibinfo{title}{Superstretchable {Dynamic} {Polymer} {Networks}},
\newblock \bibinfo{journal}{Advanced Materials} \bibinfo{volume}{31} (\bibinfo{year}{2019}) \bibinfo{pages}{1904029}. \DOIprefix\doi{10.1002/adma.201904029}.
\bibitem[{Cai et~al.(2022)Cai, Wang, Utomo, Vidavsky, and Silberstein}]{cai_highly_2022}
\bibinfo{author}{Cai H.}, \bibinfo{author}{Wang Z.}, \bibinfo{author}{Utomo N.~W.}, \bibinfo{author}{Vidavsky Y.}, \bibinfo{author}{Silberstein M.~N.},
\newblock \bibinfo{title}{Highly stretchable ionically crosslinked acrylate elastomers inspired by polyelectrolyte complexes},
\newblock \bibinfo{journal}{Soft Matter} \bibinfo{volume}{18} (\bibinfo{year}{2022}) \bibinfo{pages}{7679--7688}. \DOIprefix\doi{10.1039/D2SM00755J}.
\bibitem[{Kersey et~al.(2006)Kersey, Loveless, and Craig}]{kersey_hybrid_2006}
\bibinfo{author}{Kersey F.~R.}, \bibinfo{author}{Loveless D.~M.}, \bibinfo{author}{Craig S.~L.},
\newblock \bibinfo{title}{A hybrid polymer gel with controlled rates of cross-link rupture and self-repair},
\newblock \bibinfo{journal}{Journal of The Royal Society Interface} \bibinfo{volume}{4} (\bibinfo{year}{2006}) \bibinfo{pages}{373--380}. \DOIprefix\doi{10.1098/rsif.2006.0187}.
\bibitem[{Brochu et~al.(2011)Brochu, Craig, and Reichert}]{brochu_self-healing_2011}
\bibinfo{author}{Brochu A.~B.~W.}, \bibinfo{author}{Craig S.~L.}, \bibinfo{author}{Reichert W.~M.},
\newblock \bibinfo{title}{Self-healing biomaterials},
\newblock \bibinfo{journal}{Journal of Biomedical Materials Research Part A} \bibinfo{volume}{96A} (\bibinfo{year}{2011}) \bibinfo{pages}{492--506}. \DOIprefix\doi{10.1002/jbm.a.32987}.
\bibitem[{Li et~al.(2016)Li, Wang, Keplinger, Zuo, Jin, Sun, Zheng, Cao, Lissel, Linder, You, and Bao}]{li_highly_2016}
\bibinfo{author}{Li~C.-H.}, \bibinfo{author}{Wang C.}, \bibinfo{author}{Keplinger C.}, \bibinfo{author}{Zuo J.-L.}, \bibinfo{author}{Jin L.}, \bibinfo{author}{Sun Y.}, \bibinfo{author}{Zheng P.}, \bibinfo{author}{Cao Y.}, \bibinfo{author}{Lissel F.}, \bibinfo{author}{Linder C.}, \bibinfo{author}{You X.-Z.}, \bibinfo{author}{Bao Z.},
\newblock \bibinfo{title}{A highly stretchable autonomous self-healing elastomer},
\newblock \bibinfo{journal}{Nature Chemistry} \bibinfo{volume}{8} (\bibinfo{year}{2016}) \bibinfo{pages}{618--624}. \DOIprefix\doi{10.1038/nchem.2492}.
\bibitem[{Li et~al.(2021)Li, Yu, and Guo}]{li_shape-memory_2021}
\bibinfo{author}{Li~Z.}, \bibinfo{author}{Yu~R.}, \bibinfo{author}{Guo B.},
\newblock \bibinfo{title}{Shape-{Memory} and {Self}-{Healing} {Polymers} {Based} on {Dynamic} {Covalent} {Bonds} and {Dynamic} {Noncovalent} {Interactions}: {Synthesis}, {Mechanism}, and {Application}},
\newblock \bibinfo{journal}{ACS Applied Bio Materials} \bibinfo{volume}{4} (\bibinfo{year}{2021}) \bibinfo{pages}{5926--5943}. \DOIprefix\doi{10.1021/acsabm.1c00606}.
\bibitem[{Wojtecki et~al.(2011)Wojtecki, Meador, and Rowan}]{wojtecki_using_2011}
\bibinfo{author}{Wojtecki R.~J.}, \bibinfo{author}{Meador M.~A.}, \bibinfo{author}{Rowan S.~J.},
\newblock \bibinfo{title}{Using the dynamic bond to access macroscopically responsive structurally dynamic polymers},
\newblock \bibinfo{journal}{Nature Materials} \bibinfo{volume}{10} (\bibinfo{year}{2011}) \bibinfo{pages}{14--27}. \DOIprefix\doi{10.1038/nmat2891}.
\bibitem[{Eom et~al.(2021)Eom, Kim, Lee, Jeon, Park, Lee, Hwang, Park, and Oh}]{eom_mechano-responsive_2021}
\bibinfo{author}{Eom Y.}, \bibinfo{author}{Kim S.-M.}, \bibinfo{author}{Lee M.}, \bibinfo{author}{Jeon H.}, \bibinfo{author}{Park J.}, \bibinfo{author}{Lee E.~S.}, \bibinfo{author}{Hwang S.~Y.}, \bibinfo{author}{Park J.}, \bibinfo{author}{Oh~D.~X.},
\newblock \bibinfo{title}{Mechano-responsive hydrogen-bonding array of thermoplastic polyurethane elastomer captures both strength and self-healing},
\newblock \bibinfo{journal}{Nature Communications} \bibinfo{volume}{12} (\bibinfo{year}{2021}) \bibinfo{pages}{621}. \DOIprefix\doi{10.1038/s41467-021-20931-z}.
\bibitem[{Doolan et~al.(2023)Doolan, Alesbrook, Baker, Brown, Williams, Hilton, Tabata, Wozniakiewicz, Hiscock, and Goult}]{doolan_next-generation_2023}
\bibinfo{author}{Doolan J.~A.}, \bibinfo{author}{Alesbrook L.~S.}, \bibinfo{author}{Baker K.}, \bibinfo{author}{Brown I.~R.}, \bibinfo{author}{Williams G.~T.}, \bibinfo{author}{Hilton K.~L.~F.}, \bibinfo{author}{Tabata M.}, \bibinfo{author}{Wozniakiewicz P.~J.}, \bibinfo{author}{Hiscock J.~R.}, \bibinfo{author}{Goult B.~T.},
\newblock \bibinfo{title}{Next-generation protein-based materials capture and preserve projectiles from supersonic impacts},
\newblock \bibinfo{journal}{Nature Nanotechnology}  (\bibinfo{year}{2023}) \bibinfo{pages}{1--7}. \DOIprefix\doi{10.1038/s41565-023-01431-1}.
\bibitem[{Ge and Rubinstein(2015)}]{ge_strong_2015}
\bibinfo{author}{Ge~T.}, \bibinfo{author}{Rubinstein M.},
\newblock \bibinfo{title}{Strong {Selective} {Adsorption} of {Polymers}},
\newblock \bibinfo{journal}{Macromolecules} \bibinfo{volume}{48} (\bibinfo{year}{2015}) \bibinfo{pages}{3788--3801}. \DOIprefix\doi{10.1021/acs.macromol.5b00586}.
\bibitem[{Yount et~al.(2005)Yount, Loveless, and Craig}]{yount_small-molecule_2005}
\bibinfo{author}{Yount W.~C.}, \bibinfo{author}{Loveless D.~M.}, \bibinfo{author}{Craig S.~L.},
\newblock \bibinfo{title}{Small-{Molecule} {Dynamics} and {Mechanisms} {Underlying} the {Macroscopic} {Mechanical} {Properties} of {Coordinatively} {Cross}-{Linked} {Polymer} {Networks}},
\newblock \bibinfo{journal}{Journal of the American Chemical Society} \bibinfo{volume}{127} (\bibinfo{year}{2005}) \bibinfo{pages}{14488--14496}. \DOIprefix\doi{10.1021/ja054298a}.
\bibitem[{Zhang et~al.(2020)Zhang, Vidavsky, Aharonovich, Yang, Buche, Diesendruck, and Silberstein}]{zhang_bridging_2020}
\bibinfo{author}{Zhang X.}, \bibinfo{author}{Vidavsky Y.}, \bibinfo{author}{Aharonovich S.}, \bibinfo{author}{Yang S.~J.}, \bibinfo{author}{Buche M.~R.}, \bibinfo{author}{Diesendruck C.~E.}, \bibinfo{author}{Silberstein M.~N.},
\newblock \bibinfo{title}{Bridging experiments and theory: isolating the effects of metal–ligand interactions on viscoelasticity of reversible polymer networks},
\newblock \bibinfo{journal}{Soft Matter} \bibinfo{volume}{16} (\bibinfo{year}{2020}) \bibinfo{pages}{8591--8601}. \DOIprefix\doi{10.1039/D0SM01115K}.
\bibitem[{Huang et~al.(2022)Huang, Jayathilaka, Islam, Tanaka, Silberstein, Kilian, and Kruzic}]{huang_structural_2022}
\bibinfo{author}{Huang Y.}, \bibinfo{author}{Jayathilaka P.~B.}, \bibinfo{author}{Islam M.~S.}, \bibinfo{author}{Tanaka C.~B.}, \bibinfo{author}{Silberstein M.~N.}, \bibinfo{author}{Kilian K.~A.}, \bibinfo{author}{Kruzic J.~J.},
\newblock \bibinfo{title}{Structural aspects controlling the mechanical and biological properties of tough, double network hydrogels},
\newblock \bibinfo{journal}{Acta Biomaterialia} \bibinfo{volume}{138} (\bibinfo{year}{2022}) \bibinfo{pages}{301--312}. \DOIprefix\doi{10.1016/j.actbio.2021.10.044}.
\bibitem[{James and Guth(1943)}]{james_theory_1943}
\bibinfo{author}{James H.~M.}, \bibinfo{author}{Guth E.},
\newblock \bibinfo{title}{Theory of the {Elastic} {Properties} of {Rubber}},
\newblock \bibinfo{journal}{The Journal of Chemical Physics} \bibinfo{volume}{11} (\bibinfo{year}{1943}) \bibinfo{pages}{455--481}. \DOIprefix\doi{10.1063/1.1723785}.
\bibitem[{Flory(1985)}]{flory_molecular_1985}
\bibinfo{author}{Flory P.~J.},
\newblock \bibinfo{title}{Molecular {Theory} of {Rubber} {Elasticity}},
\newblock \bibinfo{journal}{Polymer Journal} \bibinfo{volume}{17} (\bibinfo{year}{1985}) \bibinfo{pages}{1--12}. \DOIprefix\doi{10.1295/polymj.17.1}.
\bibitem[{Tanaka and Edwards(1992)}]{tanaka_viscoelastic_1992}
\bibinfo{author}{Tanaka F.}, \bibinfo{author}{Edwards S.~F.},
\newblock \bibinfo{title}{Viscoelastic properties of physically crosslinked networks. 1. {Transient} network theory},
\newblock \bibinfo{journal}{Macromolecules} \bibinfo{volume}{25} (\bibinfo{year}{1992}) \bibinfo{pages}{1516--1523}. \DOIprefix\doi{10.1021/ma00031a024}.
\bibitem[{Arruda and Boyce(1993)}]{arruda_three-dimensional_1993}
\bibinfo{author}{Arruda E.~M.}, \bibinfo{author}{Boyce M.~C.},
\newblock \bibinfo{title}{A three-dimensional constitutive model for the large stretch behavior of rubber elastic materials},
\newblock \bibinfo{journal}{Journal of the Mechanics and Physics of Solids} \bibinfo{volume}{41} (\bibinfo{year}{1993}) \bibinfo{pages}{389--412}. \DOIprefix\doi{10.1016/0022-5096(93)90013-6}.
\bibitem[{Wu and Van Der~Giessen(1993)}]{wu_improved_1993}
\bibinfo{author}{Wu~P.~D.}, \bibinfo{author}{Van Der~Giessen E.},
\newblock \bibinfo{title}{On improved network models for rubber elasticity and their applications to orientation hardening in glassy polymers},
\newblock \bibinfo{journal}{Journal of the Mechanics and Physics of Solids} \bibinfo{volume}{41} (\bibinfo{year}{1993}) \bibinfo{pages}{427--456}. \DOIprefix\doi{10.1016/0022-5096(93)90043-F}.
\bibitem[{Miehe et~al.(2004)Miehe, Göktepe, and Lulei}]{miehe_micro-macro_2004}
\bibinfo{author}{Miehe C.}, \bibinfo{author}{Göktepe S.}, \bibinfo{author}{Lulei F.},
\newblock \bibinfo{title}{A micro-macro approach to rubber-like materials—{Part} {I}: the non-affine micro-sphere model of rubber elasticity},
\newblock \bibinfo{journal}{Journal of the Mechanics and Physics of Solids} \bibinfo{volume}{52} (\bibinfo{year}{2004}) \bibinfo{pages}{2617--2660}. \DOIprefix\doi{10.1016/j.jmps.2004.03.011}.
\bibitem[{Vernerey et~al.(2017)Vernerey, Long, and Brighenti}]{vernerey_statistically-based_2017}
\bibinfo{author}{Vernerey F.~J.}, \bibinfo{author}{Long R.}, \bibinfo{author}{Brighenti R.},
\newblock \bibinfo{title}{A statistically-based continuum theory for polymers with transient networks},
\newblock \bibinfo{journal}{Journal of the Mechanics and Physics of Solids} \bibinfo{volume}{107} (\bibinfo{year}{2017}) \bibinfo{pages}{1--20}. \DOIprefix\doi{10.1016/j.jmps.2017.05.016}.
\bibitem[{Saleh(2015)}]{saleh_perspective_2015}
\bibinfo{author}{Saleh O.~A.},
\newblock \bibinfo{title}{Perspective: {Single} polymer mechanics across the force regimes},
\newblock \bibinfo{journal}{The Journal of Chemical Physics} \bibinfo{volume}{142} (\bibinfo{year}{2015}) \bibinfo{pages}{194902}. \DOIprefix\doi{10.1063/1.4921348}.
\bibitem[{Flory(1942)}]{flory_thermodynamics_1942}
\bibinfo{author}{Flory P.~J.},
\newblock \bibinfo{title}{Thermodynamics of {High} {Polymer} {Solutions}},
\newblock \bibinfo{journal}{The Journal of Chemical Physics} \bibinfo{volume}{10} (\bibinfo{year}{1942}) \bibinfo{pages}{51--61}. \DOIprefix\doi{10.1063/1.1723621}.
\bibitem[{Hong et~al.(2009)Hong, Liu, and Suo}]{hong_inhomogeneous_2009}
\bibinfo{author}{Hong W.}, \bibinfo{author}{Liu Z.}, \bibinfo{author}{Suo Z.},
\newblock \bibinfo{title}{Inhomogeneous swelling of a gel in equilibrium with a solvent and mechanical load},
\newblock \bibinfo{journal}{International Journal of Solids and Structures} \bibinfo{volume}{46} (\bibinfo{year}{2009}) \bibinfo{pages}{3282--3289}. \DOIprefix\doi{10.1016/j.ijsolstr.2009.04.022}.
\bibitem[{Bouklas and Huang(2012)}]{bouklas_swelling_2012}
\bibinfo{author}{Bouklas N.}, \bibinfo{author}{Huang R.},
\newblock \bibinfo{title}{Swelling kinetics of polymer gels: comparison of linear and nonlinear theories},
\newblock \bibinfo{journal}{Soft Matter} \bibinfo{volume}{8} (\bibinfo{year}{2012}) \bibinfo{pages}{8194--8203}. \DOIprefix\doi{10.1039/C2SM25467K}.
\bibitem[{Leibler et~al.(1991)Leibler, Rubinstein, and Colby}]{leibler_dynamics_1991}
\bibinfo{author}{Leibler L.}, \bibinfo{author}{Rubinstein M.}, \bibinfo{author}{Colby R.~H.},
\newblock \bibinfo{title}{Dynamics of reversible networks},
\newblock \bibinfo{journal}{Macromolecules} \bibinfo{volume}{24} (\bibinfo{year}{1991}) \bibinfo{pages}{4701--4707}. \DOIprefix\doi{10.1021/ma00016a034}.
\bibitem[{Stukalin et~al.(2013)Stukalin, Cai, Kumar, Leibler, and Rubinstein}]{stukalin_self-healing_2013}
\bibinfo{author}{Stukalin E.~B.}, \bibinfo{author}{Cai L.-H.}, \bibinfo{author}{Kumar N.~A.}, \bibinfo{author}{Leibler L.}, \bibinfo{author}{Rubinstein M.},
\newblock \bibinfo{title}{Self-{Healing} of {Unentangled} {Polymer} {Networks} with {Reversible} {Bonds}},
\newblock \bibinfo{journal}{Macromolecules} \bibinfo{volume}{46} (\bibinfo{year}{2013}) \bibinfo{pages}{7525--7541}. \DOIprefix\doi{10.1021/ma401111n}.
\bibitem[{Bouklas et~al.(2015)Bouklas, Landis, and Huang}]{bouklas_effect_2015}
\bibinfo{author}{Bouklas N.}, \bibinfo{author}{Landis C.~M.}, \bibinfo{author}{Huang R.},
\newblock \bibinfo{title}{Effect of {Solvent} {Diffusion} on {Crack}-{Tip} {Fields} and {Driving} {Force} for {Fracture} of {Hydrogels}},
\newblock \bibinfo{journal}{Journal of Applied Mechanics} \bibinfo{volume}{82} (\bibinfo{year}{2015}). \DOIprefix\doi{10.1115/1.4030587}.
\bibitem[{Shen and Vernerey(2020)}]{shen_rate-dependent_2020}
\bibinfo{author}{Shen T.}, \bibinfo{author}{Vernerey F.~J.},
\newblock \bibinfo{title}{Rate-dependent fracture of transient networks},
\newblock \bibinfo{journal}{Journal of the Mechanics and Physics of Solids} \bibinfo{volume}{143} (\bibinfo{year}{2020}) \bibinfo{pages}{104028}. \DOIprefix\doi{10.1016/j.jmps.2020.104028}.
\bibitem[{Lee et~al.(2023)Lee, Lee, Chester, and Cho}]{lee_finite_2023}
\bibinfo{author}{Lee J.}, \bibinfo{author}{Lee S.}, \bibinfo{author}{Chester S.~A.}, \bibinfo{author}{Cho H.},
\newblock \bibinfo{title}{Finite element implementation of a gradient-damage theory for fracture in elastomeric materials},
\newblock \bibinfo{journal}{International Journal of Solids and Structures} \bibinfo{volume}{279} (\bibinfo{year}{2023}) \bibinfo{pages}{112309}. \DOIprefix\doi{10.1016/j.ijsolstr.2023.112309}.
\bibitem[{Bosnjak et~al.(2022)Bosnjak, Tepermeister, and Silberstein}]{bosnjak_modeling_2022}
\bibinfo{author}{Bosnjak N.}, \bibinfo{author}{Tepermeister M.}, \bibinfo{author}{Silberstein M.~N.},
\newblock \bibinfo{title}{Modeling coupled electrochemical and mechanical behavior of soft ionic materials and ionotronic devices},
\newblock \bibinfo{journal}{Journal of the Mechanics and Physics of Solids} \bibinfo{volume}{168} (\bibinfo{year}{2022}) \bibinfo{pages}{105014}. \DOIprefix\doi{10.1016/j.jmps.2022.105014}.
\bibitem[{C.~Picu(2011)}]{cpicu_mechanics_2011}
\bibinfo{author}{C.~Picu R.},
\newblock \bibinfo{title}{Mechanics of random fiber networks—a review},
\newblock \bibinfo{journal}{Soft Matter} \bibinfo{volume}{7} (\bibinfo{year}{2011}) \bibinfo{pages}{6768--6785}. \DOIprefix\doi{10.1039/C1SM05022B}.
\bibitem[{Bergström and Boyce(2001)}]{bergstrom_deformation_2001}
\bibinfo{author}{Bergström J.~S.}, \bibinfo{author}{Boyce M.~C.},
\newblock \bibinfo{title}{Deformation of {Elastomeric} {Networks}: {Relation} between {Molecular} {Level} {Deformation} and {Classical} {Statistical} {Mechanics} {Models} of {Rubber} {Elasticity}},
\newblock \bibinfo{journal}{Macromolecules} \bibinfo{volume}{34} (\bibinfo{year}{2001}) \bibinfo{pages}{614--626}. \DOIprefix\doi{10.1021/ma0007942}.
\bibitem[{Somasi et~al.(2002)Somasi, Khomami, Woo, Hur, and Shaqfeh}]{somasi_brownian_2002}
\bibinfo{author}{Somasi M.}, \bibinfo{author}{Khomami B.}, \bibinfo{author}{Woo N.~J.}, \bibinfo{author}{Hur J.~S.}, \bibinfo{author}{Shaqfeh E.~S.~G.},
\newblock \bibinfo{title}{Brownian dynamics simulations of bead-rod and bead-spring chains: numerical algorithms and coarse-graining issues},
\newblock \bibinfo{journal}{Journal of Non-Newtonian Fluid Mechanics} \bibinfo{volume}{108} (\bibinfo{year}{2002}) \bibinfo{pages}{227--255}. \DOIprefix\doi{10.1016/S0377-0257(02)00132-5}.
\bibitem[{Doyle and Underhill(2005)}]{doyle_brownian_2005}
\bibinfo{author}{Doyle P.~S.}, \bibinfo{author}{Underhill P.~T.},
\newblock \bibinfo{title}{Brownian {Dynamics} {Simulations} of {Polymers} and {Soft} {Matter}},
\newblock in: \bibinfo{editor}{Yip S.} (Ed.), \bibinfo{booktitle}{Handbook of {Materials} {Modeling}: {Methods}}, \bibinfo{publisher}{Springer Netherlands}, \bibinfo{address}{Dordrecht}, \bibinfo{year}{2005}, pp. \bibinfo{pages}{2619--2630}.
\bibitem[{Thompson et~al.(2022)Thompson, Aktulga, Berger, Bolintineanu, Brown, Crozier, in~'t Veld, Kohlmeyer, Moore, Nguyen, Shan, Stevens, Tranchida, Trott, and Plimpton}]{thompson_lammps_2022}
\bibinfo{author}{Thompson A.~P.}, \bibinfo{author}{Aktulga H.~M.}, \bibinfo{author}{Berger R.}, \bibinfo{author}{Bolintineanu D.~S.}, \bibinfo{author}{Brown W.~M.}, \bibinfo{author}{Crozier P.~S.}, \bibinfo{author}{Veld in~'t P.~J.}, \bibinfo{author}{Kohlmeyer A.}, \bibinfo{author}{Moore S.~G.}, \bibinfo{author}{Nguyen T.~D.}, \bibinfo{author}{Shan R.}, \bibinfo{author}{Stevens M.~J.}, \bibinfo{author}{Tranchida J.}, \bibinfo{author}{Trott C.}, \bibinfo{author}{Plimpton S.~J.},
\newblock \bibinfo{title}{{LAMMPS} - a flexible simulation tool for particle-based materials modeling at the atomic, meso, and continuum scales},
\newblock \bibinfo{journal}{Computer Physics Communications} \bibinfo{volume}{271} (\bibinfo{year}{2022}) \bibinfo{pages}{108171}. \DOIprefix\doi{10.1016/j.cpc.2021.108171}.
\bibitem[{Goodrich et~al.(2018)Goodrich, Brenner, and Ribbeck}]{goodrich_enhanced_2018}
\bibinfo{author}{Goodrich C.~P.}, \bibinfo{author}{Brenner M.~P.}, \bibinfo{author}{Ribbeck K.},
\newblock \bibinfo{title}{Enhanced diffusion by binding to the crosslinks of a polymer gel},
\newblock \bibinfo{journal}{Nature Communications} \bibinfo{volume}{9} (\bibinfo{year}{2018}) \bibinfo{pages}{4348}. \DOIprefix\doi{10.1038/s41467-018-06851-5}.
\bibitem[{Evans and Ritchie(1997)}]{evans_dynamic_1997}
\bibinfo{author}{Evans E.}, \bibinfo{author}{Ritchie K.},
\newblock \bibinfo{title}{Dynamic strength of molecular adhesion bonds},
\newblock \bibinfo{journal}{Biophysical Journal} \bibinfo{volume}{72} (\bibinfo{year}{1997}) \bibinfo{pages}{1541--1555}. \DOIprefix\doi{10.1016/S0006-3495(97)78802-7}.
\bibitem[{Hoy and Fredrickson(2009)}]{hoy_thermoreversible_2009}
\bibinfo{author}{Hoy R.~S.}, \bibinfo{author}{Fredrickson G.~H.},
\newblock \bibinfo{title}{Thermoreversible associating polymer networks. {I}. {Interplay} of thermodynamics, chemical kinetics, and polymer physics},
\newblock \bibinfo{journal}{The Journal of Chemical Physics} \bibinfo{volume}{131} (\bibinfo{year}{2009}) \bibinfo{pages}{224902}. \DOIprefix\doi{10.1063/1.3268777}.
\bibitem[{Amin et~al.(2016)Amin, Likhtman, and Wang}]{amin_dynamics_2016}
\bibinfo{author}{Amin D.}, \bibinfo{author}{Likhtman A.~E.}, \bibinfo{author}{Wang Z.},
\newblock \bibinfo{title}{Dynamics in {Supramolecular} {Polymer} {Networks} {Formed} by {Associating} {Telechelic} {Chains}},
\newblock \bibinfo{journal}{Macromolecules} \bibinfo{volume}{49} (\bibinfo{year}{2016}) \bibinfo{pages}{7510--7524}. \DOIprefix\doi{10.1021/acs.macromol.6b00561}.
\bibitem[{Perego and Khabaz(2020)}]{perego_volumetric_2020}
\bibinfo{author}{Perego A.}, \bibinfo{author}{Khabaz F.},
\newblock \bibinfo{title}{Volumetric and {Rheological} {Properties} of {Vitrimers}: {A} {Hybrid} {Molecular} {Dynamics} and {Monte} {Carlo} {Simulation} {Study}},
\newblock \bibinfo{journal}{Macromolecules} \bibinfo{volume}{53} (\bibinfo{year}{2020}) \bibinfo{pages}{8406--8416}. \DOIprefix\doi{10.1021/acs.macromol.0c01423}.
\bibitem[{Zhao et~al.(2022)Zhao, Wei, Fang, Gao, Yue, Zhang, Ganesan, Meng, and Liu}]{zhao_molecular_2022}
\bibinfo{author}{Zhao H.}, \bibinfo{author}{Wei X.}, \bibinfo{author}{Fang Y.}, \bibinfo{author}{Gao K.}, \bibinfo{author}{Yue T.}, \bibinfo{author}{Zhang L.}, \bibinfo{author}{Ganesan V.}, \bibinfo{author}{Meng F.}, \bibinfo{author}{Liu J.},
\newblock \bibinfo{title}{Molecular {Dynamics} {Simulation} of the {Structural}, {Mechanical}, and {Reprocessing} {Properties} of {Vitrimers} {Based} on a {Dynamic} {Covalent} {Polymer} {Network}},
\newblock \bibinfo{journal}{Macromolecules} \bibinfo{volume}{55} (\bibinfo{year}{2022}) \bibinfo{pages}{1091--1103}. \DOIprefix\doi{10.1021/acs.macromol.1c02034}.
\bibitem[{Wagner et~al.(2024)Wagner, Lamont, White, and Vernerey}]{wagner_catch_2024-1}
\bibinfo{author}{Wagner R.~J.}, \bibinfo{author}{Lamont S.~C.}, \bibinfo{author}{White Z.~T.}, \bibinfo{author}{Vernerey F.~J.},
\newblock \bibinfo{title}{Catch bond kinetics are instrumental to cohesion of fire ant rafts under load},
\newblock \bibinfo{journal}{Proceedings of the National Academy of Sciences} \bibinfo{volume}{121} (\bibinfo{year}{2024}) \bibinfo{pages}{e2314772121}. \DOIprefix\doi{10.1073/pnas.2314772121}.
\bibitem[{Huang et~al.(2023)Huang, Ramlawi, Sheridan, Chen, Ewoldt, Braun, and Evans}]{huang_dynamic_2023}
\bibinfo{author}{Huang J.}, \bibinfo{author}{Ramlawi N.}, \bibinfo{author}{Sheridan G.~S.}, \bibinfo{author}{Chen C.}, \bibinfo{author}{Ewoldt R.~H.}, \bibinfo{author}{Braun P.~V.}, \bibinfo{author}{Evans C.~M.},
\newblock \bibinfo{title}{Dynamic {Covalent} {Bond} {Exchange} {Enhances} {Penetrant} {Diffusion} in {Dense} {Vitrimers}},
\newblock \bibinfo{journal}{Macromolecules} \bibinfo{volume}{56} (\bibinfo{year}{2023}) \bibinfo{pages}{1253--1262}. \DOIprefix\doi{10.1021/acs.macromol.2c02547}.
\bibitem[{Taylor et~al.(2024)Taylor, Wang, Ge, O’Connor, and Grest}]{taylor_smoother_2024}
\bibinfo{author}{Taylor P.~A.}, \bibinfo{author}{Wang J.}, \bibinfo{author}{Ge~T.}, \bibinfo{author}{O’Connor T.~C.}, \bibinfo{author}{Grest G.~S.},
\newblock \bibinfo{title}{Smoother {Surfaces} {Enhance} {Diffusion} of {Nanorods} in {Entangled} {Polymer} {Melts}},
\newblock \bibinfo{journal}{Macromolecules} \bibinfo{volume}{57} (\bibinfo{year}{2024}) \bibinfo{pages}{2482--2489}. \DOIprefix\doi{10.1021/acs.macromol.3c01826}.
\bibitem[{Yang et~al.(2015)Yang, Yu, Mu, Shi, Wei, Guo, and Jerry Qi}]{yang_molecular_2015}
\bibinfo{author}{Yang H.}, \bibinfo{author}{Yu~K.}, \bibinfo{author}{Mu~X.}, \bibinfo{author}{Shi X.}, \bibinfo{author}{Wei Y.}, \bibinfo{author}{Guo Y.}, \bibinfo{author}{Jerry Qi H.},
\newblock \bibinfo{title}{A molecular dynamics study of bond exchange reactions in covalent adaptable networks},
\newblock \bibinfo{journal}{Soft Matter} \bibinfo{volume}{11} (\bibinfo{year}{2015}) \bibinfo{pages}{6305--6317}. \DOIprefix\doi{10.1039/C5SM00942A}.
\bibitem[{Amin and Wang(2020)}]{amin_nonlinear_2020}
\bibinfo{author}{Amin D.}, \bibinfo{author}{Wang Z.},
\newblock \bibinfo{title}{Nonlinear rheology and dynamics of supramolecular polymer networks formed by associative telechelic chains under shear and extensional flows},
\newblock \bibinfo{journal}{Journal of Rheology} \bibinfo{volume}{64} (\bibinfo{year}{2020}) \bibinfo{pages}{581--600}. \DOIprefix\doi{10.1122/1.5120897}.
\bibitem[{Zheng et~al.(2021)Zheng, Yang, Sun, Zhang, and Guo}]{zheng_molecular_2021}
\bibinfo{author}{Zheng X.}, \bibinfo{author}{Yang H.}, \bibinfo{author}{Sun Y.}, \bibinfo{author}{Zhang Y.}, \bibinfo{author}{Guo Y.},
\newblock \bibinfo{title}{A molecular dynamics simulation on self-healing behavior based on disulfide bond exchange reactions},
\newblock \bibinfo{journal}{Polymer} \bibinfo{volume}{212} (\bibinfo{year}{2021}) \bibinfo{pages}{123111}. \DOIprefix\doi{10.1016/j.polymer.2020.123111}.
\bibitem[{Agrawal et~al.(2016)Agrawal, Holzworth, Nantasetphong, Amirkhizi, Oswald, and Nemat-Nasser}]{agrawal_prediction_2016}
\bibinfo{author}{Agrawal V.}, \bibinfo{author}{Holzworth K.}, \bibinfo{author}{Nantasetphong W.}, \bibinfo{author}{Amirkhizi A.~V.}, \bibinfo{author}{Oswald J.}, \bibinfo{author}{Nemat-Nasser S.},
\newblock \bibinfo{title}{Prediction of viscoelastic properties with coarse-grained molecular dynamics and experimental validation for a benchmark polyurea system},
\newblock \bibinfo{journal}{Journal of Polymer Science Part B: Polymer Physics} \bibinfo{volume}{54} (\bibinfo{year}{2016}) \bibinfo{pages}{797--810}. \DOIprefix\doi{10.1002/polb.23976}.
\bibitem[{Liu et~al.(2020)Liu, Fu, Shao, Cai, and Chipot}]{liu_accurate_2020}
\bibinfo{author}{Liu H.}, \bibinfo{author}{Fu~H.}, \bibinfo{author}{Shao X.}, \bibinfo{author}{Cai W.}, \bibinfo{author}{Chipot C.},
\newblock \bibinfo{title}{Accurate {Description} of {Cation}--$\pi$ {Interactions} in {Proteins} with a {Nonpolarizable} {Force} {Field} at {No} {Additional} {Cost}},
\newblock \bibinfo{journal}{Journal of Chemical Theory and Computation} \bibinfo{volume}{16} (\bibinfo{year}{2020}) \bibinfo{pages}{6397--6407}. \DOIprefix\doi{10.1021/acs.jctc.0c00637}.
\bibitem[{Zhang et~al.(2023)Zhang, Dai, Tepermeister, Deng, Yeo, and Silberstein}]{zhang_understanding_2023}
\bibinfo{author}{Zhang X.}, \bibinfo{author}{Dai J.}, \bibinfo{author}{Tepermeister M.}, \bibinfo{author}{Deng Y.}, \bibinfo{author}{Yeo J.}, \bibinfo{author}{Silberstein M.~N.},
\newblock \bibinfo{title}{Understanding {How} {Metal}–{Ligand} {Coordination} {Enables} {Solvent} {Free} {Ionic} {Conductivity} in {PDMS}},
\newblock \bibinfo{journal}{Macromolecules} \bibinfo{volume}{56} (\bibinfo{year}{2023}) \bibinfo{pages}{3119--3131}. \DOIprefix\doi{10.1021/acs.macromol.2c02519}.
\bibitem[{Kremer and Grest(1990)}]{kremer_dynamics_1990}
\bibinfo{author}{Kremer K.}, \bibinfo{author}{Grest G.~S.},
\newblock \bibinfo{title}{Dynamics of entangled linear polymer melts: {A} molecular‐dynamics simulation},
\newblock \bibinfo{journal}{The Journal of Chemical Physics} \bibinfo{volume}{92} (\bibinfo{year}{1990}) \bibinfo{pages}{5057--5086}. \DOIprefix\doi{10.1063/1.458541}.
\bibitem[{Hernández~Cifre et~al.(2003)Hernández~Cifre, Barenbrug, Schieber, and van~den Brule}]{hernandez_cifre_brownian_2003}
\bibinfo{author}{Hernández~Cifre J.~G.}, \bibinfo{author}{Barenbrug T.~M. A. O.~M.}, \bibinfo{author}{Schieber J.~D.}, \bibinfo{author}{Brule van~den B.~H. A.~A.},
\newblock \bibinfo{title}{Brownian dynamics simulation of reversible polymer networks under shear using a non-interacting dumbbell model},
\newblock \bibinfo{journal}{Journal of Non-Newtonian Fluid Mechanics} \bibinfo{volume}{113} (\bibinfo{year}{2003}) \bibinfo{pages}{73--96}. \DOIprefix\doi{10.1016/S0377-0257(03)00063-6}.
\bibitem[{Sugimura et~al.(2013)Sugimura, Asai, Matsunaga, Akagi, Sakai, Noguchi, and Shibayama}]{sugimura_mechanical_2013}
\bibinfo{author}{Sugimura A.}, \bibinfo{author}{Asai M.}, \bibinfo{author}{Matsunaga T.}, \bibinfo{author}{Akagi Y.}, \bibinfo{author}{Sakai T.}, \bibinfo{author}{Noguchi H.}, \bibinfo{author}{Shibayama M.},
\newblock \bibinfo{title}{Mechanical properties of a polymer network of {Tetra}-{PEG} gel},
\newblock \bibinfo{journal}{Polymer Journal} \bibinfo{volume}{45} (\bibinfo{year}{2013}) \bibinfo{pages}{300--306}. \DOIprefix\doi{10.1038/pj.2012.149}.
\bibitem[{Kothari et~al.(2018)Kothari, Hu, Gupta, and Elbanna}]{kothari_mechanical_2018}
\bibinfo{author}{Kothari K.}, \bibinfo{author}{Hu~Y.}, \bibinfo{author}{Gupta S.}, \bibinfo{author}{Elbanna A.},
\newblock \bibinfo{title}{Mechanical {Response} of {Two}-{Dimensional} {Polymer} {Networks}: {Role} of {Topology}, {Rate} {Dependence}, and {Damage} {Accumulation}},
\newblock \bibinfo{journal}{Journal of Applied Mechanics} \bibinfo{volume}{85} (\bibinfo{year}{2018}). \DOIprefix\doi{10.1115/1.4038883}.
\bibitem[{Wagner et~al.(2021)Wagner, Hobbs, and Vernerey}]{wagner_network_2021}
\bibinfo{author}{Wagner R.~J.}, \bibinfo{author}{Hobbs E.}, \bibinfo{author}{Vernerey F.~J.},
\newblock \bibinfo{title}{A network model of transient polymers: exploring the micromechanics of nonlinear viscoelasticity},
\newblock \bibinfo{journal}{Soft Matter} \bibinfo{volume}{17} (\bibinfo{year}{2021}) \bibinfo{pages}{8742--8757}. \DOIprefix\doi{10.1039/D1SM00753J}.
\bibitem[{Wyse~Jackson et~al.(2022)Wyse~Jackson, Michel, Lwin, Fortier, Das, Bonassar, and Cohen}]{wyse_jackson_structural_2022}
\bibinfo{author}{Wyse~Jackson T.}, \bibinfo{author}{Michel J.}, \bibinfo{author}{Lwin P.}, \bibinfo{author}{Fortier L.~A.}, \bibinfo{author}{Das M.}, \bibinfo{author}{Bonassar L.~J.}, \bibinfo{author}{Cohen I.},
\newblock \bibinfo{title}{Structural origins of cartilage shear mechanics},
\newblock \bibinfo{journal}{Science Advances} \bibinfo{volume}{8} (\bibinfo{year}{2022}) \bibinfo{pages}{eabk2805}. \DOIprefix\doi{10.1126/sciadv.abk2805}.
\bibitem[{Wagner et~al.(2022)Wagner, Dai, Su, and Vernerey}]{wagner_mesoscale_2022}
\bibinfo{author}{Wagner R.~J.}, \bibinfo{author}{Dai J.}, \bibinfo{author}{Su~X.}, \bibinfo{author}{Vernerey F.~J.},
\newblock \bibinfo{title}{A mesoscale model for the micromechanical study of gels},
\newblock \bibinfo{journal}{Journal of the Mechanics and Physics of Solids}  (\bibinfo{year}{2022}) \bibinfo{pages}{104982}. \DOIprefix\doi{10.1016/j.jmps.2022.104982}.
\bibitem[{Wagner and Vernerey(2023)}]{wagner_coupled_2023}
\bibinfo{author}{Wagner R.~J.}, \bibinfo{author}{Vernerey F.~J.},
\newblock \bibinfo{title}{Coupled bond dynamics alters relaxation in polymers with multiple intrinsic dissociation rates},
\newblock \bibinfo{journal}{Soft Matter}  (\bibinfo{year}{2023}). \DOIprefix\doi{10.1039/D3SM00014A}.
\bibitem[{Richardson et~al.(2019)Richardson, Wilcox, Randolph, and Anseth}]{richardson_hydrazone_2019}
\bibinfo{author}{Richardson B.~M.}, \bibinfo{author}{Wilcox D.~G.}, \bibinfo{author}{Randolph M.~A.}, \bibinfo{author}{Anseth K.~S.},
\newblock \bibinfo{title}{Hydrazone covalent adaptable networks modulate extracellular matrix deposition for cartilage tissue engineering},
\newblock \bibinfo{journal}{Acta Biomaterialia} \bibinfo{volume}{83} (\bibinfo{year}{2019}) \bibinfo{pages}{71--82}. \DOIprefix\doi{10.1016/j.actbio.2018.11.014}.
\bibitem[{Cruz et~al.(2012)Cruz, Chinesta, and Régnier}]{cruz_review_2012}
\bibinfo{author}{Cruz C.}, \bibinfo{author}{Chinesta F.}, \bibinfo{author}{Régnier G.},
\newblock \bibinfo{title}{Review on the {Brownian} {Dynamics} {Simulation} of {Bead}-{Rod}-{Spring} {Models} {Encountered} in {Computational} {Rheology}},
\newblock \bibinfo{journal}{Archives of Computational Methods in Engineering} \bibinfo{volume}{19} (\bibinfo{year}{2012}) \bibinfo{pages}{227--259}. \DOIprefix\doi{10.1007/s11831-012-9072-2}.
\bibitem[{Sliozberg and Chantawansri(2013)}]{sliozberg_computational_2013}
\bibinfo{author}{Sliozberg Y.~R.}, \bibinfo{author}{Chantawansri T.~L.},
\newblock \bibinfo{title}{Computational study of imperfect networks using a coarse-grained model},
\newblock \bibinfo{journal}{The Journal of Chemical Physics} \bibinfo{volume}{139} (\bibinfo{year}{2013}) \bibinfo{pages}{194904}. \DOIprefix\doi{10.1063/1.4832140}.
\bibitem[{Einstein(1905)}]{einstein_uber_1905}
\bibinfo{author}{Einstein A.},
\newblock \bibinfo{title}{Über die von der molekularkinetischen {Theorie} der {Wärme} geforderte {Bewegung} von in ruhenden {Flüssigkeiten} suspendierten {Teilchen}},
\newblock \bibinfo{journal}{Annalen der Physik} \bibinfo{volume}{322} (\bibinfo{year}{1905}) \bibinfo{pages}{549--560}. \DOIprefix\doi{10.1002/andp.19053220806}.
\bibitem[{Rubinstein and Colby(2003)}]{rubinstein_polymer_2003}
\bibinfo{author}{Rubinstein M.}, \bibinfo{author}{Colby R.~H.}, \bibinfo{title}{Polymer {Physics}}, \bibinfo{edition}{- hardcover - michael rubinstein; ralph h. colby -} ed., \bibinfo{publisher}{Oxford University Press}, \bibinfo{year}{2003}.
\bibitem[{Doi(2013)}]{doi_soft_2013}
\bibinfo{author}{Doi M.}, \bibinfo{title}{Soft {Matter} {Physics}}, \bibinfo{publisher}{OUP Oxford}, \bibinfo{year}{2013}.
\bibitem[{Cohen(1991)}]{cohen_pade_1991}
\bibinfo{author}{Cohen A.},
\newblock \bibinfo{title}{A {Padé} approximant to the inverse {Langevin} function},
\newblock \bibinfo{journal}{Rheologica Acta} \bibinfo{volume}{30} (\bibinfo{year}{1991}) \bibinfo{pages}{270--273}. \DOIprefix\doi{10.1007/BF00366640}.
\bibitem[{Rector et~al.(1994)Rector, Swol, and Henderson}]{rector_simulation_1994}
\bibinfo{author}{Rector D.~R.}, \bibinfo{author}{Swol F.~V.}, \bibinfo{author}{Henderson J.~R.},
\newblock \bibinfo{title}{Simulation of surfactant solutions},
\newblock \bibinfo{journal}{Molecular Physics}  (\bibinfo{year}{1994}). \DOIprefix\doi{10.1080/00268979400100724}.
\bibitem[{Shimada et~al.(2005)Shimada, Kato, Saito, Matsuyama, and Kinugasa}]{shimada_precise_2005}
\bibinfo{author}{Shimada K.}, \bibinfo{author}{Kato H.}, \bibinfo{author}{Saito T.}, \bibinfo{author}{Matsuyama S.}, \bibinfo{author}{Kinugasa S.},
\newblock \bibinfo{title}{Precise measurement of the self-diffusion coefficient for poly(ethylene glycol) in aqueous solution using uniform oligomers},
\newblock \bibinfo{journal}{The Journal of Chemical Physics} \bibinfo{volume}{122} (\bibinfo{year}{2005}) \bibinfo{pages}{244914}. \DOIprefix\doi{10.1063/1.1948378}.
\bibitem[{Kravanja et~al.(2018)Kravanja, {\v{S}}kerget, Knez, and Knez~Hrn{\v{c}}i{\v{c}}}]{kravanja_diffusion_2018}
\bibinfo{author}{Kravanja G.}, \bibinfo{author}{{\v{S}}kerget M.}, \bibinfo{author}{Knez {\v{Z}}.}, \bibinfo{author}{Knez~Hrn{\v{c}}i{\v{c}} M.},
\newblock \bibinfo{title}{Diffusion coefficients of water and propylene glycol in supercritical {CO2} from pendant drop tensiometry},
\newblock \bibinfo{journal}{The Journal of Supercritical Fluids} \bibinfo{volume}{133} (\bibinfo{year}{2018}) \bibinfo{pages}{1--8}. \DOIprefix\doi{10.1016/j.supflu.2017.09.022}.
\bibitem[{Shi(2021)}]{shi_molecular_2021}
\bibinfo{author}{Shi Q.},
\newblock \bibinfo{title}{Molecular dynamics simulation of diffusion and separation of {CO2}/{CH4}/{N2} on {MER} zeolites},
\newblock \bibinfo{journal}{Journal of Fuel Chemistry and Technology} \bibinfo{volume}{49} (\bibinfo{year}{2021}) \bibinfo{pages}{1531--1539}. \DOIprefix\doi{10.1016/S1872-5813(21)60095-6}.
\bibitem[{Rouse(1953)}]{rouse_theory_1953}
\bibinfo{author}{Rouse P.~E., Jr.},
\newblock \bibinfo{title}{A {Theory} of the {Linear} {Viscoelastic} {Properties} of {Dilute} {Solutions} of {Coiling} {Polymers}},
\newblock \bibinfo{journal}{The Journal of Chemical Physics} \bibinfo{volume}{21} (\bibinfo{year}{1953}) \bibinfo{pages}{1272--1280}. \DOIprefix\doi{10.1063/1.1699180}.
\bibitem[{Eyring(1935)}]{eyring_activated_1935}
\bibinfo{author}{Eyring H.},
\newblock \bibinfo{title}{The {Activated} {Complex} and the {Absolute} {Rate} of {Chemical} {Reactions}.},
\newblock \bibinfo{journal}{Chemical Reviews} \bibinfo{volume}{17} (\bibinfo{year}{1935}) \bibinfo{pages}{65--77}. \DOIprefix\doi{10.1021/cr60056a006}.
\bibitem[{Hult et~al.(1999)Hult, Johansson, Malmström, Freire, Burchard, McLeish, and Milner}]{hult_advances_1999}
\bibinfo{author}{Hult A.}, \bibinfo{author}{Johansson M.}, \bibinfo{author}{Malmström E.}, \bibinfo{author}{Freire J.}, \bibinfo{author}{Burchard W.}, \bibinfo{author}{McLeish T.}, \bibinfo{author}{Milner S.}, \bibinfo{title}{Advances in {Polymer} {Science}}, Branched {Polymers} {II}, \bibinfo{publisher}{Springer}, \bibinfo{year}{1999}.
\bibitem[{Bell(1978)}]{bell_models_1978}
\bibinfo{author}{Bell G.~I.},
\newblock \bibinfo{title}{Models for the {Specific} {Adhesion} of {Cells} to {Cells}},
\newblock \bibinfo{journal}{Science} \bibinfo{volume}{200} (\bibinfo{year}{1978}) \bibinfo{pages}{618--627}. \DOIprefix\doi{10.1126/science.347575}.
\bibitem[{Song et~al.(2021)Song, Shen, Vernerey, and Cai}]{song_force-dependent_2021}
\bibinfo{author}{Song Z.}, \bibinfo{author}{Shen T.}, \bibinfo{author}{Vernerey F.~J.}, \bibinfo{author}{Cai S.},
\newblock \bibinfo{title}{Force-dependent bond dissociation explains the rate-dependent fracture of vitrimers},
\newblock \bibinfo{journal}{Soft Matter} \bibinfo{volume}{17} (\bibinfo{year}{2021}) \bibinfo{pages}{6669--6674}. \DOIprefix\doi{10.1039/D1SM00518A}.
\bibitem[{Buche and Silberstein(2021)}]{buche_chain_2021}
\bibinfo{author}{Buche M.~R.}, \bibinfo{author}{Silberstein M.~N.},
\newblock \bibinfo{title}{Chain breaking in the statistical mechanical constitutive theory of polymer networks},
\newblock \bibinfo{journal}{Journal of the Mechanics and Physics of Solids} \bibinfo{volume}{156} (\bibinfo{year}{2021}) \bibinfo{pages}{104593}. \DOIprefix\doi{10.1016/j.jmps.2021.104593}.
\bibitem[{Guo et~al.(2009)Guo, Lad, Ray, and Akhremitchev}]{guo_association_2009}
\bibinfo{author}{Guo S.}, \bibinfo{author}{Lad N.}, \bibinfo{author}{Ray C.}, \bibinfo{author}{Akhremitchev B.~B.},
\newblock \bibinfo{title}{Association {Kinetics} from {Single} {Molecule} {Force} {Spectroscopy} {Measurements}},
\newblock \bibinfo{journal}{Biophysical Journal} \bibinfo{volume}{96} (\bibinfo{year}{2009}) \bibinfo{pages}{3412--3422}. \DOIprefix\doi{10.1016/j.bpj.2009.01.031}.
\bibitem[{Bell and Terentjev(2017)}]{bell_kinetics_2017}
\bibinfo{author}{Bell S.}, \bibinfo{author}{Terentjev E.~M.},
\newblock \bibinfo{title}{Kinetics of {Tethered} {Ligands} {Binding} to a {Surface} {Receptor}},
\newblock \bibinfo{journal}{Macromolecules} \bibinfo{volume}{50} (\bibinfo{year}{2017}) \bibinfo{pages}{8810--8815}. \DOIprefix\doi{10.1021/acs.macromol.7b01742}.
\bibitem[{Marzocca et~al.(2013)Marzocca, Salgueiro, and Somoza}]{marzocca_physical_2013}
\bibinfo{author}{Marzocca A.~J.}, \bibinfo{author}{Salgueiro W.}, \bibinfo{author}{Somoza A.},
\newblock \bibinfo{title}{Physical {Phenomena} {Related} to {Free} {Volumes} in {Rubber} and {Blends}},
\newblock in: \bibinfo{editor}{Visakh P.~M.}, \bibinfo{editor}{Thomas S.}, \bibinfo{editor}{Chandra A.~K.}, \bibinfo{editor}{Mathew A.~P.} (Eds.), \bibinfo{booktitle}{Advances in {Elastomers} {II}: {Composites} and {Nanocomposites}}, Advanced {Structured} {Materials}, \bibinfo{publisher}{Springer}, \bibinfo{address}{Berlin, Heidelberg}, \bibinfo{year}{2013}, pp. \bibinfo{pages}{399--426}.
\bibitem[{Qi et~al.(2003)Qi, Joyce, and Boyce}]{qi_durometer_2003}
\bibinfo{author}{Qi~H.~J.}, \bibinfo{author}{Joyce K.}, \bibinfo{author}{Boyce M.~C.},
\newblock \bibinfo{title}{Durometer {Hardness} and the {Stress}-{Strain} {Behavior} of {Elastomeric} {Materials}},
\newblock \bibinfo{journal}{Rubber Chemistry and Technology} \bibinfo{volume}{76} (\bibinfo{year}{2003}) \bibinfo{pages}{419--435}. \DOIprefix\doi{10.5254/1.3547752}.
\bibitem[{Vatankhah-Varnosfaderani et~al.(2017)Vatankhah-Varnosfaderani, Daniel, Everhart, Pandya, Liang, Matyjaszewski, Dobrynin, and Sheiko}]{vatankhah-varnosfaderani_mimicking_2017}
\bibinfo{author}{Vatankhah-Varnosfaderani M.}, \bibinfo{author}{Daniel W.~F.~M.}, \bibinfo{author}{Everhart M.~H.}, \bibinfo{author}{Pandya A.~A.}, \bibinfo{author}{Liang H.}, \bibinfo{author}{Matyjaszewski K.}, \bibinfo{author}{Dobrynin A.~V.}, \bibinfo{author}{Sheiko S.~S.},
\newblock \bibinfo{title}{Mimicking biological stress–strain behaviour with synthetic elastomers},
\newblock \bibinfo{journal}{Nature} \bibinfo{volume}{549} (\bibinfo{year}{2017}) \bibinfo{pages}{497--501}. \DOIprefix\doi{10.1038/nature23673}.
\bibitem[{Shibayama et~al.(2019)Shibayama, Li, and Sakai}]{shibayama_precision_2019}
\bibinfo{author}{Shibayama M.}, \bibinfo{author}{Li~X.}, \bibinfo{author}{Sakai T.},
\newblock \bibinfo{title}{Precision polymer network science with tetra-{PEG} gels—a decade history and future},
\newblock \bibinfo{journal}{Colloid and Polymer Science} \bibinfo{volume}{297} (\bibinfo{year}{2019}) \bibinfo{pages}{1--12}. \DOIprefix\doi{10.1007/s00396-018-4423-7}.
\bibitem[{You et~al.(2024)You, Zheng, Li, and Lam}]{you_model_2024}
\bibinfo{author}{You H.}, \bibinfo{author}{Zheng S.}, \bibinfo{author}{Li~H.}, \bibinfo{author}{Lam K.~Y.},
\newblock \bibinfo{title}{A model with contact maps at both polymer chain and network scales for tough hydrogels with chain entanglement, hidden length and unconventional network topology},
\newblock \bibinfo{journal}{International Journal of Mechanical Sciences} \bibinfo{volume}{262} (\bibinfo{year}{2024}) \bibinfo{pages}{108713}. \DOIprefix\doi{10.1016/j.ijmecsci.2023.108713}.
\bibitem[{Herzberg(1955)}]{herzberg_molecular_1955}
\bibinfo{author}{Herzberg G.},
\newblock \bibinfo{title}{Molecular {Vibrations}. {The} theory of infrared and {Raman} vibrational spectra. {E}. {Bright} {Wilson}, {Jr}., {J}. {C}. {Decius}, and {Paul} {C}. {Cross}. {McGraw}-{Hill}, {New} {York}-{London}, 1955. xi + 371 pp. {Illus}. \$8.50.},
\newblock \bibinfo{journal}{Science} \bibinfo{volume}{122} (\bibinfo{year}{1955}) \bibinfo{pages}{422--422}. \DOIprefix\doi{10.1126/science.122.3166.422.a}.
\bibitem[{Yu et~al.(2014)Yu, Ge, and Qi}]{yu_effects_2014}
\bibinfo{author}{Yu~K.}, \bibinfo{author}{Ge~Q.}, \bibinfo{author}{Qi~H.~J.},
\newblock \bibinfo{title}{Effects of stretch induced softening to the free recovery behavior of shape memory polymer composites},
\newblock \bibinfo{journal}{Polymer} \bibinfo{volume}{55} (\bibinfo{year}{2014}) \bibinfo{pages}{5938--5947}. \DOIprefix\doi{10.1016/j.polymer.2014.06.050}.
\bibitem[{Wanasinghe et~al.(2022)Wanasinghe, Dodo, and Konkolewicz}]{wanasinghe_dynamic_2022}
\bibinfo{author}{Wanasinghe S.~V.}, \bibinfo{author}{Dodo O.~J.}, \bibinfo{author}{Konkolewicz D.},
\newblock \bibinfo{title}{Dynamic {Bonds}: {Adaptable} {Timescales} for {Responsive} {Materials}},
\newblock \bibinfo{journal}{Angewandte Chemie International Edition} \bibinfo{volume}{61} (\bibinfo{year}{2022}) \bibinfo{pages}{e202206938}. \DOIprefix\doi{10.1002/anie.202206938}.
\bibitem[{Chen et~al.(2019)Chen, Zhou, Wu, Zhao, and Zhang}]{chen_rapid_2019}
\bibinfo{author}{Chen M.}, \bibinfo{author}{Zhou L.}, \bibinfo{author}{Wu~Y.}, \bibinfo{author}{Zhao X.}, \bibinfo{author}{Zhang Y.},
\newblock \bibinfo{title}{Rapid {Stress} {Relaxation} and {Moderate} {Temperature} of {Malleability} {Enabled} by the {Synergy} of {Disulfide} {Metathesis} and {Carboxylate} {Transesterification} in {Epoxy} {Vitrimers}},
\newblock \bibinfo{journal}{ACS Macro Letters} \bibinfo{volume}{8} (\bibinfo{year}{2019}) \bibinfo{pages}{255--260}. \DOIprefix\doi{10.1021/acsmacrolett.9b00015}.
\bibitem[{Puthur and Sebastian(2002)}]{puthur_theory_2002}
\bibinfo{author}{Puthur R.}, \bibinfo{author}{Sebastian K.~L.},
\newblock \bibinfo{title}{Theory of polymer breaking under tension},
\newblock \bibinfo{journal}{Physical Review B} \bibinfo{volume}{66} (\bibinfo{year}{2002}) \bibinfo{pages}{024304}. \DOIprefix\doi{10.1103/PhysRevB.66.024304}.
\bibitem[{Lamont et~al.(2021)Lamont, Mulderrig, Bouklas, and Vernerey}]{lamont_rate-dependent_2021}
\bibinfo{author}{Lamont S.~C.}, \bibinfo{author}{Mulderrig J.}, \bibinfo{author}{Bouklas N.}, \bibinfo{author}{Vernerey F.~J.},
\newblock \bibinfo{title}{Rate-{Dependent} {Damage} {Mechanics} of {Polymer} {Networks} with {Reversible} {Bonds}},
\newblock \bibinfo{journal}{Macromolecules} \bibinfo{volume}{54} (\bibinfo{year}{2021}) \bibinfo{pages}{10801--10813}. \DOIprefix\doi{10.1021/acs.macromol.1c01943}.
\bibitem[{Buche et~al.(2022)Buche, Silberstein, and Grutzik}]{buche_freely_2022}
\bibinfo{author}{Buche M.~R.}, \bibinfo{author}{Silberstein M.~N.}, \bibinfo{author}{Grutzik S.~J.},
\newblock \bibinfo{title}{Freely jointed chain models with extensible links},
\newblock \bibinfo{journal}{Physical Review E} \bibinfo{volume}{106} (\bibinfo{year}{2022}) \bibinfo{pages}{024502}. \DOIprefix\doi{10.1103/PhysRevE.106.024502}.
\bibitem[{Mulderrig et~al.(2023)Mulderrig, Talamini, and Bouklas}]{mulderrig_statistical_2023}
\bibinfo{author}{Mulderrig J.}, \bibinfo{author}{Talamini B.}, \bibinfo{author}{Bouklas N.},
\newblock \bibinfo{title}{A statistical mechanics framework for polymer chain scission, based on the concepts of distorted bond potential and asymptotic matching},
\newblock \bibinfo{journal}{Journal of the Mechanics and Physics of Solids}  (\bibinfo{year}{2023}) \bibinfo{pages}{105244}.
\bibitem[{Treloar(1943)}]{treloar_elasticity_1943}
\bibinfo{author}{Treloar L.~R.~G.},
\newblock \bibinfo{title}{The elasticity of a network of long-chain molecules—{II}},
\newblock \bibinfo{journal}{Transactions of the Faraday Society} \bibinfo{volume}{39} (\bibinfo{year}{1943}) \bibinfo{pages}{241--246}. \DOIprefix\doi{10.1039/TF9433900241}.
\bibitem[{Okamoto et~al.(2011)Okamoto, Clayton, and Bayly}]{okamoto_viscoelastic_2011}
\bibinfo{author}{Okamoto R.~J.}, \bibinfo{author}{Clayton E.~H.}, \bibinfo{author}{Bayly P.~V.},
\newblock \bibinfo{title}{Viscoelastic properties of soft gels: comparison of magnetic resonance elastography and dynamic shear testing in the shear wave regime},
\newblock \bibinfo{journal}{Physics in Medicine \& Biology} \bibinfo{volume}{56} (\bibinfo{year}{2011}) \bibinfo{pages}{6379}. \DOIprefix\doi{10.1088/0031-9155/56/19/014}.
\bibitem[{Chen et~al.(2015)Chen, Huang, Weiss, and Colby}]{chen_viscoelasticity_2015}
\bibinfo{author}{Chen Q.}, \bibinfo{author}{Huang C.}, \bibinfo{author}{Weiss R.~A.}, \bibinfo{author}{Colby R.~H.},
\newblock \bibinfo{title}{Viscoelasticity of {Reversible} {Gelation} for {Ionomers}},
\newblock \bibinfo{journal}{Macromolecules} \bibinfo{volume}{48} (\bibinfo{year}{2015}) \bibinfo{pages}{1221--1230}. \DOIprefix\doi{10.1021/ma502280g}.
\bibitem[{Shabbir et~al.(2016)Shabbir, Javakhishvili, Cerveny, Hvilsted, Skov, Hassager, and Alvarez}]{shabbir_linear_2016}
\bibinfo{author}{Shabbir A.}, \bibinfo{author}{Javakhishvili I.}, \bibinfo{author}{Cerveny S.}, \bibinfo{author}{Hvilsted S.}, \bibinfo{author}{Skov A.~L.}, \bibinfo{author}{Hassager O.}, \bibinfo{author}{Alvarez N.~J.},
\newblock \bibinfo{title}{Linear {Viscoelastic} and {Dielectric} {Relaxation} {Response} of {Unentangled} {UPy}-{Based} {Supramolecular} {Networks}},
\newblock \bibinfo{journal}{Macromolecules} \bibinfo{volume}{49} (\bibinfo{year}{2016}) \bibinfo{pages}{3899--3910}. \DOIprefix\doi{10.1021/acs.macromol.6b00122}.
\bibitem[{Peter et~al.(2021)Peter, Deal, Zhao, He, Tang, and Lemmon}]{peter_rheological_2021}
\bibinfo{author}{Peter T.}, \bibinfo{author}{Deal H.}, \bibinfo{author}{Zhao H.}, \bibinfo{author}{He~A.}, \bibinfo{author}{Tang C.}, \bibinfo{author}{Lemmon C.},
\newblock \bibinfo{title}{Rheological characterization of poly-dimethyl siloxane formulations with tunable viscoelastic properties},
\newblock \bibinfo{journal}{RSC Advances}  (\bibinfo{year}{2021}). \DOIprefix\doi{doi.org/10.1039/D1RA03548G}.
\bibitem[{Colby et~al.(1998)Colby, Zheng, Rafailovich, Sokolov, Peiffer, Schwarz, Strzhemechny, and Nguyen}]{colby_dynamics_1998}
\bibinfo{author}{Colby R.~H.}, \bibinfo{author}{Zheng X.}, \bibinfo{author}{Rafailovich M.~H.}, \bibinfo{author}{Sokolov J.}, \bibinfo{author}{Peiffer D.~G.}, \bibinfo{author}{Schwarz S.~A.}, \bibinfo{author}{Strzhemechny Y.}, \bibinfo{author}{Nguyen D.},
\newblock \bibinfo{title}{Dynamics of {Lightly} {Sulfonated} {Polystyrene} {Ionomers}},
\newblock \bibinfo{journal}{Physical Review Letters} \bibinfo{volume}{81} (\bibinfo{year}{1998}) \bibinfo{pages}{3876--3879}. \DOIprefix\doi{10.1103/PhysRevLett.81.3876}.
\bibitem[{Chen et~al.(2013)Chen, Liang, Shiau, and Colby}]{chen_linear_2013}
\bibinfo{author}{Chen Q.}, \bibinfo{author}{Liang S.}, \bibinfo{author}{Shiau H.-s.}, \bibinfo{author}{Colby R.~H.},
\newblock \bibinfo{title}{Linear {Viscoelastic} and {Dielectric} {Properties} of {Phosphonium} {Siloxane} {Ionomers}},
\newblock \bibinfo{journal}{ACS Macro Letters} \bibinfo{volume}{2} (\bibinfo{year}{2013}) \bibinfo{pages}{970--974}. \DOIprefix\doi{10.1021/mz400476w}.
\bibitem[{Xie et~al.(2024)Xie, Wang, Zhang, Cao, Tang, and Xu}]{xie_length_2024}
\bibinfo{author}{Xie M.-J.}, \bibinfo{author}{Wang C.-C.}, \bibinfo{author}{Zhang R.}, \bibinfo{author}{Cao J.}, \bibinfo{author}{Tang M.-Z.}, \bibinfo{author}{Xu~Y.-X.},
\newblock \bibinfo{title}{Length effect of crosslinkers on the mechanical properties and dimensional stability of vitrimer elastomers with inhomogeneous networks},
\newblock \bibinfo{journal}{Polymer} \bibinfo{volume}{290} (\bibinfo{year}{2024}) \bibinfo{pages}{126550}. \DOIprefix\doi{10.1016/j.polymer.2023.126550}.
\bibitem[{Zimm et~al.(1953)Zimm, Stockmayer, and Fixman}]{zimm_excluded_1953}
\bibinfo{author}{Zimm B.~H.}, \bibinfo{author}{Stockmayer W.~H.}, \bibinfo{author}{Fixman M.},
\newblock \bibinfo{title}{Excluded {Volume} in {Polymer} {Chains}},
\newblock \bibinfo{journal}{The Journal of Chemical Physics} \bibinfo{volume}{21} (\bibinfo{year}{1953}) \bibinfo{pages}{1716--1723}. \DOIprefix\doi{10.1063/1.1698650}.
\bibitem[{Sun and Faller(2006)}]{sun_crossover_2006}
\bibinfo{author}{Sun Q.}, \bibinfo{author}{Faller R.},
\newblock \bibinfo{title}{Crossover from {Unentangled} to {Entangled} {Dynamics} in a {Systematically} {Coarse}-{Grained} {Polystyrene} {Melt}},
\newblock \bibinfo{journal}{Macromolecules} \bibinfo{volume}{39} (\bibinfo{year}{2006}) \bibinfo{pages}{812--820}. \DOIprefix\doi{10.1021/ma0514774}.
\bibitem[{Ge et~al.(2013)Ge, Pierce, Perahia, Grest, and Robbins}]{ge_molecular_2013}
\bibinfo{author}{Ge~T.}, \bibinfo{author}{Pierce F.}, \bibinfo{author}{Perahia D.}, \bibinfo{author}{Grest G.~S.}, \bibinfo{author}{Robbins M.~O.},
\newblock \bibinfo{title}{Molecular {Dynamics} {Simulations} of {Polymer} {Welding}: {Strength} from {Interfacial} {Entanglements}},
\newblock \bibinfo{journal}{Physical Review Letters} \bibinfo{volume}{110} (\bibinfo{year}{2013}) \bibinfo{pages}{098301}. \DOIprefix\doi{10.1103/PhysRevLett.110.098301}.
\bibitem[{Schieber and Andreev(2014)}]{schieber_entangled_2014}
\bibinfo{author}{Schieber J.~D.}, \bibinfo{author}{Andreev M.},
\newblock \bibinfo{title}{Entangled {Polymer} {Dynamics} in {Equilibrium} and {Flow} {Modeled} {Through} {Slip} {Links}},
\newblock \bibinfo{journal}{Annual Review of Chemical and Biomolecular Engineering} \bibinfo{volume}{5} (\bibinfo{year}{2014}) \bibinfo{pages}{367--381}. \DOIprefix\doi{10.1146/annurev-chembioeng-060713-040252}.
\bibitem[{Masubuchi(2014)}]{masubuchi_simulating_2014}
\bibinfo{author}{Masubuchi Y.},
\newblock \bibinfo{title}{Simulating the {Flow} of {Entangled} {Polymers}},
\newblock \bibinfo{journal}{Annual Review of Chemical and Biomolecular Engineering} \bibinfo{volume}{5} (\bibinfo{year}{2014}) \bibinfo{pages}{11--33}. \DOIprefix\doi{10.1146/annurev-chembioeng-060713-040401}.
\bibitem[{Ge et~al.(2018)Ge, Grest, and Rubinstein}]{ge_nanorheology_2018}
\bibinfo{author}{Ge~T.}, \bibinfo{author}{Grest G.~S.}, \bibinfo{author}{Rubinstein M.},
\newblock \bibinfo{title}{Nanorheology of {Entangled} {Polymer} {Melts}},
\newblock \bibinfo{journal}{Physical Review Letters} \bibinfo{volume}{120} (\bibinfo{year}{2018}) \bibinfo{pages}{057801}. \DOIprefix\doi{10.1103/PhysRevLett.120.057801}.
\bibitem[{Kim et~al.(2021)Kim, Zhang, Shi, and Suo}]{kim_fracture_2021}
\bibinfo{author}{Kim J.}, \bibinfo{author}{Zhang G.}, \bibinfo{author}{Shi M.}, \bibinfo{author}{Suo Z.},
\newblock \bibinfo{title}{Fracture, fatigue, and friction of polymers in which entanglements greatly outnumber cross-links},
\newblock \bibinfo{journal}{Science} \bibinfo{volume}{374} (\bibinfo{year}{2021}) \bibinfo{pages}{212--216}. \DOIprefix\doi{10.1126/science.abg6320}.
\bibitem[{Steck et~al.(2023)Steck, Kim, Kutsovsky, and Suo}]{steck_multiscale_2023}
\bibinfo{author}{Steck J.}, \bibinfo{author}{Kim J.}, \bibinfo{author}{Kutsovsky Y.}, \bibinfo{author}{Suo Z.},
\newblock \bibinfo{title}{Multiscale stress deconcentration amplifies fatigue resistance of rubber},
\newblock \bibinfo{journal}{Nature} \bibinfo{volume}{624} (\bibinfo{year}{2023}) \bibinfo{pages}{303--308}. \DOIprefix\doi{10.1038/s41586-023-06782-2}.
\bibitem[{Shi et~al.(2023)Shi, Kim, Nian, and Suo}]{shi_highly_2023}
\bibinfo{author}{Shi M.}, \bibinfo{author}{Kim J.}, \bibinfo{author}{Nian G.}, \bibinfo{author}{Suo Z.},
\newblock \bibinfo{title}{Highly entangled hydrogels with degradable crosslinks},
\newblock \bibinfo{journal}{Extreme Mechanics Letters} \bibinfo{volume}{59} (\bibinfo{year}{2023}) \bibinfo{pages}{101953}. \DOIprefix\doi{10.1016/j.eml.2022.101953}.
\bibitem[{Ahlawat et~al.(2021)Ahlawat, Rajput, and Patil}]{ahlawat_elasticity_2021}
\bibinfo{author}{Ahlawat V.}, \bibinfo{author}{Rajput S.~S.}, \bibinfo{author}{Patil S.},
\newblock \bibinfo{title}{Elasticity of single flexible polymer chains in good and poor solvents},
\newblock \bibinfo{journal}{Polymer} \bibinfo{volume}{230} (\bibinfo{year}{2021}) \bibinfo{pages}{124031}. \DOIprefix\doi{10.1016/j.polymer.2021.124031}.
\bibitem[{Liese et~al.(2017)Liese, Gensler, Krysiak, Schwarzl, Achazi, Paulus, Hugel, Rabe, and Netz}]{liese_hydration_2017}
\bibinfo{author}{Liese S.}, \bibinfo{author}{Gensler M.}, \bibinfo{author}{Krysiak S.}, \bibinfo{author}{Schwarzl R.}, \bibinfo{author}{Achazi A.}, \bibinfo{author}{Paulus B.}, \bibinfo{author}{Hugel T.}, \bibinfo{author}{Rabe J.~P.}, \bibinfo{author}{Netz R.~R.},
\newblock \bibinfo{title}{Hydration {Effects} {Turn} a {Highly} {Stretched} {Polymer} from an {Entropic} into an {Energetic} {Spring}},
\newblock \bibinfo{journal}{ACS Nano} \bibinfo{volume}{11} (\bibinfo{year}{2017}) \bibinfo{pages}{702--712}. \DOIprefix\doi{10.1021/acsnano.6b07071}.
\bibitem[{Lee et~al.(2008)Lee, Venable, MacKerell, and Pastor}]{lee_molecular_2008}
\bibinfo{author}{Lee H.}, \bibinfo{author}{Venable R.~M.}, \bibinfo{author}{MacKerell A.~D.}, \bibinfo{author}{Pastor R.~W.},
\newblock \bibinfo{title}{Molecular {Dynamics} {Studies} of {Polyethylene} {Oxide} and {Polyethylene} {Glycol}: {Hydrodynamic} {Radius} and {Shape} {Anisotropy}},
\newblock \bibinfo{journal}{Biophysical Journal} \bibinfo{volume}{95} (\bibinfo{year}{2008}) \bibinfo{pages}{1590--1599}. \DOIprefix\doi{10.1529/biophysj.108.133025}.
\bibitem[{Consolati et~al.(2023)Consolati, Nichetti, and Quasso}]{consolati_probing_2023}
\bibinfo{author}{Consolati G.}, \bibinfo{author}{Nichetti D.}, \bibinfo{author}{Quasso F.},
\newblock \bibinfo{title}{Probing the {Free} {Volume} in {Polymers} by {Means} of {Positron} {Annihilation} {Lifetime} {Spectroscopy}},
\newblock \bibinfo{journal}{Polymers} \bibinfo{volume}{15} (\bibinfo{year}{2023}) \bibinfo{pages}{3128}. \DOIprefix\doi{10.3390/polym15143128}.
\bibitem[{Tao et~al.(2023)Tao, He, Arbaugh, McCutcheon, and Li}]{tao_machine_2023}
\bibinfo{author}{Tao L.}, \bibinfo{author}{He~J.}, \bibinfo{author}{Arbaugh T.}, \bibinfo{author}{McCutcheon J.~R.}, \bibinfo{author}{Li~Y.},
\newblock \bibinfo{title}{Machine learning prediction on the fractional free volume of polymer membranes},
\newblock \bibinfo{journal}{Journal of Membrane Science} \bibinfo{volume}{665} (\bibinfo{year}{2023}) \bibinfo{pages}{121131}. \DOIprefix\doi{10.1016/j.memsci.2022.121131}.
\bibitem[{Boerner et~al.(2023)Boerner, Deems, Furlani, Knuth, and Towns}]{boerner_access_2023}
\bibinfo{author}{Boerner T.~J.}, \bibinfo{author}{Deems S.}, \bibinfo{author}{Furlani T.~R.}, \bibinfo{author}{Knuth S.~L.}, \bibinfo{author}{Towns J.},
\newblock \bibinfo{title}{{ACCESS}: {Advancing} {Innovation}: {NSF}’s {Advanced} {Cyberinfrastructure} {Coordination} {Ecosystem}: {Services} \& {Support}},
\newblock in: \bibinfo{booktitle}{Practice and {Experience} in {Advanced} {Research} {Computing}}, {PEARC} '23, \bibinfo{publisher}{Association for Computing Machinery}, \bibinfo{address}{New York, NY, USA}, \bibinfo{year}{2023}, pp. \bibinfo{pages}{173--176}. \DOIprefix\doi{10.1145/3569951.3597559}.

\end{thebibliography}
\bibliographystyle{elsarticle-num-names} 

\end{document}